\documentclass[%
 reprint,
superscriptaddress,
nofootinbib,
bibnotes,
 amsmath,amssymb,
 aps,
]{revtex4-1}

\usepackage{latexsym,mathrsfs,bbm}
\usepackage{tikz}
\usetikzlibrary{patterns,patterns.meta} 
\usetikzlibrary{backgrounds,fit,decorations.pathreplacing,calc}
\usepackage{graphicx}
\usepackage{dcolumn}
\usepackage{braket}
\usepackage{gensymb}
\usepackage{bm}
\usepackage{amsmath}
\usepackage{amsthm}
\usepackage{float}
\usepackage{cancel}
\usepackage{mathtools}
\usepackage{color}
\usepackage{comment}
\usepackage{stmaryrd}
\usepackage[utf8]{inputenc}
\usepackage{csquotes}

\usepackage{framed} 
\usepackage{thmtools}
\usepackage[normalem]{ulem}

\UseRawInputEncoding

\usepackage{diagbox}
\usepackage{adjustbox}

\theoremstyle{plain}

\DeclareMathOperator{\Tr}{Tr}
\newcommand{\Te}{{\theta_{\mathrm{e}}}}

\newcommand{\pps}{p_{\theta}^{\mathrm{PS}}}

\newcommand{\Var}{\mathrm{Var}}
\newcommand*{\ketbra}[2]{\lvert #1 \rangle\!\langle #2 \rvert}

\newcommand{\Dave}[1]{\textcolor{black}{#1} }

\newcommand{\Stephan}[1]{\textcolor{black}{#1} }

\newcommand{\nicole}[1]{\textcolor{black}{#1} }

\newcommand{\cket}[1]{|#1\rangle}
\newcommand{\bracket}[2]{\langle #1|#2\rangle}

\newcommand{\Acal}{{\mathcal A}}
\newcommand{\Bcal}{{\mathcal B}}

\newcommand{\EcalKDC}{{\mathcal E}_{\mathrm{KD+}}}
\newcommand{\EcalKDCpu}{{\mathcal E}_{\mathrm{KD+}}^{\mathrm{pure}}}
\newcommand{\EcalKDCext}{{\mathcal E}_{\mathrm{KD+}}^{\mathrm{ext}}}

\newcommand{\Hcal}{{\mathcal H}}

\newcommand{\PiAcal}{{\hat{\Pi}_{\Acal}}}
\newcommand{\PiBcal}{{\hat{\Pi}_{\Bcal}}}

\newcommand{\R}{\mathbb{R}}

\newcommand{\nab}{n_{\Acal,\Bcal}}

\newcommand{\na}{n_{\Acal}}
\newcommand{\nb}{n_{\Bcal}}
\newcommand{\mab}{m_{\Acal,\Bcal}}
\newcommand{\Mab}{M_{\Acal, \Bcal}}

\newcommand{\conv}[1]{\mathrm{conv}\left(#1\right)}

\renewcommand{\Re}[1]{\mathrm{Re}\left(#1\right)}
\renewcommand{\Im}[1]{\mathrm{Im}#1}

\newcommand{\convAB}{\conv{\Acal \cup \Bcal}}
\renewcommand{\epsilon}{\varepsilon}

\newtheorem{Theorem}{Theorem}[section]

\newtheorem{Lemma}[Theorem]{Lemma}

\def\la{\langle}
\def\ra{\rangle}

\def\be{\begin{equation}}
\def\ee{\end{equation}}

\newcommand{\im}{\mathrm{Im}}
\newcommand{\Pt}{\partial_{\theta}}
\newcommand{\PsiT}{\Psi_{\theta}}
\newcommand{\PsT}{\Psi_{\theta}^{\mathrm{PS}}}

\newcommand{\LParen}{ \bm{(} }
\newcommand{\RParen}{ \bm{)} }

\usepackage{hyperref}

\allowdisplaybreaks

\begin{document}

\preprint{APS/123-QED}


\title{Properties and Applications of the Kirkwood--Dirac Distribution}

\author{David R.  M.  Arvidsson-Shukur}
\thanks{\Dave{Author to whom any correspondence should be addressed: drma2@cam.ac.uk.}}
\affiliation{
Hitachi Cambridge Laboratory, J. J. Thomson Avenue,  Cambridge,  CB3 0HE,  United Kingdom
}

\author{William F. Braasch, Jr.}
\affiliation{Joint Center for Quantum Information and Computer Science, NIST and University of Maryland, College Park, MD 20742, USA}

\author{Stephan De Bi\`evre}
\affiliation{
Univ. Lille, CNRS, Inria, UMR 8524, Laboratoire Paul Painlev\'e, F-59000 Lille, France
}

\author{Justin Dressel}
\affiliation{Institute for Quantum Studies, Chapman University, Orange, CA 92866, USA}
\affiliation{Schmid College of Science and Technology, Chapman University, Orange, CA 92866, USA}

\author{Andrew N. Jordan}
\affiliation{The Kennedy Chair in Physics, Chapman University, Orange, CA 92866, USA}
\affiliation{Institute for Quantum Studies, Chapman University, Orange, CA 92866, USA}
\affiliation{Schmid College of Science and Technology, Chapman University, Orange, CA 92866, USA}
\affiliation{Department of Physics and Astronomy, University of Rochester, Rochester, NY 14627, USA}

\author{Christopher Langrenez}
\affiliation{
Univ. Lille, CNRS, Inria, UMR 8524, Laboratoire Paul Painlev\'e, F-59000 Lille, France
}

\author{Matteo Lostaglio}
\affiliation{
PsiQuantum, 700 Hansen Way, Palo Alto, CA 94304, USA
}

\author{Jeff S.  Lundeen}
\affiliation{Nexus for Quantum Technologies, University of Ottawa, Department of
Physics, Ottawa, Canada }

\author{Nicole Yunger Halpern}
\affiliation{Joint Center for Quantum Information and Computer Science, NIST and University of Maryland, College Park, MD 20742, USA}
\affiliation{Institute for Physical Science and Technology, University of Maryland, College Park, MD 20742, USA}

\date{\today}

\begin{abstract}
There are several mathematical formulations of quantum mechanics. The Schr\"odinger picture expresses 
quantum states in terms of wavefunctions over, e.g., position or momentum.
Alternatively, phase-space formulations represent 
states with quasi-probability distributions over{, e.g.,} position and momentum. 
A quasi-probability distribution resembles 
a probability distribution but may have negative and non-real entries. The most famous quasi-probability distribution, the Wigner function, has played a pivotal role in the development of a continuous-variable quantum theory
that has clear analogues of position and momentum. However, the Wigner function is ill-suited for much modern quantum-information research, which is focused on finite-dimensional systems and general observables.
Instead, recent years have  seen the Kirkwood--Dirac (KD) distribution come to the forefront as a powerful quasi-probability distribution for analysing quantum mechanics. The KD distribution allows tools from statistics and probability theory to be applied to problems in quantum-information processing. A notable  difference to the Wigner function is that the KD distribution can represent a quantum state in terms of arbitrary observables. This  paper reviews the KD distribution, in three parts. First, we present definitions and basic properties of the KD distribution and its generalisations. Second, we summarise the KD distribution’s extensive usage in the study or development of measurement disturbance; quantum metrology; weak values; direct measurements of quantum states; quantum thermodynamics; quantum scrambling and out-of-time-ordered correlators; and the foundations of quantum mechanics, including Leggett--Garg inequalities, the consistent-histories interpretation and contextuality.  We emphasise connections between operational quantum advantages and negative or non-real KD quasi-probabilities.  
Third, we delve into the KD distribution's mathematical structure. We summarise the current knowledge regarding the geometry of KD-positive states (the states for which the KD distribution is a classical probability distribution), describe how to witness and quantify KD non-positivity, and outline relationships between KD non-positivity, coherence and observables’ incompatibility.
\end{abstract}

\maketitle

\onecolumngrid

\tableofcontents

\section{Introduction}

\begin{displayquote}
\textit{Negative energies and probabilities should not be considered as nonsense.
They are well-defined concepts mathematically, like a negative sum of money.}

-- Paul Dirac in 1942 ~\cite{Dirac42}.
\end{displayquote}

Phenomena such as non-commutation, coherence, and entanglement fundamentally distinguish quantum mechanics from classical physics. However, knowing exactly when an experiment  lacks a classical analogue is notoriously difficult.   \Dave{If, and only if,  
\nicole{joint probability distributions describe  a system's preparation, manipulation and measurement,}
a corresponding experiment can be \nicole{modeled} classically~\cite{Dirac26, Dirac26-2, Wigner32,  Dirac45, Cohen66, Hudson74, Srinivas75, Griffiths84, Hartle04, spekkens2008negativity, Allahverdyan15}.   }

Common strategies for pinpointing, understanding and developing  quantum phenomena rely on quasi-probability distributions.   A quasi-probability distribution enables the mathematical representation of a quantum state in terms of a joint distribution over the eigenvalues of 
possibly  
incompatible observables. Quasi-probability distributions satisfy some, but not all,  of Kolmogorov's axioms of joint probability functions~\cite{Kolmogorov33}. \Dave{Some quasi-probability distributions can \nicole{contain negative or even non-real elements.}
}
Such `anomalous' values enable a probability-like description of quantum experiments and often \Dave{herald} non-classical phenomena.

In 1932, Wigner discovered his eponymous function, which has become the best-known quasi-probability distribution~\cite{Wigner32,Ville48, moyal49, Wootters87}. 
Cousins of the Wigner function include the Glauber--Sudarshan ~\cite{Glauber63, Sudarshan63} and Husimi ~\cite{Husimi40} representations.
All three represent quantum states with  distributions over the eigenvalues of   conjugate observables that have continuous spectra,  such as position and momentum, or an electromagnetic field's real and imaginary components.  In many settings, the Wigner function assumes negative components when there is no classical description of the quantum state in a measurement scenario~\cite{Mari12, Booth22}.  Thus, the Wigner function has become a popular tool for studying quantum phenomena in continuous-variable systems---most notably in quantum optics~\cite{Mandel95}.

Most modern quantum-information research, particularly quantum computing, concerns discrete-variable systems (e.g., \nicole{qubit systems)}
and 
\nicole{diverse}
observables 
\nicole{(not merely position and momentum).}
The Wigner function is \Dave{defined 
\nicole{in terms of} two fixed conjugate observables such as position and momentum, }
\nicole{rather than} arbitrary observables.  \Dave{(The discrete Wigner function \cite{bjork2008discrete} is defined in terms of two fixed maximally non-commuting observables.)}
\nicole{When needing greater flexibility, one can employ the Kirkwood--Dirac (KD) distribution.} 
 Before we introduce the KD distribution and summarise its use cases, we briefly introduce its history.

In 1933, John Kirkwood was inspired by the then-one-year-old Wigner function. He suggested a phase-space methodology for calculating thermodynamic partition functions~\cite{Kirkwood33}. Doing so, he introduced the quasi-probability distribution now known as the KD distribution.   Shortly afterwards,  in 1937,  Terletsky  independently introduced the real part of this distribution~\cite{Terletsky37, Margenau61}.  
In 1945, Paul Dirac published an article that highlighted an `analogy between classical and quantum mechanics'~\cite{Dirac45}.   In this work, he independently rediscovered the KD distribution.  Dirac argued that the only difference between classical and quantum mechanics stems from non-commutation~\cite{Dirac26, Dirac26-2, Dirac45}. Furthermore, he showed how the KD distribution can be used to calculate expectation values of  functions of observables in a (quasi-)probabilistic framework~\cite{Dirac45}.

After its original construction, the KD distribution received little attention. It was rediscovered again in 1968, by Rihaczek, 
within a classical signal-processing framework~\cite{Rihaczek68}.  The KD distribution's real part, too, was rediscovered, by Margenau and Hill in 1961~\cite{Margenau61, Johansen04, Johansen04-2}. 
Not until  the recent development of  quantum information theory did the KD distribution  enjoy a substantial revival (e.g.,~\cite{Johansen04-2,NYH_jarzynski_2017,allahverdyan_nonequilibrium_2014}).  
It is now a \Dave{common} mathematical tool in quantum theory.  The reason relates to Kirkwood's and Dirac's original motivations: The KD distribution provides a statistically-inspired framework for quantum calculations that are burdensome within the Schr\"odinger and Heisenberg formalisms.  

Recent years have seen the KD distribution deployed to study or develop several areas of quantum mechanics.  This Article's purpose is to review these past results.   In quantum metrology (Sec.~\ref{Sec:Metrology}), non-real KD quasi-probabilities are essential for accessing unknown information encoded in quantum states.    Further, it is sometimes useful to distil metrological information from several quantum states into a few.  If an underlying KD distribution contains negative values, the rate of this distillation can exceed classical limits (Sec.~\ref{SubSec:PSMetrology}).  An example of such a technique is \textit{weak-value amplification} (Sec.~\ref{Sec:WeakValues}),  which can improve the signal-to-noise ratio of measurements of weak-coupling parameters.  Weak-value amplification achieves this improvement via the ability of KD negativity to boost  a pre- and post-selected `average' of an observable, such that the average lies outside the observable's spectrum.    As an alternative to tomography,  measurements of KD distributions (Sec.~\ref{Sec:StateMeas}) enable direct and/or efficient reconstructions of quantum states.  In quantum thermodynamics (Sec.~\ref{Sec:QThermo}),   \Dave{engines can operate in ways forbidden by classical theories} 
\nicole{when a certain KD distribution exhibits non-positivity.}
 Furthermore, classical thermodynamics entails probability distributions over the possible amounts of work or entropy exchanged during a stochastic process. A KD distribution replaces such probability distributions in quantum thermodynamics. In quantum chaos (Sec.~\ref{Sec:OTOC}), a generalised KD distribution \Dave{can signal} scrambling, the spreading of information about a local perturbation via many-body entanglement. 
Finally, the KD distribution surfaces also in foundational settings (Sec.~\ref{Sec:Foundations}). `Temporal Bell inequalities', or Leggett--Garg inequalities, 
can be violated only when underlying KD distributions are non-positive (Sec.~\ref{SubSec:LG}). Moreover, in the consistent-histories interpretation of quantum mechanics, cross-terms in a KD distribution  govern whether a classical phase-space history can be ascribed to a quantum state (Sec.~\ref{SubSec:Consistent}). A rigorous notion of non-classicality is quantum contextuality. Several proofs of experimental contextuality 
\nicole{rely on non-positive KD distributions  directly or indirectly}
(Sec.~\ref{SubSec:Context}).

Thanks to this growing interest, it has become crucial to
understand the mathematical properties of the KD distribution. Sufficient and necessary requirements for the Wigner function to be negative have been known since 1976~\cite{Hudson74}. However, such requirements for pure states' KD distributions were mapped out only recently~\cite{ArvidssonShukur2021,DeBievre2021, debievre2023}. The extension to mixed states is not yet fully understood~\cite{langrenez2023characterizing,ADBLSLT24}.  Section~\ref{Sec:NonPos} reviews mathematical results about positive and non-positive KD distributions. We summarise the current understanding of the geometry of KD-positive states. Also, we outline how to witness and quantify KD non-positivity and coherence.

\nicole{Different readers will benefit from using this review differently. Reading the entire paper, one gains a holistic view of the KD distribution. Reading individual sections in a modular fashion, one can target the distribution's most relevant applications and properties.
}
\nicole{We introduce the KD distribution and outline its basic properties before reviewing \Dave{use cases.}}

\section{Definition and basic properties}
\label{Sec:DefProp}

In this section, we 
define the KD distribution. The KD distribution is a versatile and adaptable 
\nicole{object that can assume multiple forms.}
For simplicity, we introduce,  first,  the most common form of the KD distribution. Then, we describe how to construct a general KD distribution. The KD distribution is applicable to discrete and continuous-variable systems. However, this review centres on discrete systems.   Indeed, the KD distribution has been applied most to discrete systems.  Nevertheless, 
\nicole{Sec.~\ref{SubSec:ContinuousKD} }
covers continuous-variable KD distributions, \Dave{as needed for our review of wavefunction measurements 
\nicole{(Sec.~\ref{Sec:StateMeas}).}} Section~\ref{Sec:NonPos} introduces further 
properties of the KD distribution.

\subsection{\Dave{Quasi-probability representations of quantum mechanics}}

\nicole{Here, we briefly overview
quasi-probability representations of quantum mechanics. 
(See~\cite{Ferrie2011} and references therein for an extensive discussion.)
Then, we define the KD representation of quantum mechanics and the KD quasi-probability distribution.  
Less mathematically inclined readers may skip to the next subsection.  
}  

\Dave{To introduce quasi-probability representations of quantum mechanics, we must first provide some definitions.}
  \nicole{Denote by $\mathcal H$ a $d$-dimensional Hilbert space, by $\hat{\rho}$ an arbitrary density operator defined on $\mathcal H$ and by $\hat C$ an arbitrary linear operator defined on $\mathcal H$.   Additionally, let us} choose a set $\Lambda$. 
  \nicole{What $\Lambda$ consists of depends on the specific quasi-probability representation, as illustrated across this review. Examples include tuples of incompatible observables' eigenvalues (Sec. \ref{SubSec:Standard}),  hidden variables (Sec. \ref{SubSec:Context}) and position--momentum phase space (Sec. \ref{SubSec:ContinuousKD} and Ref. \cite{Wigner32}). 
  (For \Dave{notational} simplicity, we assume $\Lambda$ to be finite from now on.)
  We associate to each density operator $\hat{\rho}$ a complex-valued function 
  $\lambda \to Q_\lambda({\hat\rho})$ on $\Lambda$,
  and to each linear operator $\hat C$ a function $\lambda \to T_\lambda({\hat C})$ on $\Lambda$, such that
\begin{equation}\label{eq:overlapformula}
\Tr (\hat C^\dagger \hat\rho)
=\sum_{\lambda} T_\lambda^*({\hat C}) \: Q_\lambda({\hat\rho}).
\end{equation}
The function $Q(\hat\rho): \lambda \to Q_\lambda({\hat\rho})$ is a \emph{quasi-probability distribution} (or quasi-probability density) for $\hat{\rho}$, and \Dave{$T(\hat C):\lambda\to T_\lambda({\hat C })$} a \emph{symbol} for $\hat C$. Together, the maps $\hat \rho\to Q({\hat\rho})$ and $\hat C\to T({\hat C})$ form a quasi-probability representation of quantum mechanics on $\mathcal H$. 
}

\nicole{Depending on the physical setup, quasi-probability representations may obey extra requirements.
One typically assumes that $\hat{\rho} \to Q({\hat\rho})$ is convex-linear and that $\hat C \to T({\hat C})$ is linear. Additionally, the identity operator's symbol is often a constant, equal to one, in $\lambda$:
\begin{equation}\label{eq:normalize1}
\Dave{ T_{\lambda}({\hat{\mathbbm{1}}})=1, \; \; \forall \lambda \in \Lambda . }
\end{equation}
This equality, with Eq. \eqref{eq:overlapformula}, implies 
\nicole{the quasiprobability distribution's normalisation:}
\begin{equation}\label{eq:normalize2}
\Tr \left( \hat{\rho} \right) = \sum_{\lambda} Q_\lambda({\hat{\rho}}) =1, \; \forall \hat{\rho} .
\end{equation}
An occasional requirement is the reality of every observable's symbol:
\begin{equation}\label{eq:adjointness}
T_{\lambda} ({\hat C^\dagger})=T^*_{\lambda}({\hat C}), \; \; \forall \lambda \in \Lambda .
\end{equation}
Several quasi-probability distributions \Dave{satisfy Eq. \eqref{eq:adjointness},}
as explained in~\cite{Ferrie2011}. KD distributions (defined below) can be non-real. Thus, they fall outside the framework in \cite{Ferrie2011}, which concerns real-valued quasi-probabilities.  \Dave{Nevertheless, much of the framework in \cite{Ferrie2011} can be straightforwardly extended to encompass also complex-valued quasi-probability distributions.} 
}

\nicole{Equation~\eqref{eq:overlapformula} forms the \Dave{quasi-probability representation's} central ingredient. Using it, one can \Dave{calculate} quantum expectation values by calculating averages of random variables $T({\hat C})$ 
with respect to complex quasi-probability measures $Q({\hat{\rho}})$.
\Dave{Thus, the quasi-probability distribution's role in expectation-value calculations} 
parallels a positive probability measure's \Dave{role in} classical mechanics. Here, however, the measures need not be positive; whence the terminology ``quasi-probability''.  \Dave{Moreover,} quasi-probability \Dave{representations} 
can be straightforwardly enriched to represent operations and evolutions~\cite{Bartlett2015, schmid2024}. 
}

\Stephan{We now define a quantum state's KD distribution  and show how it fits into the above-outlined framework.}   
We consider two orthonormal bases in $\mathcal{H}$:  $\{ \ket{a_i} \} $ and $ \{ \ket{b_j} \}$, wherein $i,j = 1, \ldots d$.  \nicole{Without loss of generality, we can view $\{ \ket{a_i} \} $ and $ \{ \ket{b_j} \}$ as eigenbases of Hermitian observables $\hat{A}$ and $\hat{B}$.}
For now, we assume that these observables are non-degenerate. They, and the identity operator, eigendecompose as
\begin{align}\label{eq:ABobs}
    \hat{A} = \sum_i a_i \,\ketbra{a_i}{a_i}, \quad 
    \hat{B} = \sum_j b_j\,\ketbra{b_j}{b_j} \quad \text{and} \quad \hat{\mathbbm{1}} = \sum_i \ketbra{a_i}{a_i} = \sum_j \ketbra{b_j}{b_j}.
\end{align} 
We write discrete sums here for simplicity. However, one can generalise the equations to continuous bases \Stephan{and/or infinite-dimensional Hilbert spaces:} One replaces the summations with integrals over
an appropriate measure (see Sec.~\ref{SubSec:ContinuousKD}).
We will suppose that $\braket{a_i|b_j} \neq 0$ for all $i,j$. One can then express an arbitrary operator $\hat{C}$ in terms of both bases:
\begin{align}\label{eq:reconstruction1}
    \hat{C} &= \left(
    \sum_i \ketbra{a_i}{a_i}\right) \hat{C}
    \left(\sum_j\ketbra{b_j}{b_j}\right) 
    = \sum_{i,j} \bra{a_i} \hat{C} \ket{b_j}\, \ketbra{a_i}{b_j} 
    = \sum_{i,j} \braket{b_j|a_i} \!\bra{a_i}\hat{C}\ket{b_j}\, \frac{\ketbra{a_i}{b_j}}{\braket{b_j|a_i}} 
    \equiv \sum_{i,j} C(a_i,b_j) \hat{G}_{a_i,b_j} \, .
\end{align}
The coefficients $C(a_i,b_j)$ and  normalised outer products $\hat{G}_{a_i,b_j}$ are 
\begin{align}
\label{Eq:Frame}
    C(a_i,b_j) \equiv \braket{b_j|a_i}\!\bra{a_i}\hat{C}\ket{b_j}  \quad \text{and} \quad  
    \hat{G}_{a_i,b_j} \equiv \frac{\ketbra{a_i}{b_j}}{\braket{b_j|a_i}} \, .
\end{align}
The set $\{\hat{G}_{a_i,b_j}\mid 1\leq i,j\leq d\}$ forms a  basis 
for the space of linear operators defined on $\mathcal H$. \Dave{However,  the $\hat{G}_{a_i, b_j}$ are generally not \Dave{orthogonal} projectors or self-adjoint. 
Nevertheless, $\{\hat{G}_{a_i,b_j} \}$ forms an orthogonal basis for the space of operators defined on $\mathcal{H}$, with respect to the Hilbert-Schmidt inner product.}
In terms of this basis, $\hat{C}$ has the  component representation $C(a_i,b_j)$.

\nicole{Suppose that the operator $\hat{C}$ is a quantum state $\hat{\rho}$. The components $C(a_i,b_j) $ form the \textit{standard} KD quasi-probability distribution of $\hat \rho$. We denote it by $Q(\hat{\rho})$:   
\begin{equation}
\label{Eq:KDStand}
   Q_{a_i,b_j} \left( \hat{\rho} \right) 
   \equiv  \braket{b_j | a_i } \bra{a_i} \hat{\rho} \ket{b_j} .
\end{equation}
\Dave{The KD symbol of an arbitrary linear operator $\hat{C}$ is}
\begin{equation}\label{eq:KDsymbol}
T_{a_i,b_j}(\hat C)
\equiv \frac{\langle a_i|\hat C|b_j\rangle}{\langle a_i|b_j\rangle} \, .
\end{equation}
If $\hat C$ is an observable, the $T_{a_i,b_j}(\hat C)$ are weak values (see Sec.~\ref{sec:wv}). Using Eqs.~\eqref{Eq:KDStand} and~\eqref{eq:KDsymbol}, one can verify Eq.~\eqref{eq:overlapformula}:
\nicole{ $\Tr(\hat C^\dagger\hat \rho)=\sum_{a_i,b_j} T_{a_i,b_j}(\hat C)^* Q_{a_i,b_j}(\hat \rho).$  }
Consequently, $(Q,T)$ forms a quasi-probability representation of quantum mechanics over the set $\Dave{ \Lambda=\{(a_i,b_j)\mid 1\leq i,j\leq d\}}$. We call this representation \Dave{the} \emph{standard KD representation}, which \Dave{depends}
on $\hat{A}$ and $\hat{B}$. This representation satisfies Eqs.~\eqref{eq:normalize1}--\eqref{eq:normalize2} but not Eq.~\eqref{eq:adjointness}.
}

\Dave{We now detail further the structure of the standard KD representation of quantum mechanics.}
\nicole{Define the projector $\hat\Pi_{a_i} = \ketbra{a_i}{a_i}$ and the projector product}
\nicole{ $\hat F_{a_i, b_j}  =  \hat\Pi_{a_i}\hat\Pi_{b_j} \, .$  }
The \Dave{ set  
$\{ \hat F_{a_i, b_j} \}$} forms an orthogonal basis
for the space of operators defined on $\Hcal$:
\nicole{ $\Tr \left( \hat F^\dagger_{a_{i'}, b_{j'}} \hat F_{a_i, b_j} \right)
=|\langle a_i|b_j\rangle|^2\delta_{i,i'}\delta_{j,j'}.$ } 
\nicole{This  basis is a particular instance of a \emph{frame}.
}
\nicole{Dual to $ \{ \hat F_{a_i, b_j} \}$ is the \emph{dual basis} $\{ \hat{G}_{a_i, b_j} \}$, since }
\nicole{ $\Tr \left( \hat{G}^\dagger_{a_i, b_j}\hat F_{a_{i'}, b_{j'}} \right) 
=\delta_{i,i'}\delta_{j,j'} \, .$  }
\nicole{The dual basis' elements have the form}
\nicole{ $\hat{G}_{a_i, b_j}
=  \hat \Pi_{a_i}\hat\Pi_{b_j}  /  |\langle a_i|b_j\rangle|^2 \, .$  }
\nicole{Concise expressions for $Q(\hat \rho)$ and $T(\hat C)$ follow:}
\begin{equation} 
Q_{a_i,b_j}({\hat\rho})  
=  \Tr \left(\hat F^\dagger_{a_i, b_j}\hat \rho  \right) , 
\quad \text{and} \quad 
T_{a_i,b_j}({\hat C})
=  \Tr \left(\hat{G}^\dagger_{a_i, b_j}\hat C  \right).
\end{equation}
\nicole{Now, we can express states $\hat{\rho}$ in terms of quasi-probabilities $Q(\hat{\rho})$
and observables $\hat{C}$ in terms of symbols $T(\hat{C})$.
The corresponding \emph{reconstruction formulae} are}
\begin{equation}\label{eq:reconstruction2}
   \hat \rho=\sum Q_{a_i,b_j}({\hat \rho}) \hat G_{a_i, b_j},
   \quad \text{and} \quad
   \hat C = \sum_{a_i,b_j} T_{a_i,b_j}({\hat C}) \hat F_{a_i, b_j} \, .
\end{equation}
\nicole{Reference~\cite{Ferrie2011} details the construction of quasi-probability representations using frames  and their duals.
In no quasi-probability representation are all $Q(\hat\rho)$ positive measures and all $T_{a_i,b_j}(\hat C)\geq 0$ for all $i,j$ and all $\hat C\geq 0$. For a more precise statement, see, e.g., Theorem 4 of~\cite{Ferrie2011}.
}

\nicole{
 In \cite{Dirac45},  Dirac
 \nicole{aimed to define} a function of non-commuting observables $\hat{A}$ and $\hat{B}$. The ``ordering problem'' of quantum mechanics motivated Dirac: The position operator $\hat x$ and momentum operator $\hat p$ do not commute. Therefore, how can one associate an operator $\hat F(\hat x,\hat p)$ to a function $f(x,p)$ of the classical position $x$ and momentum $p$?  That is, how does one ``quantise'' functions of classical observables? If $f$ is a polynomial, one can,
 e.g., left-order the variables: One would place all powers of $\Dave{\hat{x}}$ leftward of all powers of $\Dave{{\hat{p}}}$. An alternative is right-ordering. Wigner proposed a symmetric ordering, embodied in the  Wigner function~\cite{Wigner32}. Dirac 
\Dave{introduced} left-ordering and right-ordering \Dave{of} general \Dave{non-}commuting operators $\hat A$ and $\hat B$. 
\nicole{We can rephrase Dirac's proposal as follows.}
Consider non-commuting observables $\hat A$ and $\hat B$, as well as a function $h(a_i,b_j)$ of the eigenvalue pairs $(a_i,b_j)\in\R^2$. 
\nicole{Define the operator}
$\hat h(\hat A,\hat B)$ \nicole{as}
 \begin{equation}\label{eq:Dirac_functionAB}
     \hat h(\hat A,\hat B) 
     \nicole{\equiv}  \sum_{a_i,b_j} h(a_i, b_j) \hat F_{a_i, b_j} \, .
 \end{equation}
\nicole{This definition resembles}
the second equation in Eqs.~\eqref{eq:reconstruction2}. In Dirac's language, every operator $\hat C$ 
\nicole{is} a function $\hat h(\hat A,\hat B)$ of 
$\hat A$ and $\hat B$. The symbol $T(\hat C)$ plays the role of $h$:
\begin{equation}\label{eq:diracformula}
    \hat C=\hat h(\hat A,\hat B), 
    \quad \textrm{wherein} \quad  
    h(a_i, b_j)=T_{a_i,b_j}(\hat C).
\end{equation}
When $h(a_i, b_j)=f(a_i)g(b_j)$, \,
$\hat h(\hat A,\hat B)=\hat f(\hat A)\hat g(\hat B)$. In other words, Dirac's prescription amounts to an ordering ``$\hat A$ to the left of $\hat B$''. 
\nicole{One can order the operators oppositely instead: One replaces}
$\hat F_{a_i,b_j}$ by its adjoint, $\hat F^\dagger_{a_i,b_j}=\hat\Pi_{b_j}\hat \Pi_{a_i}$. 
}

\subsection{The standard KD distribution}
\label{SubSec:Standard}

\nicole{The previous subsection introduced the standard KD distribution $Q(\hat{\rho})$ for a quantum state  $\hat{\rho}$ with respect to bases $ \{ \ket{a_i} \} $ and $ \{ \ket{b_j} \}$. We can cast this distribution as a $d \times d$-dimensional matrix with entries}
\begin{equation}
\label{Eq:KDStand2}
   Q_{i,j} \left( \hat{\rho} \right) 
   \equiv  \braket{b_j | a_i } \bra{a_i} \hat{\rho} \ket{b_j} .
\end{equation}
\nicole{In the left-hand side, we have replaced $a_i$ with $i$, and $b_j$ with $j$, for notational simplicity. Equation~\eqref{Eq:KDStand2} does not depend upon
}
whether $ \{ \ket{a_i} \} $ and $ \{ \ket{b_j} \}$ are discrete or continuous.  As 
\nicole{mentioned above,} this review focuses on finite dimensions; the continuous KD distribution \Dave{is introduced in Sec. \ref{SubSec:ContinuousKD} and used} only in Sec.~\ref{Sec:StateMeas}.  Suppose that $\braket{b_j|a_i} \neq 0$ for all $i,j$. 
\Stephan{By the reconstruction formulae Eqs.~\eqref{eq:reconstruction2},} $Q(\hat{\rho})$ enables an informationally complete description of $\hat{\rho}$:
\begin{equation}
\label{Eq:StateDecom}
\hat{\rho} = \sum_{i,j} \hat{G}_{a_i,b_j}  Q_{i,j} \left( \hat{\rho} \right)  .
\end{equation}
As \Stephan{explained} above,  $\{ \hat{G}_{a_i,b_j} \} =  \{ {\ketbra{a_i}{b_j}} / {\braket{b_j|a_i}} \} $ forms a  basis.  
Now, suppose that some $\braket{b_j|a_i} = 0$. The KD distribution can still convey useful, albeit partial, information about $\hat{\rho}$. Examples surface in
weak-value experiments 
(\nicole{Sec.}~\ref{Sec:WeakValues}), where
one need not necessarily apply an entire KD distribution.

The KD distribution satisfies several of Kolmogorov's axioms~\cite{Kolmogorov33} for joint probability distributions:
\begin{align}
\label{Eq:KDProps}
\sum_{i,j} Q_{i,j} \left( \hat{\rho} \right)  =1 , \quad \sum_j Q_{i,j} \left( \hat{\rho} \right)  = \braket{a_i | \hat{\rho} | a_i}  \quad \text{and} \quad  \sum_i Q_{i,j} \left( \hat{\rho} \right) = \braket{b_j | \hat{\rho} | b_j} .
\end{align}
\Stephan{
Nevertheless, $Q(\hat \rho)$ is not a joint probability distribution:  Although 
$| Q_{i,j} \left( \hat{\rho} \right) | \in [0,1]$, 
$Q_{i,j} \left( \hat{\rho} \right) $ can assume negative or non-real values.  Still, the marginals of $Q(\hat\rho)$ are probability distributions. Also, they reproduce the Born-rule probabilities associated to $\hat A$ and $\hat B$. 
This simple reproduction is not a requirement for quasi-probability representations generally, but a special and important feature of the KD distributions. 
\nicole{The Wigner function, too, reproduces the Born-rule probabilities via marginalization~\cite{Wigner32}. To reproduce the probabilities with the Husimi function or Glauber-Sudarshan distribution, one must undertake a more extended calculation~\cite{Glauber63, Sudarshan63,Husimi40}.}
\nicole{Furthermore, the KD representation of quantum mechanics does not obey Eq.~\eqref{eq:adjointness}. Consequently, self-adjoint operators do not necessarily have real KD symbols.}
}

The negative and non-real values of $Q_{i,j}(\hat\rho)$ are sometimes called `non-classical'.  What is considered non-classical differs from setting to setting (and author to author). Therefore, we refer to negative or non-real values simply as  \emph{non-positive}. 
We call $Q \left( \hat{\rho} \right) $ \emph{positive} if all its entries are positive or zero.  The most common measure of KD non-positivity is~\cite{Alonso_OTOC_2019}
\begin{equation}
\label{Eq:TotKDNonPos1}
\mathcal{N} \LParen Q (\hat{\rho})  \RParen = \sum_{i,j}  \left| Q_{i,j} (\hat{\rho}) \right|  .
\end{equation}
$\mathcal{N} \LParen Q (\hat{\rho})  \RParen=1$ if, and only if, the KD distribution is a classical joint probability distribution. Section~\ref{Sec:NonPos} outlines further properties of $\mathcal{N} \LParen Q (\hat{\rho})  \RParen$.

\subsection{\Dave{Continuous-variable KD distribution}}
\label{SubSec:ContinuousKD}

\nicole{Throughout this article, we denote discrete-variable projectors by $\hat{\Pi}$ and continuous-variable projectors by $\hat{\pi}$.
We denote by $P(a)$ a probability distribution over a discrete variable $a$ and by $\mathbb{P}(x)$ a probability density over a continuous variable $x$.
}
For continuous-variable quantum states, the KD distribution is typically constructed in terms of the phase space's canonical conjugate observables: Position, 
\begin{eqnarray}
    \label{eq_position}
    \hat{x}=\int x \, \hat{\pi}_{x}dx 
    \equiv \int x \ketbra{x}{x}dx ,
\end{eqnarray}
and momentum,
\begin{equation}
    \label{eq_momentum}
    \hat{p}=\int p\hat{\pi}_{p}dp 
    \equiv \int p\ketbra{p}{p}dp  . 
\end{equation}
The operators satisfy $[\hat{x},\hat{p}]=i\hbar\hat{\mathbbm{1}}$. The corresponding 
\Dave{basis} elements of Eq.~\eqref{Eq:Frame} (with $\hat{A} = \hat{x}$ and $\hat{B} = \hat{p}$) have the form 
\begin{align}\label{eq:hatlambdaxp}
\hat{G}(x,p) & =\frac{\ketbra{x}{p}}{\braket{p|x}}
=\sqrt{2\pi\hbar} \: e^{ipx/\hbar} \, \ketbra{x}{p} 
=\Stephan{2\pi\hbar  \,  \hat\pi_x\hat \pi_p} \, .
\end{align}
In terms of them, we can expand the density operator:
\begin{align}
\hat{\rho} & =
\iint Q(x,p;\hat{\rho}) \, \hat{G}(x,p) \, dx \, dp.
\end{align}
$Q(x,p;\hat{\rho})$ denotes the continuous-variable KD distribution:
\begin{equation}
Q(x,p;\hat{\rho})
\equiv\braket{p|x}\bra{x}\hat{\rho}\ket{p}
=\mathrm{Tr}\left(  \hat{\pi}_{p}\hat{\pi}_{x}\hat{\rho}\right) .\label{eq:KD_continuous}
\end{equation}
\Dave{We interpret this formula in Sec. \ref{Sec:StateMeas}. } As 
for discrete KD distributions, any marginalisation (an integration over momentum or position) completes the Fourier transform. The position probability density, 
$\mathbb{P}(x)=\mathrm{Tr} \left( \hat{\pi}_{x}\hat{\rho} \right)$,
or the momentum probability density, 
$\mathbb{P}(p)  =  \mathrm{Tr}  \left(  \hat{\pi}_{p}\hat{\rho}  \right)$, results.

Consequently, a pure state $\ket{\psi}$ has a KD distribution
of the form
\begin{align}
Q(x,p;\psi) & \equiv\braket{p|x}\!\braket{x|\psi}\!\braket{\psi|p}
=\frac{e^{-ipx/\hbar}} {\sqrt{2\pi\hbar}} 
\psi(x)\widetilde{\psi}^{*}(p).
\end{align}
$Q(x,p;\psi) \equiv\braket{p|x}\!\braket{x|\psi}\!\braket{\psi|p}$ combines the position wavefunction $\psi(x)=\braket{x|\psi}$
and the momentum wavefunction $\widetilde{\psi}(p)=\braket{p|\psi}$
with the Fourier kernel that connects the two bases. 
In summary, despite 
subtle differences, the continuous KD distribution 
straightforwardly extends the discrete KD distribution. 

\nicole{We can now specify how to quantise functions of non-commuting observables. Equation~\eqref{eq:Dirac_functionAB} becomes
\begin{equation}
\label{Eq:DiracContForm}
\hat h(\hat x,\hat p)=\int h(x,p) \hat\pi_x  \hat\pi_p \: dx \; dp.
\end{equation}
\Dave{Equation \eqref{Eq:DiracContForm}} \Dave{constitutes a (Kohn Nirenberg)} quantisation~\Dave{\cite{Kohn65}} in which a polynomial in $p$, with $x$-dependent coefficients, is transformed into a differential operator. 
Interchanging the $\hat{\pi}_x$ and $\hat{\pi}_p$ yields the opposite ordering.
}

\subsection{Quasi-probabilistic Bayesian update}
\label{SubSec:Bayes}

\nicole{In classical statistics, Bayes' theorem dictates how one updates a joint probability distribution upon acquiring new information.}
The KD distribution obeys a quasi-probabilistic version of Bayes' theorem~\cite{Steinberg1995, Johansen07, Hofmann12, NYH_quasiprobability_2018}.
\nicole{This version updates the distribution when one learns a restriction on \Dave{$i$ or} $j$. }
We define the conditional quasi-probabilities $\tilde Q_{i\mid j}(\hat \rho)$ and $\tilde Q_{j\mid i}(\hat \rho)$ by 
\begin{equation}
\label{Eq:KDBayes}
\tilde{Q}_{i|j} \left( \hat{\rho} \right) 
\equiv \frac{Q_{i,j} \left( \hat{\rho} \right) }{P(b_{j}|\hat{\rho})}  \quad \text{and} \quad
\tilde{Q}_{j|i} \left( \hat{\rho} \right) \equiv \frac{Q_{i,j} \left( \hat{\rho} \right) }{P(a_{i}|\hat{\rho})} \, .
\end{equation}
\Dave{We interpret $\tilde{Q}_{i|j^{\star}} \left( \hat{\rho} \right) $ as the KD distribution ${Q}_{i,j} \left( \hat{\rho} \right) $ conditioned on $j=j^{\star}$. }
We have defined
$P(b_j|\hat{\rho}) = \sum_i Q_{i,j} \left( \hat{\rho} \right) =\braket{b_j|\hat{\rho}|b_j}$
as the probability of obtaining the outcome $b_j$ upon measuring the state $\hat \rho$ in the  $\{ \ket{b_j}\}$ basis. 
We have defined $P(a_i|\hat\rho)$ analogously. If $Q(\hat{\rho})$ is positive, then $| \tilde{Q}_{i|j}(\hat{\rho}) | \in [0,1]$ for all $i,j$, as expected from classical probability theory. However, a KD distribution can assume negative values. Therefore, the denominator in Eq.~\eqref{Eq:KDBayes} can have a lesser magnitude than the numerator. Consequently, $| \tilde{Q}_{i|j}(\hat{\rho}) | \in [0,\infty)$,  in general. As we review below, such anomalous  Bayesian updates explain several non-classical advantages in quantum-information processing.  Moreover, the conditional-KD-distribution formulae suffice for reformulating a quantum evolution as a quasi-probability update for classical (eigen)values~\cite{Hofmann12,Hofmann14}. Bayesian evolution of the KD distribution was experimentally demonstrated via direct measurement~\cite{Bamber2014}, a method described in Sec.~\ref{Sec:StateMeas}.

More generally, one can condition $Q_{i,j} \left( \hat{\rho} \right) $ 
such that $j$ assumes a value from a subset $\mathcal{F}$~\cite{ArvidssonShukur20, Jenne22}: 
\begin{equation}
\label{Eq:KDBayesExt}
\tilde{Q}_{i,j|j\in \mathcal{F} } \left( \hat{\rho} \right) =  \frac{    Q_{i,j} \left( \hat{\rho} \right)  }{ 
\nicole{ \sum_{ i^{\prime} }  \sum_{ j^{\prime} \in \mathcal{F} }
}
Q_{i^{\prime},j^{\prime}} \left( \hat{\rho} \right)  } \, ,
\end{equation}
if $j$ is in $\mathcal{F}$. \Dave{Otherwise,  $\tilde{Q}_{i,j|j\in \mathcal{F} } \left( \hat{\rho} \right) = 0$.} 
The denominator is the probability that, if $\hat{\rho}$ is prepared and $\{ \ket{b_j} \}$ is measured, the outcome $j$ will be in $\mathcal{F}$.

\subsection{Generalisations of the KD distribution}
\label{SubSec:GeneralKD}

The distribution in Eq.~\eqref{Eq:KDStand2} is, from a fundamental perspective, the most studied and mathematically understood KD distribution. Nevertheless, 
generalisations of the standard KD distribution
have been used frequently 
in applications. The first generalisation~\cite{Kirkwood33, Dirac45} allows one to decompose a quantum state with respect to more than two sets of measurement operators. The second generalisation~\cite{NYH_quasiprobability_2018, ArvidssonShukur2021, Lupu22} allows for these sets to be not only rank-$1$ projectors, but general positive-operator-valued  measures.  

Consider $k$ non-degenerate observables $\hat{A}^{(l)}=\sum_{i_l=1}^d a_{i_l}^{(l)} \ketbra{a_{i_l}^{(l)}} {a_{i_l}^{(l)}}$, where $l = 1, 2, \ldots, k$. 
We denote the corresponding eigenbases by $ \{ \ket{a_{i_l}^{(l)}}\}$.  The $k$-extended KD distribution~\cite{NYH_quasiprobability_2018} is 
\begin{equation}
\label{Eq:ExtKD}
Q_{i_1,\ldots,i_k} (\hat{\rho}) = \braket{a_{i_k}^{(k)} | a_{i_{k-1}}^{(k-1)}} \braket{a_{i_{k-1}}^{(k-1)} | a_{i_{k-2}}^{(k-2)}} \cdots \bra{a_{i_{1}}^{(1)}} \hat{\rho} \ket{a_{i_{k}}^{(k)}} .
\end{equation}
\Dave{Equation \eqref{Eq:ExtKD} shows that, to write an extended KD distribution, we must choose how to order the observables. The physical problem of interest should determine the ordering. For examples, see the extended KD distributions constructed in Sec.  \ref{SubSec:PSMetrology} [Eq. \eqref{Eq:PSMetKD}] and Sec. \ref{Sec:OTOC} [Eq.  \eqref{Eq:OTOCKD}].}

The second generalisation of the KD distribution involves 
measurement operators that are not necessarily rank-1 projectors, but form general positive-operator-valued measures.  A positive-operator-valued measure is a set $\{ \hat{M}_i \}$ of 
positive-semidefinite operators $\hat{M}_i \geq 0$ that are normalised: $\sum_i \hat{M}_i = \hat{\mathbbm{1}}$~\cite{Nielsen11}.
Positive-operator-valued measures represent arbitrary quantum measurements, not only projective measurements.
In terms of $k$ positive-operator-valued measures $\mathcal{M}^{(l)} = \left\{  \hat{M}^{(l)}_{i_l} \right\}$, where $l = 1, 2, \ldots, k$, the measurement-generalised KD distribution is
\begin{equation}
\label{Eq:GenKD}
Q_{i_1,\ldots,i_k} (\hat{\rho}) = \Tr \left( \hat{M}^{(k)}_{i_k} \hat{M}^{(k-1)}_{i_{k-1}} \cdots \hat{M}^{(1)}_{i_1} \hat{\rho}    \right) .
\end{equation}
Equation~\eqref{Eq:GenKD} shows the 
most general form of a KD distribution (yet defined). 
The extended and measurement-generalised KD distributions satisfy trivially extended versions of the properties listed in Eqs.~\eqref{Eq:KDProps} and~\eqref{Eq:KDBayes}. The non-positivity of $\{ Q_{i_1,\ldots,i_k} (\hat{\rho}) \} $ is often quantified by $\mathcal{N} \LParen Q (\hat{\rho})  \RParen = \sum_{i_1,\ldots,i_k}  \left| Q_{i_1,\ldots,i_k} (\hat{\rho}) \right|$, as outlined in Sec.~\ref{SubSec:KDNonclas}.  In this review, we denote by $Q(\hat{\rho})$ standard, extended and measurement-generalised KD distributions. The context and the number of indices will specify a distribution further.

\subsection{Optimisation with the KD distribution}
\label{SubSec:Useage}

A common use of the KD distribution is the evaluation and optimisation of operational formulae.
Consider some physical formula $F(\hat{\rho})$ of interest.  
One might recast it in terms of a KD distribution:
$\tilde{F} \LParen Q(\hat{\rho}) \RParen$.  
If so, one can optimise $\tilde{F} \LParen Q(\hat{\rho}) \RParen$ with respect to a  classical probability distribution  or a general KD distribution:
\begin{align}
F^{\mathrm{p}}  \nicole{\equiv} 
\underset{ Q_{i,j} \in [0,1] }{\mathrm{opt}} \left\{ \tilde{F}(Q) \right\} ,  
\quad \quad \mathrm{or} \quad \quad
 F^{\mathrm{np}}  \nicole{\equiv} 
 \underset{Q(\hat{\rho})  }{\mathrm{opt}} 
 \left\{ \tilde{F}\LParen Q(\hat{\rho}) \RParen \right\}  ,
\end{align}
The $\mathrm{opt}$ could entail a maximisation, a minimisation or some other optimisation procedure. If $F^{\mathrm{np}} $ differs from $F^{\mathrm{p}} $, then  non-commutation
(in the form of KD non-positivity) can break the bound
on the value
achievable in scenarios  
describable with classical probability distributions. 
Moreover, the optimised KD distribution, 
\begin{equation}
Q^{\star}(\hat{\rho}) = \mathrm{arg \, } \underset{Q(\hat{\rho})  }{\mathrm{ opt}} \left\{ \tilde{F} \LParen Q(\hat{\rho}) \RParen \right\} ,
\end{equation}
guides the construction of an optimal experiment. In the next sections, we will see several examples. 

\section{The KD distribution and quantum metrology}

\label{Sec:Metrology}

Metrology is the science of measuring or estimating unknown physical parameters.  When one estimates parameters 
that characterise quantum processes, it is natural to use quantum systems as probes.
Measurements of unknown parameters, using quantum systems, 
fall under the domain of quantum metrology. In quantum metrology, non-classical phenomena, such as coherence, entanglement and non-commutation, can boost estimation abilities beyond classical bounds~\cite{Helstrom76,Braunstein94,bCover06,Giovanetti06, Giovanetti11, Krischek11, Demkowicz14, Smith22, Smith23, ArvidssonShukur20}.   Below, we review how non-real entries in  KD distributions play a fundamental role in quantum metrology. We also show \Dave{that one} can break classical bounds on metrological quantum-information distillation \Dave{only in the presence of KD non-positivity}~\cite{ArvidssonShukur20}.  For simplicity, we focus on pure states, Stone's-encoded unitaries  (explained below) and single-parameter metrology. However, generalisations  extend the results beyond these restrictions~\cite{ Jenne22,Lupu22,Salvati23}.  Further applications lie beyond the scope of this review: For example,  KD negativity can quantify resources in interaction-free measurements \cite{Hance_2024}. 

\subsection{Measurement disturbance}
\label{SubSec:Disturb}

\begin{figure}
\includegraphics[scale=0.25]{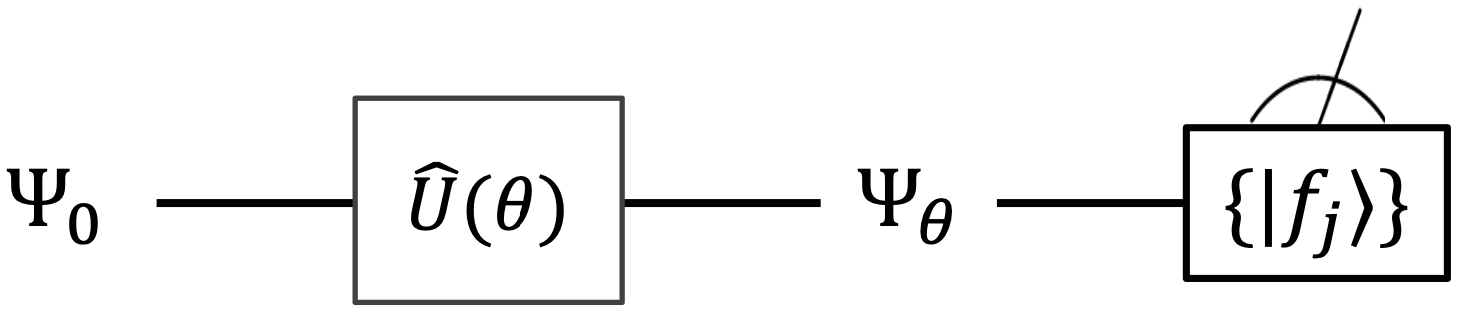}

\caption{
\label{fig:MetroOne}
\textbf{Standard quantum metrology.} The goal is to use a quantum system as a probe to learn an unknown parameter of a unitary.  Initially, the probe state is $\ket{\Psi_0}$. The probe undergoes a unitary $\hat{U}(\theta)$, which encodes the unknown parameter $\theta$ in the probe state: $\ket{\Psi_0} \rightarrow \ket{\Psi_{\theta}} = \hat{U}(\theta) \ket{\Psi_0}$. 
Measuring the state in a suitable basis yields information about $\theta$.
 }

\end{figure}

Before we review the KD distribution's connection to quantum metrology, we describe how imaginary KD quasi-probabilities encode the disturbance of a quantum state. 
Let $\theta$ denote a real parameter that we wish to measure.
Suppose that a unitary $\hat{U}(\theta)$ obeys Stone's theorem~\cite{Stone32}: 
$\hat{U}(\theta) = e^{-i \hat{A} \theta}$,
wherein $\hat{A} = \sum_i a_i \ketbra{a_i}{a_i}$ denotes a Hermitian generator. We assume that the eigenvalues $a_i$ are non-degenerate.
Consider  evolving a state $\ket{\Psi_0}$ under $\hat{U}(\theta)$
to the output state $\ket{\Psi_{\theta}} = \hat{U}(\theta) \ket{\Psi_0}$.  Finally, consider measuring some basis $\{ \ket{f_j} \}$. (See Fig.~\ref{fig:MetroOne}.) This process yields a probability distribution $\{ P(f_j | \Psi_{\theta} ) = |\braket{f_j|\Psi_{\theta} } |^2 \}$. 

A natural question to ask is \emph{how much do changes in $\theta$ disturb the measurement-outcome probabilities?} The question invites us to differentiate:
\begin{align}
 \Pt  P(f_j | \Psi_{\theta} ) & =  \Pt |\braket{f_j |\Psi_{\theta} } |^2  \nonumber \\
 & =-i \braket{f_j |\hat{A}|\Psi_{\theta} } \braket{\Psi_{\theta}|f_j}  + i \braket{f_j|\Psi_{\theta} } \braket{\Psi_{\theta}|\hat{A}|f_j} \nonumber  \\
 & = 2 \im \left( \braket{f_j|\hat{A}|\Psi_{\theta}} \braket{\Psi_{\theta}|f_j }  \right)
  \nonumber  \\
 & = \sum_i 2 a_i \im \left( \braket{f_j|\Psi_{\theta} } \braket{\Psi_{\theta}\ket{a_i} \bra{a_i} f_j} \right) 
  \nonumber  \\
 & = \sum_i 2 a_i \im \bm{(} Q_{i,j}(\PsiT)  \bm{)} .
 \label{Eq:ImagDist}
\end{align}
We have defined the KD quasi-probability
\Dave{$Q_{i,j}(\PsiT) \equiv \braket{f_j|\Psi_{\theta}}\braket{\Psi_{\theta}|a_i} \braket{a_i|f_j }  $}.  A generator's ability to disturb a quantum state is related to  a KD distribution's imaginary part. For further results on this topic, see~\cite{Dressel12-2}. This reference describes how imaginary weak values (which stem from imaginary KD distributions, as outlined in Sec.~\ref{Sec:WeakValues}) \Dave{signal} a von Neumann measurement's disturbance of a quantum state.  Moreover,  \cite{Budiyono_2023_as,Budiyono_PRA_23} discuss connections between the imaginary part of a KD distribution and a quantum state's asymmetry. 

\subsection{Standard quantum metrology}
\label{SubSec:StandardMetrology}

Again, we consider a unitary evolution $\hat{U}(\theta) = e^{-i \hat{A} \theta}$, where $\theta$  is now an unknown parameter.   To estimate $\theta$, one can prepare a probe state $\ket{\Psi_0}$, evolve it under $\hat{U}(\theta)$ and measure the final state, $\ket{\Psi_{\theta}} = \hat{U}(\theta) \ket{\Psi_0}$. 
Again, the measurements are of some basis $\{ \ket{f_j} \}$.
Repeating this process across 
many trials provides metrologically useful statistics (see Fig.~\ref{fig:MetroOne}).  
The measurement outcomes obey a probability distribution $\{ P(f_j | \Psi_{\theta} ) = |\braket{f_j|\Psi_{\theta} } |^2 \}$.

Sampling from this distribution, one observes the probabilities $\{ P_{\mathrm{o}}(f_j | \Psi_{\theta} ) \}$. An estimator $\Te$ is a function that maps the sample space (the 
set of possible outcomes 
that may be outputted during the
experiment)
to a sample estimate (the $\theta$ estimate inferred from the observations).  
In other words, given observed data, one can construct an estimate 
$\Te \LParen \{ P_{\mathrm{o}}(f_j | \Psi_{\theta} ) \} \RParen$ 
of the unknown parameter $\theta$.  

The precision of any unbiased estimator, classical or quantum,  is lower-bounded by the Cramér-Rao inequality:
\begin{equation}
\label{Eq:ClasCR}
\Var \left( \Te \right) \geq \frac{1}{N I(\theta)} .
\end{equation}
$N$ denotes the number of samples (measurements) drawn from the probability distribution,  and $I(\theta)$ denotes the Fisher information~\cite{Cramer16, Rao92}. The Fisher information of a probability distribution $\{ P(f_j | \Psi_{\theta} ) \}$  is
\begin{equation}
\label{Eq:ClasFish}
I(\theta) = \sum_j \frac{\left[ \Pt  P(f_j | \Psi_{\theta} )  \right]^2}{P(f_j | \Psi_{\theta}) } \, .
\end{equation}
Inequality~\eqref{Eq:ClasCR} 
saturates for large $N$ and reasonable estimators. Thus, one can improve an estimate in two ways.
First, one can increase the number $N$ of measurements. Second, one can choose the initial state $\ket{\Psi_0}$ and the final-measurement basis $\{ \ket{f_j} \}$ such that the Fisher information is large. 

For any metrological estimation to be possible, the Fisher information about $\theta$, obtained from measuring $\ket{\Psi_{\theta}}$, must be non-zero. Such
 a non-zero Fisher information  is directly connected to the KD distribution. In Eq.~\eqref{Eq:ClasFish}, the numerator is the square of the derivative of the outcome probability~\eqref{Eq:ImagDist}. Since 
 $\tilde{Q}_{i|j} \left( \Psi_{\theta} \right) 
\equiv \frac{Q_{i,j} \left( \Psi_{\theta} \right) }{P(f_{j}|\Psi_{\theta})}$ [Eq. \eqref{Eq:KDBayes}], we can re-express Eq.~\eqref{Eq:ClasFish} as
\begin{equation}
I(\theta) 
= 4 \sum_j \frac{\left[ \sum_i a_i \im \bm{(} Q_{i,j}(\PsiT)  \bm{)}   \right]^2}{P(f_{j}|\Psi_{\theta}) }  = 4 \sum_j P(f_j | \Psi_{\theta} ) \left[ \sum_i a_i \im \bm{(} \tilde{Q}_{i|j} \left( \Psi_{\theta} \right) \bm{)}   \right]^2 
.
\end{equation}
Thus, a non-zero Fisher information (obtained by measuring $\ket{\PsiT}$) requires \emph{non-real} components  in a conditional KD distribution $\tilde{Q}_{i|j} \left( \Psi_{\theta} \right)$.
Consequently, $\{ \ket{f_j} \}$ cannot equal $\{ \ket{a_i} \}$, if an experiment is to extract information about $\theta$: Such equality would lead to a positive KD distribution. 
\Dave{In other words, if the measurement basis equals the basis of $\hat{A}$, then the final probabilities are unchanged by the unitary evolution and hence do not depend on $\theta$.}

Reference~\cite{Hofmann2011} discusses further links among the Fisher information, KD distributions and weak values. 

\subsection{Post-selected quantum metrology}
\label{SubSec:PSMetrology}

In some quantum experiments,  probe systems are more easily prepared than measured. This is the case, for example, in many optics experiments, where heralded single photons can be created 
significantly faster than they can be measured~\cite{Tomm21, Lupu22}. 
The difficulty in measuring single photons stems from particle detectors' saturation-intensity limit, or dead time: After a detector detects a particle, it experiences a time lag (dead time) until it can detect another particle. Moreover, there is a maximum energy intensity under which a detector can operate (the saturation intensity).  In metrology experiments where probes can be prepared more `cheaply' than they can be measured, it can be advantageous to distil the quantum information from many particles into a few, prior to measurement~\cite{Dressel14,  Harris17, Xu20, ArvidssonShukur20,Lupu22, Jenne22,Salvati23, ArvShuk23}.  Ideally, all the information from a high-intensity beam of probes would be distilled into a weak beam, without the loss of any information.  This feat was recently shown to be possible, with KD negativity being the enabler~\cite{Jenne22,Salvati23}.

\begin{figure}
\includegraphics[scale=0.25]{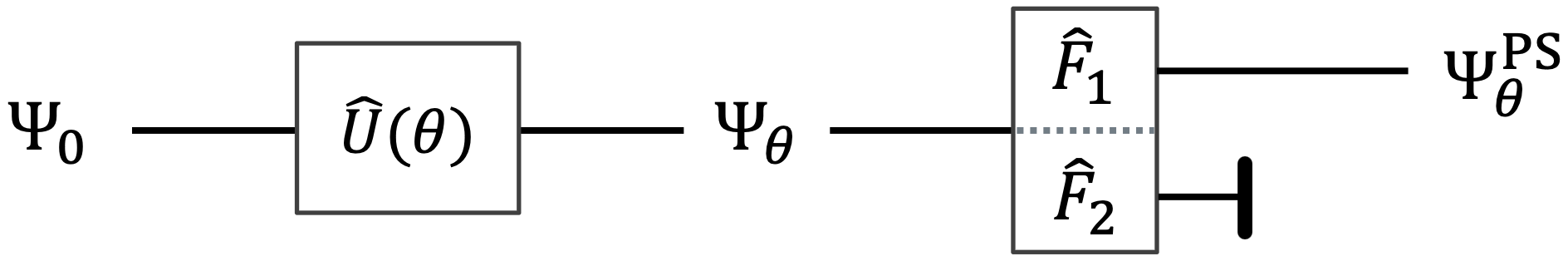}

\caption{
\label{fig:MetroTwo}
\textbf{Post-selected quantum metrology.} 
The goal is to distil the metrological information from many quantum systems into a few.  Information about an unknown parameter $\theta$ is encoded in quantum states, as in Fig.~\ref{fig:MetroOne}. 
The information-carrying states 
then undergo a two-outcome positive-operator-valued measure $\{ \hat{F}_1\equiv \hat{F}, \, \hat{F}_2\equiv \hat{\mathbbm{1}} - \hat{F} \}$, which acts as a filter.  Probes that pass the 
post-selection, yielding the $\hat{F}_1$ outcome, are retained (post-selected). Probes that do not are discarded. If $\hat{F}$ is chosen appropriately, the quantum Fisher information of a post-selected state $\ket{\PsT}$ 
can be significantly higher than that of $\ket{\PsiT}$.  The post-selection probability ensures that, on average, the post-selective filter creates no information. 
Moreover, consider operating on $n$ states with the post-selective filter, and suppose that $m$ states pass the post-selection. In certain cases, the amount of information retained in the post-selected states can nearly equal the total amount of information available initially \cite{Jenne22}.}

\end{figure}

Consider altering the general protocol outlined in Fig.~\ref{fig:MetroOne}. Between the unitary 
and the 
measurement, we insert a post-selective filter. 
The filter discards or retains an experimental trial, depending on a positive-operator-valued measure's outcome (Fig.~\ref{fig:MetroTwo}). 
The measure has the form
$\mathcal{F} = \{\hat{F}_1 \equiv \hat{F},  \hat{F}_2 \equiv \hat{\mathbbm{1}} - \hat{F} \}$.  Only if the filter yields the $\hat{F}$ outcome is the probe sent to the final detector. Otherwise, the probe is discarded. After the post-selection, the (renormalised) quantum state is
\begin{equation}
\ket{\PsT} = \frac{1}{\sqrt{\pps}} \,  \hat{K} \ket{\PsiT} .
\end{equation}
$\pps = \bra{\PsiT} \hat{F} \ket{\PsiT}$ denotes the post-selection probability (the probe's probability of passing the filter). We have introduced a Kraus operator $\hat{K}$ such that $\hat{F} = \hat{K}^{\dagger} \hat{K}$. 

Consider any parameterised quantum state $\hat{\rho}_{\theta}$. The maximum amount of Fisher information one can extract from it (by making the optimal final measurement) is the \textit{quantum} Fisher information:
\begin{equation}
\mathcal{I}_Q \left( \hat{\rho}_{\theta} \right) = \Tr \left( \hat{\Gamma}^2_{\hat{\rho}_{\theta}} \hat{\rho}_{\theta} \right) = \max_{\mathrm{meas.}} \left\{ I\left( \theta|\hat{\rho}_{\theta} \right) \right\} .
\end{equation}
$\hat{\Gamma}_{\hat{\rho}_{\theta}}$ is the symmetric logarithmic derivative, implicitly defined through $\Pt \hat{\rho}_{\theta} = \left( \hat{\Gamma}_{\hat{\rho}_{\theta}} \hat{\rho}_{\theta} + \hat{\rho}_{\theta} \hat{\Gamma}_{\hat{\rho}_{\theta}} \right)/2$~\cite{Helstrom76, Braunstein94}.  
The quantum Fisher information for $\ket{\PsT}$ is
\begin{align}
\label{Eq:PSFish}
\mathcal{I}_Q \left( \PsT \right) 
=  4 \bigg( \frac{1}{\pps} \bra{\PsiT} \hat{A} \hat{F} \hat{A} \ket{\PsiT} - \frac{1}{\left( \pps \right)^2} \left| \bra{\PsiT} \hat{A} \hat{F}  \ket{\PsiT} \right|^2 \bigg) .
\end{align}

The post-selected quantum Fisher information [Eq.~\eqref{Eq:PSFish}] can be recast in terms of an extended KD distribution. To this end, we define the $2$-extended KD distribution
\begin{equation}
\label{Eq:PSMetKD}
Q_{i,j,k}(\PsiT) \equiv \braket{\PsiT | a_i} \bra{a_i} \hat{F}_k \ket{a_j} \braket{a_j|\PsiT} .
\end{equation}
The $i$ and $j$ indices label two instances of the eigenbasis of the generator $\hat{A}$. The $k$ index labels the elements of the post-selective positive-operator-valued measure.
To fully incorporate the post-selection into the KD distribution, we implement a quasi-probabilistic Bayesian update (see Sec.~\ref{SubSec:Bayes}):
\begin{align}
\tilde{Q}_{i,j}^{\mathrm{PS}}(\PsiT) 
& = \frac{Q_{i,j,k=1}(\PsiT) }{\sum_{i,j} Q_{i,j,k=1}(\PsiT) }  
= \braket{\PsiT | a_i} \bra{a_i} \hat{F} \ket{a_j} \braket{a_j|\PsiT} / \pps .
\end{align}
Equipped with the extended KD distribution $\tilde{Q}_{i,j}^{\mathrm{PS}}(\PsiT) $, we can rewrite Eq.  \eqref{Eq:PSFish}:
\begin{align}
\label{Eq:PSFishKD}
\mathcal{I}_Q \left( \PsT \right) = 4 \left[ \sum_{i,j} a_i a_j \tilde{Q}_{i,j}^{\mathrm{PS}}(\PsiT) - \bigg| \sum_{i,j} a_i \tilde{Q}_{i,j}^{\mathrm{PS}}(\PsiT) \bigg|^2  \right] .
\end{align}
$\mathcal{I}_Q \left( \PsT \right)$ is (four times) 
an element of a quasi-probabilistic covariance matrix~\cite{Jenne22}.
If $\tilde{Q}^{\mathrm{PS}}(\PsiT) $ is a classical probability distribution, 
then $\mathcal{I}_Q \left( \PsT \right)$ is upper-bounded by four times the maximum variance, in any state, of $\hat{A}$:
\begin{equation}
\max_{\tilde{Q}_{i,j} \in [0,1] } \mathcal{I}_Q \left( \PsT \right) \leq 4 \max_{\hat{\rho}} \left\{ \mathrm{Var}_{\hat{\rho}}\left( \hat{A} \right) \right\}= \left( \Delta a \right)^2 \, .
\end{equation}
$\Delta a$ denotes the spectral gap of $\hat{A}$.  

On the other hand, if $\tilde{Q}^{\mathrm{PS}}(\PsiT) $ contains negative elements, $\mathcal{I}_Q \left( \PsT \right)$ can be arbitrarily large.  
There is no fundamental bound on how much Fisher information can be distilled from many quantum states into a few, if the probability of obtaining information-dense states is correspondingly small. Moreover, Refs.~\cite{Jenne22,Salvati23} show that this  distillation can be approximately lossless, such that $\mathcal{I}_Q \left( \PsT \right) \times \pps \approx \mathcal{I}_Q \left( \PsiT \right)$. 
In summary, the quantum Fisher information about $\theta$, encoded in $\ket{\Psi_{\theta}}^{\otimes n}$, can be losslessly compressed into $\ket{\PsT}^{\otimes m}$, where $m/n$ can be made arbitrarily small. In practice, systematic errors in the post-selective filter prevent unbounded information distillation~\cite{Salvati23}. 
These results can be extended to multiparameter metrology, where $\theta \rightarrow (\theta_1, \theta_2,    \ldots,\theta_M)$, and to mixed states~\cite{Jenne22,Salvati23}.

The exact relation between the KD non-positivity  [Eq.~\eqref{Eq:TotKDNonPos1}]  and information distillation has not been mapped out.  Nevertheless,  Ref.~\cite{ArvShuk24} derives 
the relation in the context of optimal post-selected metrology, which we now outline. Consider preparing a quantum system  in the optimal state for probing a unitary $\hat{U}(\theta)$ parameterised by an unknown $\theta$.
Assume that the post-selection filter is optimal. The rate of Fisher-information distillation is directly proportional to the KD non-positivity [Eq.~\eqref{Eq:TotKDNonPos1}]: 
$\mathcal{I}_Q \left( \PsT \right) \propto \mathcal{N} \LParen \tilde{Q}^{\mathrm{PS}}(\PsiT)  \RParen$.

\begin{figure}
\includegraphics[scale=0.25]{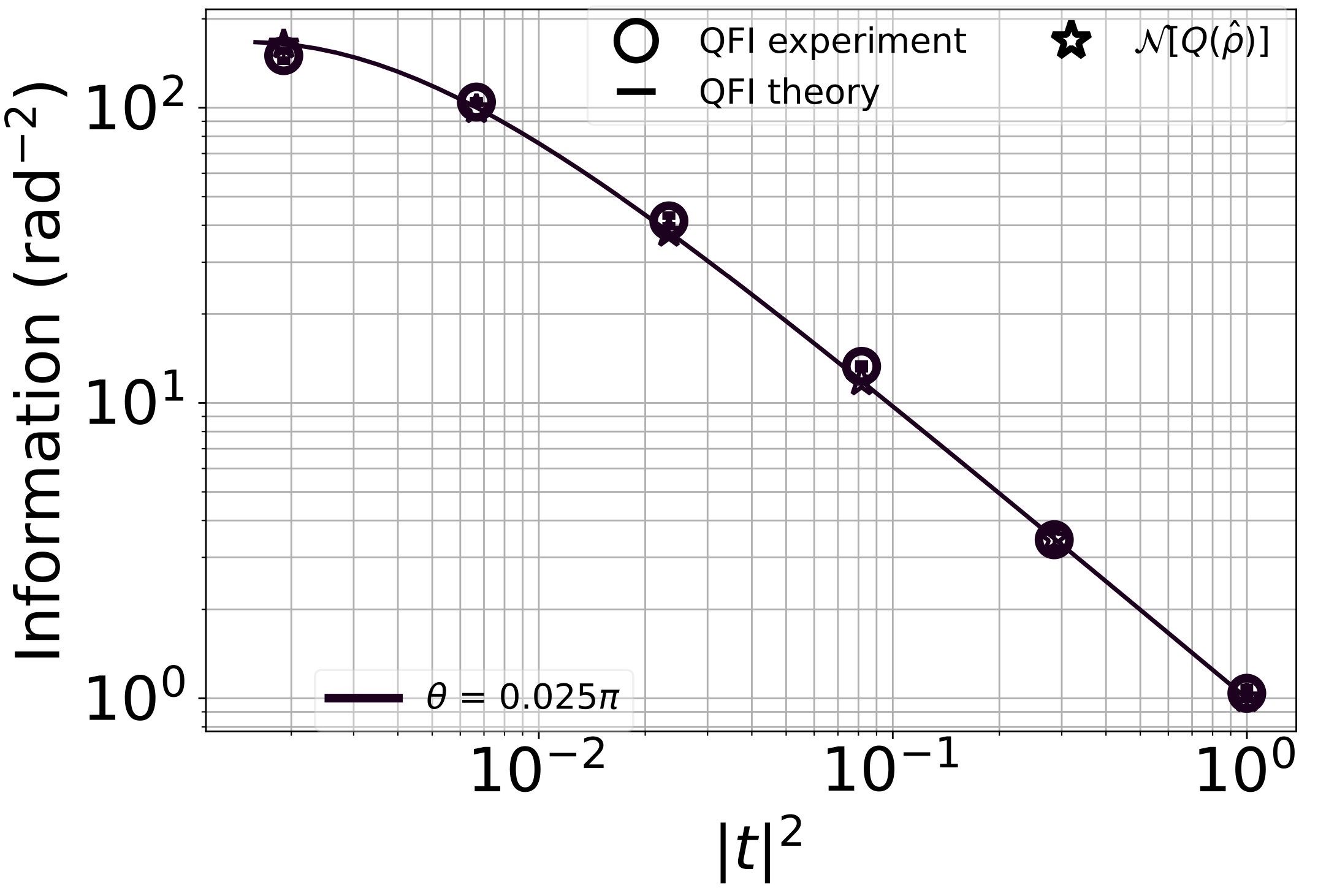}

\caption{
\label{fig:MetroThree}
\textbf{Post-selected quantum metrology in optics experiment.} 
The figure is reproduced from data 
collected during an optical realisation~\cite{Lupu22} of Fig.~\ref{fig:MetroTwo}. On the horizontal axis is $|t|^2$---the filter's post-selection probability, modulo small correction terms.
The vertical axis shows the Fisher information, per measured photon, about an unknown birefringent phase $\theta$. If no filtering is applied (if $|t|^2=1$), the maximum Fisher information (for any input state and final measurement) is $1 \, \mathrm{rad}^{-2}$ per photon. As the filtering strengthens
(as $|t|^2\rightarrow 0$), the retained photons' information content increases.
The solid line shows theoretical predictions, the circles show experimental data points, and the stars show the underlying KD distribution's 
non-positivity [Eq.~\eqref{Eq:TotKDNonPos1}]. The stars and circles  are hard to distinguish because of their overlaps. The true value of $\theta$ is $ 0.025 \pi \, \mathrm{rad}$.
}

\end{figure}

Post-selected quantum metrology has been realised experimentally. Examples of such realisations include weak-value-amplification experiments~\cite{Hosten08, Dixon09}, which we describe in Sec.~\ref{SubSec:WVA}, and experiments with partially post-selected filters. In a recent proof-of-principle demonstration~\cite{Lupu22},  Lupu-Gladstein \textit{et al.} used a partially post-selective filter to improve the single-photon measurement of a wave plate's birefringent phase. They increased the Fisher information per measured photon by over two orders of magnitude. Thus, the single-photon detectors could measure a low-intensity beam of single photons (a beam below the detector's saturation threshold), whilst 
garnering Fisher information at a high rate. Figure \ref{fig:MetroThree} shows data from their experiment.

\section{Weak Values} \label{sec:wv}
\label{Sec:WeakValues}

Several quantum-mechanical concepts involve post-selection. One of the earliest, and the most famous, example concerns \textit{weak values}~\cite{Aharanov88,Duck89,Tamir13, Dressel14}. The weak value has the form of  an average of an observable, conditioned on a pre-selected initial state and a post-selected final state. Weak values are related to several concepts in the foundations of quantum mechanics (see, e.g., Sec.~\ref{SubSec:GenContext}). Below, we define the weak value and describe a protocol for measuring it (Sec.~\ref{SubSec:WVDefs}). Then, we outline its close relation to the KD distribution (Sec.~\ref{SubSec:WVKD}). Finally, we describe how weak-value-measurement experiments 
can boost the signal-to-noise ratio in metrological estimation of  small unknown parameters (Sec.~\ref{SubSec:WVA}). 

\subsection{Definition}
\label{SubSec:WVDefs}

We first review the theory of weak  values~\cite{Aharanov88,Duck89, Tamir13, Dressel14}, from the perspective of  von Neumann's measurement model~\cite{Neumann18}. 
Consider a meter that interacts with a system of interest. For simplicity, we assume that the meter interacts impulsively with the system  via a unitary generated by the Hamiltonian $H_{\mathrm{int}} = g \, \delta(t) \, {\hat p}\otimes {\hat A}$. ${\hat p}$ denotes the meter's momentum, and $\hat A = \sum_i a_i \ketbra{a_i}{a_i}$ is 
an arbitrary Hermitian system operator. The parameter $g$ denotes the interaction strength, which controls the measurement strength. The system--meter interaction results in an entangling unitary,
\be
\label{Eq:WVUni}
{\hat U}_{\mathrm{SM}} = e^{-i g {\hat p} {\hat A} }.
\ee
We define the initial system and meter state to be $ |\psi_{\mathrm{i}}\ra |\phi\ra$.
We take the weak measurement limit, assuming that $g$ is much less than the meter state's width (in position space).  
We will express the meter state in terms of the position basis $\{ |x_{\mathrm{m}} \ra\} $. After the interaction, the joint system-meter state assumes the form 
\be
\int dx_{\mathrm{m}}  
\ketbra{x_{\mathrm{m}}}{ x_{\mathrm{m}} } {\hat U}_{\mathrm{SM}} |\psi_{\mathrm{i}}, \phi\ra = \sum_i c_i \int dx_{\mathrm{m}}  \phi(x_{\mathrm{m}} - g a_i) |a_i \ra  \ket{x_{\mathrm{m}}} , \label{normshift}
\ee
where $\phi(x_m - g a_i) = \braket{x_m - g a_i | \phi}$ and $c_i = \braket{a_i | \psi_{\mathrm{i}}}$.
The joint state is a superposition of system-and-meter products.
\nicole{The momentum ${\hat p}$ in Eq.  \eqref{Eq:WVUni} spatially translates the meter state by an amount dependent on the system state. The meter's position-basis wavefunction, $\phi(x_m - g a_i)$, encodes a shift by an amount $g a_i$ dependent on the system state.  
$\phi$ has a width much larger than the shifts $g a_i$ in this weak-measurement regime.
The opposite regime involves a narrow $\phi$: The meter state shifts by an amount large compared to the width. In this regime, textbook projective measurement occurs: By measuring the meter, one can discriminate system states unambiguously.   }

\begin{figure}
\includegraphics[scale=0.39]{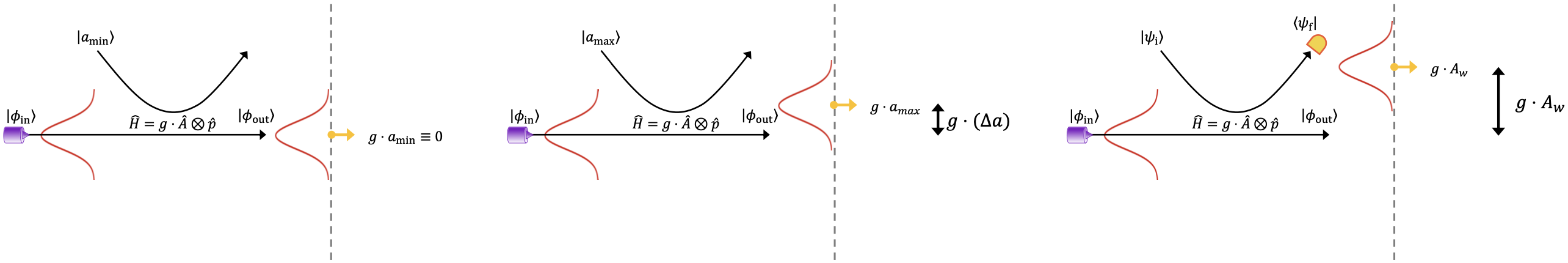}

\caption{
\label{fig:WeakValue}
\textbf{Weak-value amplification.} In (a) and (b), the system particle is initialised in an eigenstate corresponding to the minimum or maximum  eigenvalue of $\hat{A}$. Initially, the meter particle is moving. The meter has a spatial degree of freedom associated with an axis perpendicular to the motion. That degree of freedom is prepared in a state represented, relative to the position eigenbasis, by a Gaussian.  
The system interacts with the meter. A screen detector then measures the meter's position wavefunction. The difference between the maxima in (a) and (b) can be used to estimate the coupling strength $g$. In (c), the system is prepared in $\ket{\psi_{\mathrm{i}}}$, and 
$\bra{\psi_{\mathrm{f}}}$ is post-selected on.
Only if the post-selection succeeds is the meter measured. Then, the shift in the meter's position can lie outside the range defined by the spectral gap of $\hat{A}$. We have re-normalised the position profile in (c) by the post-selection probability.
 }

\end{figure}

After the weak interaction, the system is post-selected.
The weak value is defined as the 
average shift undergone by the meter during the post-selection.  Assume that the meter's initial wavefunction, relative to the position basis,
is Gaussian: $\phi(x) \propto e^{-\frac{x_{\mathrm{m}}^2}{2\sigma^2}} $. [Although we illustrate with a Gaussian example, the following derivation holds for many forms of $\phi(x)$.]
We model the system's post-selection as an arbitrary projective measurement, 
${\hat \Pi}_{\psi_{\mathrm{f}}} 
= \ketbra{\psi_{\mathrm{f}}}{\psi_{\mathrm{f}}}$. 
The conditional meter state $\phi^{\mathrm{PS}}$ has the form
\begin{align}
\phi^{\mathrm{PS}}(x_{\mathrm{m}}|\psi_{\mathrm{f}}) &= \la \psi_{\mathrm{f}} |\la x_{\mathrm{m}} |e^{-i g {\hat p}{\hat A}} |\psi_{\mathrm{i}}\ra |\phi\ra  \nonumber \\
&\approx \la \psi_{\mathrm{f}} |\la x_{\mathrm{m}} |
 \left( \hat{\mathbbm{1}} - i g {\hat p}{\hat A} \right)  
|\psi_{\mathrm{i}}\ra |\phi\ra   \label{approx1}\\
&= \la \psi_{\mathrm{f}} |\psi_{\mathrm{i}}\ra \phi(x_{\mathrm{m}}) -i g \la \psi_{\mathrm{f}} |{\hat A}| \psi_{\mathrm{i}}\ra \la x_{\mathrm{m}} | {\hat p}| \phi\ra \nonumber \\
&\approx \la \psi_{\mathrm{f}} |\psi_{\mathrm{i}}\ra \la x_{\mathrm{m}}| e^{-i g {\hat p} A_{\mathrm{w}}} |\phi\ra \label{approx2}\\
&= \la \psi_{\mathrm{f}} |\psi_{\mathrm{i}}\ra \phi(x_{\mathrm{m}} - g A_{\mathrm{w}}).~\nonumber
\end{align}
In line (\ref{approx1}), we have Taylor-expanded to first order in $g$.  In line (\ref{approx2}), we have factored out the quantity $\la \psi_{\mathrm{f}} |\psi_{\mathrm{i}}\ra$ and re-approximated the linear-order Taylor expansion as an exponential function.  The weak value\index{weak value} of the operator $\hat{A}$ is defined as 
\be
A_{\mathrm{w}} = \frac{\la \psi_{\mathrm{f}} |{\hat A} | \psi_{\mathrm{i}}\ra}{\la \psi_{\mathrm{f}} |\psi_{\mathrm{i}}\ra}.  \label{wv}
\ee
Under the approximations above, the meter's position-basis wavefunction
remains Gaussian 
after the post-selected weak measurement. Thus, the meter's state undergoes an update
\begin{equation}
    \phi(x_{\mathrm{m}}) \propto e^{-\frac{x_{\mathrm{m}}^2}{2\sigma^2}} \rightarrow \phi^{\mathrm{PS}}(x_{\mathrm{m}}|\psi_{\mathrm{f}}) \propto   e^{\frac{-(x_{\mathrm{m}}- A_{\mathrm{w}})^2}{2\sigma^2}} \propto   e^{-\frac{[x_{\mathrm{m}} - g\Re{A_{\mathrm{w}}}]^2}{2\sigma^2}}   e^{i \frac{x_{\mathrm{m}} g \Im{(A_{\mathrm{w}})}}{\sigma^2}} + \mathcal{O}(g^2) .
    \label{meter-final}
\end{equation}
One recognises, in the right-hand side, a position shift of  $g\Re{A_{\mathrm{w}}}$ and momentum shift $g \Im{(A_{\mathrm{w}})}/\sigma^2$ applied to the original wavefunction.  Figure~\ref{fig:WeakValue} provides a schematic overview of a weak-value experiment.

The weak value [Eq.~\eqref{wv}] is defined in terms of a `pre-selected' initial state $\ket{\psi_{\mathrm{i}}}$ and a `post-selected' final state $\bra{\psi_{\mathrm{f}}}$. 
To facilitate the interpretation of the weak value, we let the post-selected-on state $\ket{\psi_{\mathrm{f}}} = \ket{b_{j^{\star}}}$ be one state in the basis $\{\ket{b_j}\}$ of an observable $\hat{B}$. The weak value can  be rewritten as
\begin{align}
    \label{eq_weakhelp1}
    A_{\mathrm{w}}(\psi_{\mathrm{i}},b_{j^{\star}}) &= \frac{\bra{b_{j^{\star}}}\hat{A}\ket{\psi_{\mathrm{i}}}}{\braket{b_{j^{\star}}|\psi_{\mathrm{i}}}} = \frac{\braket{\psi_{\mathrm{i}}|b_{j^{\star}}}\!\bra{b_{j^{\star}}}\hat{A}\ket{\psi_{\mathrm{i}}}}{|\!\braket{b_{j^{\star}}|\psi_{\mathrm{i}}}\!|^2} \, .
\end{align}
To interpret the denominator, we imagine that the system is prepared in $\ket{\psi_{\mathrm{i}}}$, the weak interaction occurs, and $\hat{B}$ is measured. In the limit as $g \rightarrow 0$, we obtain outcome $b_{j^{\star}}$ with a probability $P(b_{j^{\star}}|\psi_{\mathrm{i}}) = |\!\braket{b_{j^{\star}}|\psi_{\mathrm{i}}}\!|^2$.
The weak value satisfies the summation condition
\begin{align}
    \sum_j P(b_{j^{\star}} |\psi_{\mathrm{i}})\,A_{\mathrm{w}}(\psi_{\mathrm{i}},b_{j^{\star}}) = \sum_j \braket{\psi_{\mathrm{i}}|b_{j^{\star}}}\!\bra{b_{j^{\star}}}\hat{A}\ket{\psi_{\mathrm{i}}} = \bra{\psi_{\mathrm{i}}}\hat{A}\ket{\psi_{\mathrm{i}}}.
\end{align}
Thus, the weak value $A_{\mathrm{w}}(b_{j^{\star}},\psi_{\mathrm{i}})$ is a  \emph{conditioned expectation value} of $\hat{A}$. 

$A_{\mathrm{w}}$  has several other interesting properties:  
\begin{itemize}

\item $A_{\mathrm{w}}$ [Eq.  \eqref{wv}] is symmetric with respect to the exchange of $\ket{\psi_{\mathrm{i}}}$ and $\ket{\psi_{\mathrm{f}}}$, 
up to a complex conjugation related to the weak value's time-reversal symmetry.

\item $A_{\mathrm{w}}$ can be a non-real number. As is apparent from Eqs.~\eqref{approx2} and \eqref{meter-final}, the real part of $A_{\mathrm{w}}$ shifts the meter's position-basis wavefunction.
The imaginary part of $A_{\mathrm{w}}$ shifts the meter's conjugate, momentum-basis wavefunction~\cite{Dressel12-2}.  Further properties of the imaginary part of $A_{\mathrm{w}}$ are discussed in \cite{Budiyono_2023_as,Budiyono_PRA_23}.

\item $\text{Re}(A_{\mathrm{w}})$ can lie outside the 
spectrum of $\hat{A}$.

\item The  amplitude of $\phi^{\mathrm{PS}}(x_{\mathrm{m}}|\psi_{\mathrm{f}})$ is reduced by the overlap between the system's initial and final states~\cite{Jordan23}. 
Post-selecting on a final state re-normalises the amplitude.
\end{itemize}

\subsection{Connection between weak values and the KD quasi-probability distribution}
\label{SubSec:WVKD}
Whilst the above-outlined 
explanation of the  weak value is commonly found in the literature, the connection to the KD distribution is not obvious. To further unpack the weak value's structure and reveal the connection to the KD distribution, we recall the spectral expansion 
$\hat{A} = \sum_i a_i \,\ketbra{a_i}{a_i}$. Substituting it into Eq.~\eqref{eq_weakhelp1}, we expand the weak value to
\begin{align}\label{eq:ca}
    A_{\mathrm{w}}(\psi_{\mathrm{i}},b_{j^{\star}}) &= \frac{\bra{b_{j^{\star}}}\hat{A}\ket{\psi_{\mathrm{i}}}}{\braket{b_{j^{\star}}|\psi_{\mathrm{i}}}} = \sum_i a_i \,\frac{\braket{b_{j^{\star}}|a_i}\!\braket{a_i|\psi_{\mathrm{i}}}}{\braket{b_{j^{\star}}|\psi_{\mathrm{i}}}} = \sum_i a_i\,\frac{\braket{b_{j^{\star}}|a_i}\!\braket{a_i|\psi_{\mathrm{i}}}\!\braket{\psi_{\mathrm{i}}|b_{j^{\star}}}}{|\!\braket{b_{j^{\star}}|\psi_{\mathrm{i}}}\!|^2} = \sum_i a_i \,\frac{ Q_{i,j^{\star}}(\psi_{\mathrm{i}}) }{P(b_{j^{\star}}|\psi_{\mathrm{i}})} = \sum_i a_i \tilde{Q}_{i|j^{\star}}(\psi_{\mathrm{i}}) . 
\end{align}
We have defined the KD distribution $Q_{i,j}(\psi_{\mathrm{i}}) \equiv \braket{b_{j}|a_i}\!\braket{a_i|\psi_{\mathrm{i}}}\!\braket{\psi_{\mathrm{i}}|b_{j}}$. Conditioning on $j=j^{\star}$ yields $\tilde{Q}_{i|j^{\star}}(\psi_{\mathrm{i}})$, in accordance with Eq.~\eqref{Eq:KDBayes}. 
The last equality in Eq.~\eqref{eq:ca} shows that the weak value is an average of the $\hat{A}$ eigenvalues with respect to a conditional quasi-probability distribution~\cite{Dressel15}. 
Marginalising the joint distribution $Q_{i,j}(\psi_{\mathrm{i}})$ over $i$ yields the post-selection probability. 
Furthermore, suppose that $\hat{A}$ is a rank-$1$ projector:  $\hat{A} = \ketbra{a_1}{a_1}$. The weak value equals a conditional KD quasi-probability: $\hat{A}_\mathrm{w} (\psi_{\mathrm{i}},b_{j^{\star}}) = \tilde{Q}_{i=1|j^{\star}}(\psi_{\mathrm{i}}) $.

An intuitive \emph{operational meaning} of the  weak value is a conditioned average of the outcomes of weak measurements 
of the observable $\hat{A}$. This intuition is supported by the weak value's formal equivalence to both a conditioned expectation value and a conditioned quasi-probabilistic average of the $\hat{A}$ eigenvalues. Conditioning a von Neumann measurement in the weak-coupling (small-disturbance) limit yields  the KD distribution as the  joint distribution over the outcomes of
the two ordered 
measurements performed. Moreover,  the last equality in Eq.~\eqref{eq:ca} highlights how $A_{\mathrm{w}}(\psi_{\mathrm{i}},b_{j^{\star}}) $ lies between the minimum and maximum $\hat{A}$ eigenvalue if $Q_{i,j}(\psi_{\mathrm{i}})$ is a classical probability distribution. However,  if $Q_{i,j}(\psi_{\mathrm{i}})$ is non-positive, $A_{\mathrm{w}}(\psi_{\mathrm{i}},b_{j^{\star}}) $ can lie outside this range.  In some experiments, such \textit{anomalous} weak values  witness the non-classical phenomenon of contextuality (Sec. \ref{SubSec:GenContext}).
Moreover, anomalous weak values have  been used  to amplify signal-to-noise ratios in experiments (see Sec.~\ref{SubSec:WVA}).

Importantly, the connection between conditional KD quasi-probabilities and measurable weak values allows the quasi-probabilities to be experimentally measured. Thus, despite their anomalous behaviour, KD quasi-probabilities are empirically meaningful as testable predictions. Moreover,  by Eq. \eqref{Eq:KDBayes},  a quantum state's KD representation decomposes into conditional quasi-probabilities
\begin{align}
    Q_{i,j}(\psi_{\mathrm{i}}) &= \tilde{Q}_{i|j}(\psi_{\mathrm{i}}) P(b_j|\psi_{\mathrm{i}}).
\end{align}
Consequently, to infer 
$Q_{i,j}(\psi_{\mathrm{i}})$ one can measure $\tilde{Q}_{i|j}(\psi_{\mathrm{i}}) $ and $P(b_j|\psi_{\mathrm{i}})$ independently. 
As we saw above, we can infer $\tilde{Q}_{i|j}(\psi_{\mathrm{i}}) $ and $P(b_j|\psi_{\mathrm{i}})$ via a weak measurement of $\hat{A}=\ketbra{a_i}{a_i}$, by initialising the system in $\ket{\psi_{\mathrm{i}}}$ and post-selecting on $\ket{b_j}$. To infer an entire KD distribution ${Q}(\psi_{\mathrm{i}}) $, one can perform a set of weak-measurement experiments, scanning through the values of $i$ and $j$. If $\braket{a_i|b_j}\neq 0$ for all $i$ and $j$, then $\tilde{Q}(\psi_{\mathrm{i}}) $ is informationally complete about $\ket{\psi_{\mathrm{i}}}$ [Eq. \ref{Eq:StateDecom}].
This quasi-probability method for  
measuring quantum-state amplitudes has allowed states to be determined and even experimentally tracked through successive dynamical updates via Bayes' rule~\cite{Bamber2014}. Thus, the connection to measurable weak values elevates the KD distribution from an abstract state representation to an experimentally significant and practical representation. Section~\ref{Sec:StateMeas} contains further details about the KD distribution in relation to quantum-state measurements.

\subsection{Weak-value amplification}
\label{SubSec:WVA}

Perhaps the best-known application of weak values is weak-value amplification (Fig.~\ref{fig:WeakValue}).  
Consider the weak-value experiment discussed in Sec.~\ref{SubSec:WVDefs}. Assume that the interaction strength $g$ is small and unknown. Weak-value amplification improves measurements of $g$.  
The fact that the meter degree of freedom can be shifted arbitrarily far may be viewed as an amplified response of the combined system. If  the resulting signal is re-scaled by the weak value, the parameter $g$ can be measured precisely. This technique is now widely used and is applied to many record-breaking metrology experiments ~\cite{Hosten08,Dixon09,Viza15}.

Weak-value amplification falls under the above-reviewed topic of post-selected metrology (Sec.~\ref{SubSec:PSMetrology}). However, in its simplest form, weak-value amplification is most easily understood not in terms of Fisher information, but in terms of the signal-to-noise ratio (SNR).   

We can compare the final meter state (described in Sec.~\ref{SubSec:WVDefs}) realised in experiments that involve post-selection and experiments that do not.
We can estimate $g$ by  measuring the meter's position on a detector.  We assume that experimental repetitions are uncorrelated. 
The SNR quantifies the estimation's sensitivity:
\be
{\cal R}^{\mathrm{SNR}} = \sqrt{N} \frac{|\la x_{\mathrm{m}} \ra|}{\sqrt{{{\rm Var}(x_{\mathrm{m}} ) }}} \, .
\ee
${\rm Var} (x_{\mathrm{m}}) = \la x_{\mathrm{m}}^2\ra - \la x_{\mathrm{m}}\ra^2$ is the variance in the position-measurement outcomes.
$N \gg 1$ denotes the number of experimental trials.  ${\cal R}^{\mathrm{SNR}}$ quantifies our ability to  distinguish 
the detected signal's mean from the measurement noise.

We can  apply these concepts to the weak-value-amplification protocol and contrast it with a weak-value-free alternative.
Let us return to the state in Eq.~\eqref{normshift}. Let the system be prepared in the eigenstate associated with the maximum $A$ eigenvalue $a_{\mathrm{max}} = 1$. Let the meter distribution, be Gaussian with a width $\la x_{\mathrm{m}}^2\ra = \sigma^2$.  The SNR has the form
\be
{\cal R}^{\mathrm{SNR}} _{\mathrm{s}} = \frac{ \sqrt{N} g}{\sigma} \, . \label{snr}
\ee
We have defined $N$ as the number of weak-measurement trials (Fig.~\ref{fig:WeakValue}). The trials begin with the same initial system and meter states. We assume that all meter--system interactions have the same strength $g$, for simplicity. 

In contrast, consider the weak-value-amplification protocol. For simplicity, suppose that the weak value is real and positive:  
$A_{\mathrm{w}} = \text{Re}(A_{\mathrm{w}}) > 0$.
The SNR almost has the form of~(\ref{snr}).
However, we must replace $g$ with $A_{\mathrm{w}} g$ and $N$ with  $N_{\rm PS}$, the number of trials that pass the post-selection
(the meter Gaussian's width remains unchanged in the simplest weak-value amplification):
\be
{\cal R}^{\mathrm{SNR}} _{\mathrm{wv}} = \frac{ \sqrt{N_{\rm PS}} \, A_{\mathrm{w}} g}{\sigma} \, .
\label{rwv}
\ee 
By Eq.~\eqref{meter-final}, the meter wavefunction's peak is shifted by an amount $A_{\mathrm{w}} g \gg g$, exhibiting signal amplification.  
The meter's final wave function has the same width as its initial wave function.  The post-selection's probabilistic nature implies that the number $N_{\mathrm{PS}}$ of data points, collected from $N$ trials of the experiment, will vary from batch of trials to batch of trials, if we run multiple batches of $N$ trials.   
We calculate the post-selected SNR by replacing $N_{\mathrm{PS}}$
with the expected number of data points, $ N |\la \psi_{\mathrm{f}} | \psi_{\mathrm{i}}\ra|^2 + \mathcal{O}(g^2)$, the post-selection probability times the number of trials (to within a small correction).   The SNR becomes
\begin{equation}
    {\cal R}^{\mathrm{SNR}} _{\mathrm{wv}} = {\cal R}^{\mathrm{SNR}} _{\mathrm{s}} \la  \psi_{\mathrm{f}} | {\hat A}| \psi_{\mathrm{i}}\ra
=
{\cal R}^{\mathrm{SNR}} _{\mathrm{s}} \la  \psi_{\mathrm{f}} |{\hat A} | \psi_{\mathrm{i}} \ra.
\end{equation}
In many cases, the coefficient 
$\la  \psi_{\mathrm{f}} |{\hat A} | \psi_{\mathrm{i}} \ra$
of proportionality can be made to lie close to $1$, matching the standard measurement's SNR.  This calculation was first published (to the best of our knowledge) in Ref.~\cite{Starling09}.

The conclusion of this calculation is that, at best, weak-value amplification achieves the standard technique's precision.  However, something remarkable has happened:  We can obtain the same precision as in the standard case, but by 
performing final measurements in
a tiny fraction $N_{\mathrm{PS}} \ll N$ of the trials.
This calculation relied on an assumption: Initiating trials is easy, whereas the final measurement is a valuable resource.
Often in optics, for example, the limiting resource
is not the amount of power that a laser can emit, but the amount of power that a detector can receive.  From this perspective, if we regard the number of detected photons as the limiting resource, weak-value amplification achieves a huge advantage: Consider matching the number of photons detected in the standard scheme to the number detected in the weak-value-amplification scheme. We boost  the SNR by a factor equal to the weak value, $A_{\mathrm{w}} \gg 1$.

Additionally, weak-value amplification offers many other advantages.  Some are:
\begin{itemize}
    \item Often, 
    the weak-value experiment's setup is more robust with respect to technical noise than a standard experiment's setup is.  A simple example involves optical-beam-deflection measurements. In such measurements, 
    slight turbulence can 
    limit the precision of the measurement of a beam-deflection angle.  The standard measurement technique focuses down an optical beam, to reduce the surface area of the exposed transverse region of the detector.
    A slight beam deflection translates the beam laterally by the greatest possible amount, relative to the beam profile, giving the measurement the greatest sensitivity. 
    The weak-value method (in this case, relying on the weak value's imaginary part) makes the beam profile as large as possible. In principle, both techniques offer the same shot-noise-limited precision. However, the weak-value-amplification method vastly outperforms the standard method in the presence of turbulence~\cite{Jordan14}. 
    
    \item 
    If the system suffers from {\it systematic error}, averaging over more trials does not suppress it. 
    The reason is that systematic error affects accuracy, not precision. Weak-value amplification can benefit also this situation~\Dave{\cite{feizpour2011amplifying, Jordan14, magana2014amplification, Pang16, sinclair2017weak}}.  Consider amplifying the signal by the weak value. 
    Dividing by the weak value, to estimate the parameter $g$, can suppress the systematic error by the weak value. 
    A simple example concerns bias offset. Suppose
    that we aim to measure the origin $g$ of a meter.
    Let the meter's true origin differ from the assumed origin by a small amount $\epsilon$.  
    Consider averaging over measurement outcomes from the set $\{x_{\mathrm{m}}^{(i)}\}$. No matter how much data you have, the average
    will deviate from the true $g$ value 
    by $\epsilon$, which affects each measurement. Under the weak-value-amplification scheme, however, the measured average is $A_{\mathrm{w}} g$, which has the same error, $\epsilon$.  Consider dividing the measured data by $A_{\mathrm{w}}$ to estimate $g$. The systematic error is reduced from $\epsilon$ to $\epsilon/A_{\mathrm{w}} \ll \epsilon$,  improving the accuracy.  We stress that this reduction is independent of the amount of data.  However, if 
    a systematic error obscures the to-be-measured parameter $g$,
    the weak-value-amplification method will not help: The systematic error will be amplified along with $g$.
    
    \item 
    In weak-value amplification, one post-selects the data that satisfy the post-selection criterion. Yet you need not discard the remaining particles, which need not even be measured destructively.  They can be reused in another context~\cite{Starling10} or recycled.  Indeed, optimal experiments have realised the following outcome: Even in the ideal case, free of technical noise, re-injecting non-post-selected photons into an optical interferometer has improved the SNR~\cite{Dressel13,Lyons15,Krafczyk21,Wang16}.
   \end{itemize}

\section{The KD distribution and direct measurements of quantum states}

\label{Sec:StateMeas}

\nicole{
We now discuss ways of measuring a quantum state's KD distribution, focusing on \emph{direct measurement}. Means of experimentally determining quantum states has practical and foundational applications. The problem's history encompasses contributions by von Neumann~\cite{Birkhoff1936}, Fano~\cite{Fano1957}, Pauli~\cite{Pauli1980} and others.}
\nicole{\emph{Prima facie}, the quantum state seems} abstract---a complex-valued amplitude distribution, of which one could glimpse only limited features in a measurement. Not until the 1990s was quantum tomography invented~\cite{Vogel1989} and demonstrated~\cite{Smithey1993}. 
The first studies centred on a light wave's quantum state, but tomography of other systems followed quickly. 

The term \emph{quantum tomography} is sometimes used to mean quantum-state determination. However, it is actually a specific procedure. It involves measuring in diverse bases (of the Hilbert space directly associated with a system) 
that, together, span the density-operator space. Performing these measurements on an  ensemble of identical systems, 
one can infer real-valued probability distributions that 
resemble shadows of the complex state. From these distributions, one can reconstruct the state~\cite{jordan2016mapping}.\footnote{The word \emph{tomography} stems from \emph{tomos} (slice) and \emph{graphy} (writing). In medicine, two-dimensional X-ray shadows (slices), taken at different angles, are used in the computer-aided tomographic reconstruction (a CAT scan) of a three-dimensional object---for example, a skull.} A measurement that directly outputs the KD distribution (up to a normalisation factor) removes the need for such a reconstruction. We show that this elimination \Dave{has practical and fundamental implications.}

Direct measurements of quantum states are 
used to 
infer several quasi-probability distributions, not only  the KD distribution. Direct measurement comprises a wide
range of measurement procedures, which can be `direct' in some or
all of these senses: 
\begin{enumerate}
   \item The procedure does not require a complicated mathematical
reconstruction (as described just above). 

   \item The procedure is local: It measures the state's quasi-probability amplitude [e.g., $Q_{i,j}(\hat{\rho})$] at a given point in phase space $(a_{i},b_{j})$. 
   
   \item That amplitude's value appears directly on a measurement apparatus. 
   
   \item The experimental procedure is simple and general. For example, one might measure $\hat{A}$ and then $\hat{B}$. Tomography, in contrast, requires many measurements, potentially in exotic bases that are not easily accessible experimentally.  
\end{enumerate}
Direct measurement based on weak measurement has all these properties.
The laws of quantum physics (e.g., the no-cloning theorem \cite{Wootters82}) require that, to determine an arbitrary quantum state, we must measure an  ensemble of identical systems. 
Thus all direct-measurement procedures involve averages over measurement outcomes from an ensemble, similarly to quantum tomography. However, direct measurement offer advantages over tomography.

Direct measurements of discrete and continuous KD distributions determine quantum states.
The most prominent results have concerned direct measurements of continuous-variable quantum states. We summarised a few properties of the continuous KD distribution in Sec.~\ref{SubSec:ContinuousKD}. Now, in Sec.~\ref{SubSec:KDMeas}, we outline how to measure quantum states directly. We describe two strategies: One involves the Husimi distribution and no weak measurements, and the other involves weak measurements and the KD distribution.
In  Sections~\ref{SubSec:general_direct} and~\ref{SubSec:procedure_direct}, we review generalisations of, and other procedures, for direct measurements.

\subsection{Direct quantum-state measurements}
\label{SubSec:KDMeas}

Measuring a 
classical point particle's position and momentum directly determines its state. 
In quantum physics, one cannot precisely and simultaneously measure a particle's position [Eq.~\eqref{eq_position}] and momentum [Eq.~\eqref{eq_momentum}]. Nor can one even ascribe precise values to $\hat{x}$ and $\hat{p}$ simultaneously.
One can measure position and momentum simultaneously~\cite{Arthurs1965}, if the information obtained about both quantities is imprecise. Consider imprecision in the form of uncertainty: $x\pm\Delta x$. Reference~\cite{Leonhardt1993} concerns minimal measurement-device uncertainty, which is
balanced between simultaneous measurements of position and momentum: $\Delta x=\Delta p$. The outcome $(x^{\prime},p^{\prime})$ corresponds to a projection onto  a coherent state $\left|\alpha^{\prime}=x^{\prime}+ip^{\prime}\right\rangle $. 
This outcome's probability is proportional to the Husimi quasi-probability distribution: $\mathbb{P}(x^{\prime},p^{\prime})
=\mathrm{Tr}  \left(  \ketbra{ \alpha^{\prime} }{ \alpha^{\prime} } \hat{\rho}  \right) 
\propto H_{x^{\prime},p^{\prime}}(\hat{\rho})$. 
The Husimi distribution $H_{x,p}(\hat{\rho})$ is a faithful representation of the quantum state $\hat{\rho}$~\cite{Leonhardt1993}. In this sense, 
one can extract information about the quantum system, as from a classical particle, at a phase-space point $(x^{\prime},p^{\prime})$, using imprecise measurements. One thereby directly measures the state.

Similarly to the Husimi distribution, the KD distribution corresponds to the measurement of a point in phase space. However, the continuous-variable KD distribution  \Dave{$ Q(x,p;\psi)$ (see Sec. \ref{SubSec:ContinuousKD})} does not correspond to projections onto a state such as $\left|\alpha\right\rangle $,
which has finite uncertainties, $\Delta x$ and $\Delta p$. 
\nicole{Recall the rightmost side of Eq.~\eqref{eq:KD_continuous}, 
\begin{equation}
 Q(x,p;\psi)=\mathrm{Tr}\left(\hat{\pi}_{p}\hat{\pi}_{x}\hat{\rho}\right) ,
\end{equation}
wherein $\hat{\pi}_{x} \equiv \ketbra{x}{x}$ and $\hat{\pi}_{p} \equiv \ketbra{p}{p}$.
This equation appears to encode an expectation value of 
$\hat{\pi}_{p}\hat{\pi}_{x}.$  }
This observation suggests that the KD distribution corresponds to projecting onto 
 a position eigenstate $\ket{x}$ and a momentum eigenstate $\ket{p}$  (such that $\Delta x=\Delta p=0$),  
at least intuitively. However, $\hat{\pi}_{p}\hat{\pi}_{x}$ is not Hermitian and thus not an observable. If it were an observable, a measurement of $\hat{\pi}_{p}\hat{\pi}_{x}$  would violate
uncertainty principles, as it would simultaneously fix  position and momentum. But a direct measurement of the KD distribution need only output the \textit{expectation value} of $\hat{\pi}_{p}\hat{\pi}_{x}$---output 
$\mathrm{Tr}\left(\hat{\pi}_{p}\hat{\pi}_{x}\hat{\rho}\right)$. 
This value need not be determined from a precise measurement in a particular trial.
Rather, one can determine the value
by averaging the imprecise results of a sequence of measurements performed on an ensemble of identically prepared quantum systems. In this sense, as we discuss next, the non-Hermitian operator $\hat{\pi}_{p}\hat{\pi}_{x}$ can be can be `measured'. In turn, the KD distribution can be directly measured experimentally. 

One way to measure the average of $\hat{\pi}_{p}\hat{\pi}_{x}$, and
directly determine the KD distribution, is to use weak measurements.
(For details about weak measurements, see Sec.~\ref{Sec:WeakValues}.)
Weakly measuring $\hat{\pi}_{x}$ leaves a quantum state $\hat{\rho} = \ketbra{\psi}{\psi}  $ 
mostly unchanged. Therefore, one can perform a subsequent strong (ordinary) measurement of 
$\hat{\pi}_{p}$, obtaining information about the initial state. Post-selected on a momentum value $p$ (in a sub-ensemble of the ensemble of all momentum states), 
the position-projector weak measurement's average result is the weak value, Eq.~\eqref{eq:ca}:
\begin{equation}
\pi_{x}{}_{\mathrm{w}} (\psi, p) =\frac{\left\langle p\right|\hat{\pi}_{x}\left|\psi\right\rangle }{\left\langle p|\psi\right\rangle } = \frac{\left\langle p\right|\hat{\pi}_{x}\left|\psi\right\rangle \braket{\psi | p} }{|\left\langle p|\psi\right\rangle |^2} \equiv \tilde{Q}(x;\psi|p) 
. \label{eq:weak_value_KD}
\end{equation}
[We have reproduced Eq.~\eqref{wv}, with $\hat{A} = \hat{\pi}_x$ and $\bra{\psi_{\mathrm{f}}} = \bra{p}$.] In what follows, we  denote the 
system's momentum and position by $x$ and $p$. We denote the meter's position by $x_{\mathrm{m}}$. As outlined in Sec.~\ref{SubSec:WVKD}, the conditional KD distribution is related to an observable's weak value. When the observable is a projector, as here, the weak value equals a pure state's conditional KD-distribution density: $\tilde{Q}(x;\psi|p)$. 

There is an even tighter connection between the KD distribution and weak measurements. Consider performing a weak measurement, without post-selecting on the sub-ensemble that realises outcome $p$. This non-conditioned weak measurement's average result is sometimes called the \emph{weak average}. We denote it by $\left\langle \hat{\pi}_{p}\hat{\pi}_{x}\right\rangle _{\mathrm{w}}$ and define it below, in Eqs.~\eqref{eq:wa} and \eqref{eq:weak_avg_KD}. Recall that the weak value probes whether the system first had the position $x$, \textit{given that} it subsequently had momentum $p$. In contrast, the weak average probes whether the system had properties $x$ \textit{and} then $p$.
In Sec.~\ref{SubSec:WVDefs}, we derived an expression [Eq.~\eqref{meter-final}] for a meter's late-time position-basis wavefunction. The meter's wavefunction was initially a Gaussian, and then the system was pre- and post-selected. We can apply this result to calculate the meter's final spatial probability density in a weak-value experiment intended to measure   
$\pi_{x}{}_{\mathrm{w}} (\psi, p)$: 
$|\phi^{\mathrm{PS}} (x_{\mathrm{m}} | p) |^2 
\propto \nicole{ \exp \big(
[x_{\mathrm{m}} - g \mathrm{Re} \LParen \pi_{x}{}_{\mathrm{w}} (\psi, p) \RParen ]^2 / \sigma^2 \big)
}
+ \mathcal{O}(g^2) $.
Again, $g$ denotes the weak measurement's strength. In other words, $\mathbb{P}(x_{\mathrm{m}}|p) \equiv |\phi^{\mathrm{PS}} (x_{\mathrm{m}}|p) |^2$ is the probability density associated with the meter's being at position $x_{\mathrm{m}}$, if the system's momentum was post-selected on being $p$. For simplicity, we assume that the weak value is real. 
One experimentally infers the weak value from the shifted meter's mean final position,
$\bar{x}_{\mathrm{m}}$ (see Sec.~\ref{SubSec:WVDefs} and Fig.~\ref{fig:WeakValue}).  
The average is calculated over the conditional joint system--meter measurement distribution
$\mathbb{P}(x_{\mathrm{m}}|p)$.
That is, $\frac{1}{g} \pi_{x}{}_{\mathrm{w}} (\psi, p) \approx \bar{x}_{\mathrm{m}}=\left\langle \hat{x}_{\mathrm{m}}\right\rangle 
=\int x_{\mathrm{m}}  \,  \mathbb{P}(x_{\mathrm{m}}|p)  \, d x_{\mathrm{m}}$.  
The weak average manifests similarly but depends on the unconditional joint distribution, $\mathbb{P}(x_{\mathrm{m}},p) 
\equiv \mathbb{P}(x_{\mathrm{m}}|p)  \,  \mathbb{P}(p) 
= |\phi^{\mathrm{PS}} (x_{\mathrm{m}}|p) |^2\left|\left\langle p|\psi\right\rangle \right|^{2}$.
That is, $\left\langle \hat{\pi}_{p}\hat{\pi}_{x}\right\rangle _{\mathrm{w}}\propto\left\langle \hat{x}_{\mathrm{m}}\hat{\pi}_{p}\right\rangle 
=\int x_{\mathrm{m}}  \,  \mathbb{P}(x_{\mathrm{m}},p)  \,  d x_{\mathrm{m}}$.   Consequently, the weak average is 
\begin{equation}
    \label{eq:wa}
    \left\langle \hat{\pi}_{p}\hat{\pi}_{x}\right\rangle _{\mathrm{w}} \equiv \pi_{x}{}_{\mathrm{w}} (\psi, p) \cdot\mathbb{P}(p)=\left\langle p\right|\hat{\pi}_{x}\left|\psi\right\rangle \left\langle \psi|p\right\rangle 
    =\mathrm{Tr}\left(\hat{\pi}_{p}\hat{\pi}_{x}  \ketbra{\psi}{\psi}\right) .
\end{equation}
The generalisation 
\nicole{beyond pure} states  follows from the linearity of $\hat{\rho}$: 
\begin{equation}
\left\langle \hat{\pi}_{p}\hat{\pi}_{x}\right\rangle _{\mathrm{w}}
\equiv\mathrm{Tr}\left(\hat{\pi}_{p}\hat{\pi}_{x}\hat{\rho}\right)
=Q(x,p;\hat{\rho}).\label{eq:weak_avg_KD}
\end{equation}
These results were proved in~\cite{Johansen2008,Lundeen2012}. In summary, a weak measurement of $\hat{\pi}_{x}$, followed by strong measurement of $\hat{\pi}_{p}$, directly yields the KD representation of the state. 

How, then, are weak measurements enabling the simultaneous measurement of position and momentum? As alluded to above, one must forfeit precision. However, unlike in the direct measurement of the Husimi distribution, the imprecision must not be
in the form of position or momentum uncertainty, $\Delta x$ or $\Delta p$. After all, we are now nominally measuring 
$\hat{\pi}_{p}$ and $\hat{\pi}_{x}$ eigenstates, which respectively have zero momentum and position uncertainties.
The answer was 
presented in~\cite{Thekkadath2018}. It comes from the fact that, in the weak-measurement regime, the average meter shift is small, compared to the meter's original position uncertainty (i.e., the system-meter coupling $g$ is small). Consequently, in any given trial, little information is gained from a weak measurement: 
Weak measurements are inherently imprecise. For example, 
measuring a system observable $\hat{\pi}_{x}$ weakly reduces
a type of precision called \emph{predictability},
$\mathbb{P}(x|x_{\mathrm{m}})$~\cite{Thekkadath2018}. (Recall that $x$ denotes the system's position and $x_{\mathrm{m}}$ denotes the meter's position.) 
If $x$ and $x_{\mathrm{m}}$ are such that $\mathbb{P}(x|x_{\mathrm{m}})\ll1$, the meter's position  reveals little about the system's position in any one trial. Only by averaging over many trials can the reduction in per-trial information be overcome and can 
$\left\langle \hat{\pi}_{p}\hat{\pi}_{x}\right\rangle _{\mathrm{w}}=Q(x,p;\hat{\rho})$
be determined with little statistical uncertainty. 

\begin{figure}
\includegraphics[scale=0.5]{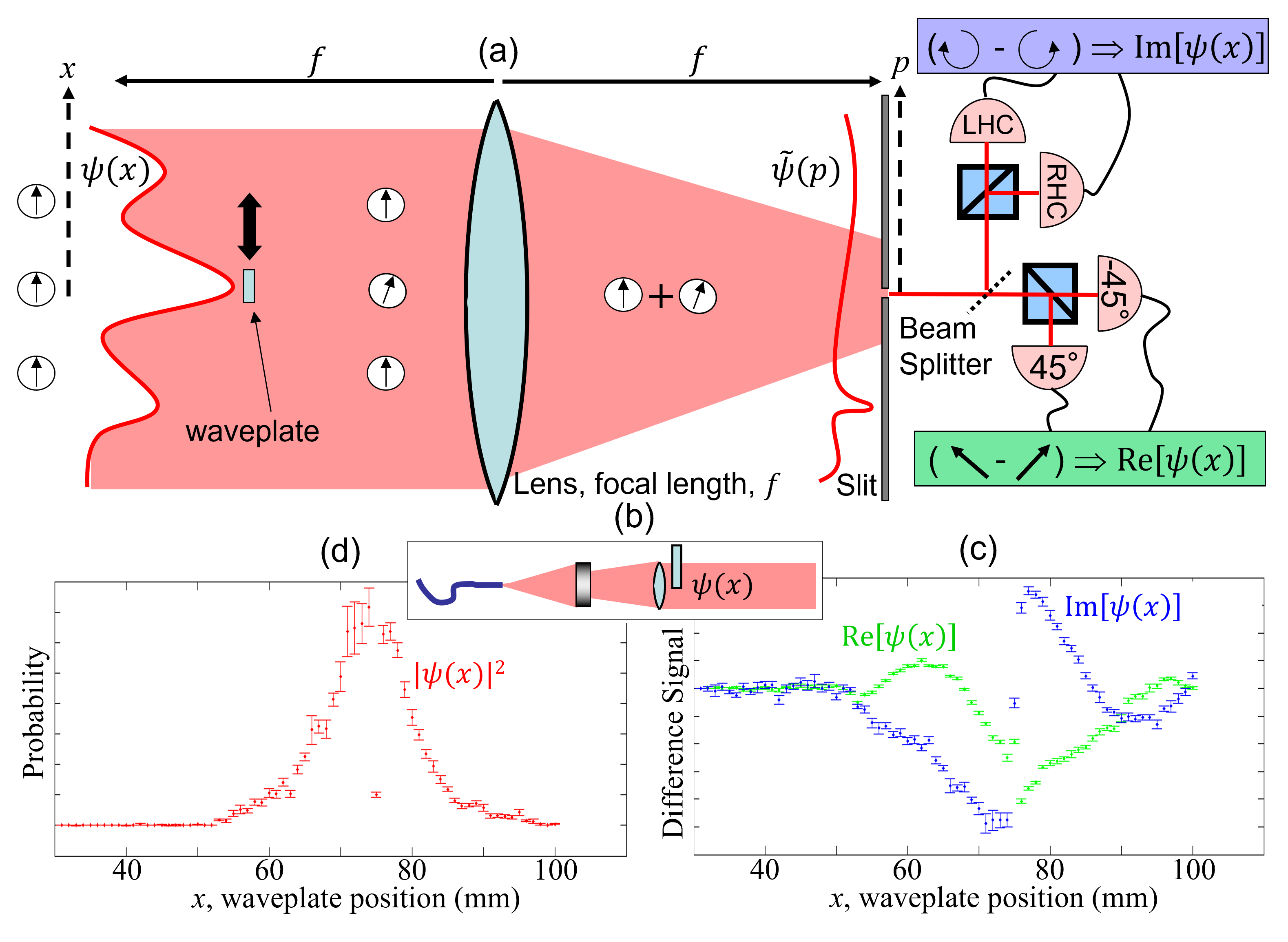}

\caption{\label{fig:Exp-Direct-Meas-1} 
\textbf{Experimental direct measurement of a photon's spatial wavefunction with weak measurement. } 
See~\cite{Lundeen2011} for details. 
(a) The photon has an unknown transverse-position wavefunction $\psi(x)$. Instead of using another particle as the weak measurement's meter, one can use an internal degree of freedom of the photon: The photon's spin (polarisation).
(See~\cite{Lundeen2005,Wu2009} for details about using spin as a meter.) Let $\hat{\sigma}_{i}$ denote the Pauli-$i$ operator, for $i = x, y, z$. The meter begins in the vertical polarisation state, in which $\langle \sigma_{z} \rangle = +1$.
One couples the observable 
$\hat{\pi}_{x}= \ketbra{x}{x}$
to the photon spin by placing a waveplate at the position $x$. The waveplate rotates the polarisation by a small amount, implementing the system--meter measurement interaction $\hat{U}_{\mathrm{SM}}=\exp\left(ig\hat{\pi}_{x}\hat{\sigma}_{y}\right)$. 
Next, the photon passes through a Fourier-transform lens. At the lens' focal plane, a slit effectively projects the momentum with
$\hat{\pi}_{p}=\ketbra{p}{p}$, where $p=0$. 
Consider the rotation of the polarisation from the vertical, averaged across the photons that pass through the slit. This average rotation
is proportional to the weak value $\pi_{x}{}_{\mathrm{w}} (\psi, p)$~\cite{Lundeen2005,Wu2009} (analogously to the position shift of a position meter). Particularly sensitive to this rotation 
is the \emph{difference signal}, the number of $45^{\circ}$-polarised photons, minus the number of $-45^{\circ}$-polarised photons. Indeed, Eq.~\eqref{eq:weak_value_KD} implies that $\mathrm{Re} \LParen \psi(x) \RParen
\propto\mathrm{Re}  \LParen  \pi_{x}{}_{\mathrm{w}} (\psi, p)  \RParen
\propto\left\langle \hat{\sigma}_{x}\right\rangle =P(45^{\circ})-P(-45^{\circ})$.
Similarly, the meter rotation in the conjugate basis is the difference signal between the right-hand circular (RHC) polarised photons and the left-hand circularly (LHC) polarised photons: $\mathrm{Im} \LParen \psi(x) \RParen
\propto  \mathrm{Im}\LParen  \pi_{x}{}_{\mathrm{w}} (\psi, p)\RParen  
\propto\left\langle \hat{\sigma}_{y}\right\rangle =P(\mathrm{RHC})-P(\mathrm{LHC})$.
These two difference signals 
are proportional to the wavefunction's real and imaginary parts at $x$. Suppose that the waveplate's position, $x$, is scanned along the $x$-axis. The two signals reveal the full complex distribution of the wavefunction $\psi(x)$. (b) Various $\psi(x)$ distributions were measured directly. In one example, experimentalists prepared a $\psi(x)$ by beginning with photons that were in the fundamental spatial mode (approximately Gaussian in shape) of an optical fiber. The photons were then transmitted through a reverse bulls-eye filter (narrowing the Gaussian), then through a collimating lens, and lastly through a glass plate that covered roughly half the extent of $\psi(x)$. (c) $\psi(x)$, as given by the two measured difference signals. The measured $\psi(x)$ exhibits an abrupt phase shift at $x\approx75$ mm. This shift resulted from the extra phase accumulated during the passage through the glass plate. This demonstrates that direct measurement is sensitive to phase, unlike a strong measurement of position:
$\left\langle \hat{\pi}_{x}\right\rangle =\mathbb{P}(x)$. 
(d) Despite this phase discontinuity, the inferred probability distribution, $\left|\psi(x)\right|^{2}=\mathrm{Re} \LParen \psi(x) \RParen^{2}
+\mathrm{Im} \LParen \psi(x) \RParen^{2}$,
is smooth. One should expect this smoothness, since the glass transmits the photons. For examples of other directly measured $\psi(x)$ distributions, see~\cite{Lundeen2011}.}
\end{figure}

The direct measurement of the quantum wavefunction was first
proposed as a process involving weak measurements. It was first experimentally demonstrated for a photon's transverse spatial wavefunction $\psi(x)$ (Fig.~\ref{fig:Exp-Direct-Meas-1}) 
\cite{Lundeen2011}. If one weakly measures $\hat{\pi}_{x}$ and  post-selects on $p=0$, Eq.~\eqref{eq:weak_value_KD} gives the weak value:  $\pi_{x}{}_{\mathrm{w}} (\psi, p{=}0) =\braket{p{=}0|x}\!\braket{x|\psi}/\braket{p{=}0|\psi}
=\psi(x)/[\psi_{\mathrm{f}}(0)\sqrt{2\pi\hbar} \, ]
=K\psi(x)$. 
We have combined the constant factors into $K$. Thus, the weak value is proportional to the complex wavefunction. Unlike in quantum tomography, however, the wavefunction appears directly on the measurement apparatus here. 
In the experiment, the photon's spin degree of freedom (polarisation) served as a meter. The spin was coupled to the photon's transverse position $x$ as follows with help from a small waveplate. The waveplate was placed so as, at a chosen position $x$, to slightly rotate the photon's linear polarisation. Then, a Fourier-transform lens and slit measured $\hat{\pi}_{p=0}$. As the waveplate was moved along $x$, the weak value was recorded for each value of $x$, giving the real and imaginary parts of $\psi(x)$. In this sense, one can directly observe the complex wavefunction associated with an  ensemble of identically prepared  quantum particles, up to the constant normalisation factor $K$.

\subsection{Generalisations of direct measurement }
\label{SubSec:general_direct}

Soon after the first demonstration of direct measurement, generalisations began appearing. 
Different photon quantum states have been measured directly; examples include the polarisation~\cite{Salvail2013,Thekkadath2016,Kim2018}, 2D-spatial~\cite{Zhang2020}, orbital-angular-momentum~\cite{Malik2014,Bolduc2016} and time-frequency states~\cite{Ogawa2021}. 
Non-photonic systems have also been measured directly; examples include matter waves~\cite{Denkmayr2017} and a nuclear-magnetic-resonance quantum processor~\cite{Lu2014}. 

Alternatively, one can generalise the post-selection. Consider measuring the weak average for every $p$ value, rather than solely at $p=0$. (One could, for example, use a camera instead of a slit.) Then, as Eq.~\eqref{eq:weak_avg_KD} shows, one would directly measure the KD distribution, 
rather than the wavefunction~\cite{Lundeen2012}.  \Dave{Further work has established a relationship between weak values and \nicole{probability distributions over simultaneous measurements' possible outcomes~\cite{ozawa2011universal}.  }  }

Moreover, direct measurement generalises straightforwardly to discrete quantum states. Hence the KD representation of a photon's polarisation state was soon measured directly~\cite{Salvail2013}. 
Using all $p$  enables the direct measurement of mixed quantum states.

\begin{figure}
\includegraphics[scale=0.5]{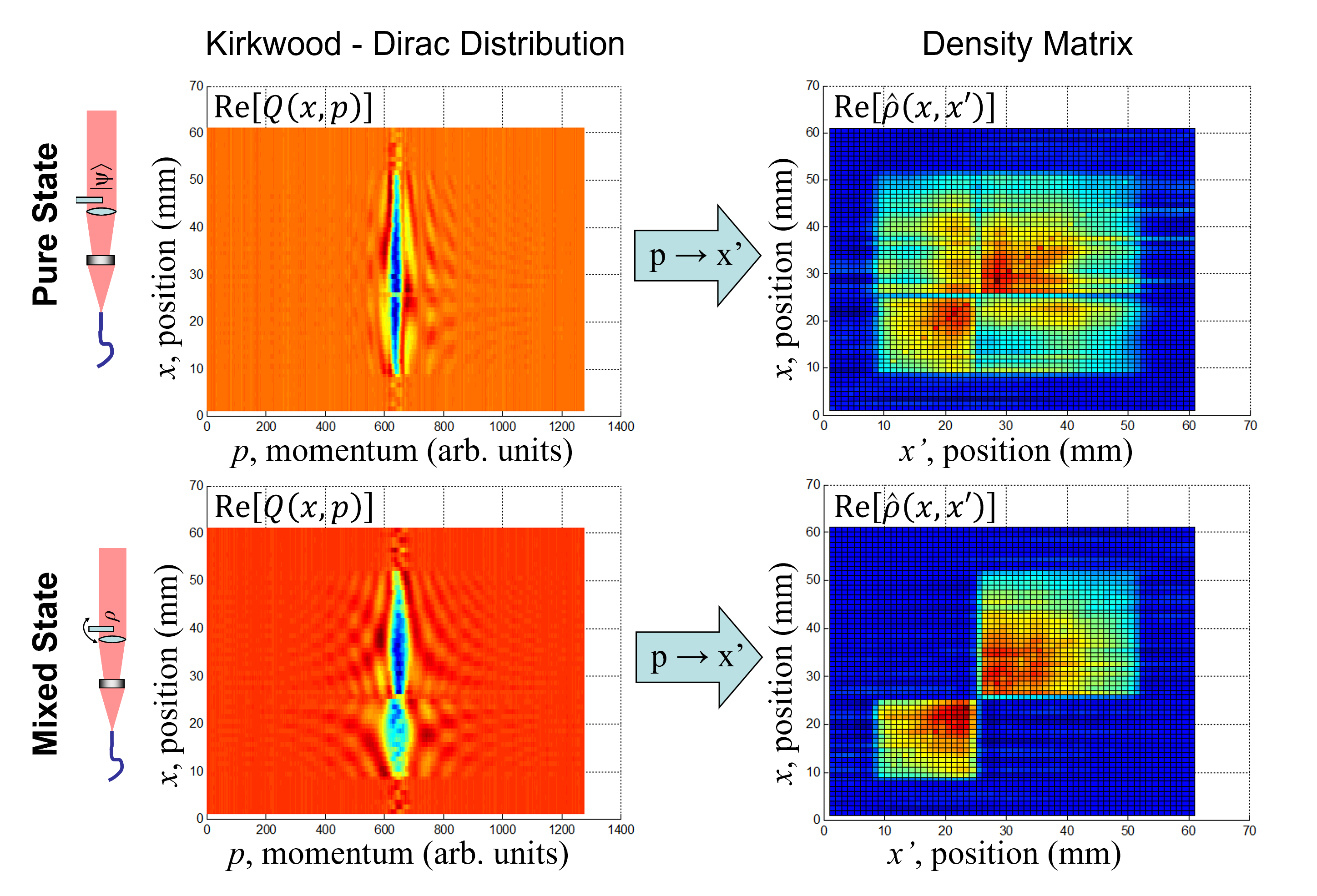}

\caption{\label{fig:KD-Dist-1} 
\textbf{Experimentally measured KD distributions of a photon's the transverse-position state. } 
See~\cite{Bamber2014} for details.  The experimental setup resembles that in Fig.~\ref{fig:Exp-Direct-Meas-1}. However, the slit is replaced by a camera, so that the polarisation differences are observed at every momentum value $p$. The observed KD distributions for the pure state in Fig.~\ref{fig:Exp-Direct-Meas-1}b (top) and a mixed state (bottom). To prepare the mixed state, one vibrates the glass plate in Fig.~\ref{fig:Exp-Direct-Meas-1}.
The vibration creates phase decoherence between the spatial photon state's two halves. On the right-hand side is the density matrix calculated from the KD distribution. As expected, the mixed state's density matrix exhibits no coherence between the two spatial regions (the off-diagonal elements vanish). Only the real components are shown, for simplicity.}
\end{figure}

A central concept in imaging and optical coherence is a photon's transverse position. One of the first determinations, by any method, of the mixed state of a photon's transverse position was accomplished via direct measurement of the KD distribution~\cite{Bamber2014}. Figure~\ref{fig:KD-Dist-1} shows the procedure, which leveraged weak measurement. By extending the concept to a sequence of $k$ weak measurements, Ref.~\cite{Lundeen2012} showed that a $k$-extended KD distribution could be measured. 
Reference~\cite{Lundeen2012} reported on a sequence of three weak measurements of projectors in two complementary bases, $\left\{ \ket{ a_{i}} \right\} $ and $\left\{ \ket{b_{k}} \right\} $.
The following density matrix was obtained:
\begin{equation}
\label{eq:direct_rho}
\rho_{a_{i},a_{j}}\equiv\braket{a_i | \hat{\rho}|a_j}\propto\left\langle\hat{\pi}_{a_{i}}\hat{\pi}_{b_{k}}\hat{\pi}_{a_{j}}\right\rangle _{\mathrm{w}}
\equiv \mathrm{Tr}\left(\hat{\pi}_{a_{i}}\hat{\pi}_{b_{k}}\hat{\pi}_{a_{j}}\hat{\rho}\right)
\equiv Q_{i,k,j}(\hat{\rho}). 
\end{equation}
Reference~\cite{Yokota2009,Lundeen2009} reported on the measurement of the real parts of weak values of two-photon entangled states.
Although these papers' procedure and results 
corresponded to direct measurement, the paper predate the concept and, thus, do not refer to it. Not until~\cite{Pan2019} were entangled states' full complex amplitudes explicitly measured directly. In summary, 
diverse states and systems can be measured directly.

On the practical side, unlike tomography, direct measurement requires only two bases. The modesty of this requirement can significantly simplify experiments. The simplification is particularly valuable in the context of quantum states whose dimension $d$ is high: State tomography typically requires measurements of $O(d)$ bases. States with dimensions as high as a billion have been experimentally determined with generalisations of direct measurement~\cite{Howland2014a,Malik2014,Shi2015,Knarr2018}. Many of these experiments benefited from numerous practical refinements of direct measurement: Increases in weak measurements' low SNRs~\cite{Haapasalo2011,Vallone2016,Cohen2018,Calderaro2018,Wen2022}; enhancements of the procedure's directness~\cite{Hariri2019,Li2021}, particularly for multipartite systems~\cite{MartinezBecerril2021}; and simplifications of the procedure (e.g., an avoidance of the scanning of the position measurement)~\cite{Shi2015,Yang2020,Zhu2021}. These improvements have enabled further applications, as we now discuss.

Like quantum tomography, direct measurement has been generalised to fully characterise detectors and processes. Direct measurement was used to determine the projection 
effected by each outcome of a detector~\cite{Xu2021}. The number of bases measured is important for quantum-process determination. The number scales as $O(d^{2})$ in standard process tomography: In different trials, one must input each element of each of the $d$ bases
and, for each state, measure all $d$ bases. Direct measurement allows one to circumvent this rigmarole. 
As few as two measurements are required for each process parameter~\cite{Zhang2017,Kim2018,Gaikwad2023}. 
More recently, the ubiquitous Feynman propagator was observed via direct measurement~\cite{Wen2023}. Fundamentally, as for the wavefunction, direct measurement allows us to experimentally observe quantities that have been typically regarded as abstract mathematical elements of quantum physics.

\subsection{Other direct-measurement procedures}
\label{SubSec:procedure_direct}

Other ways of measuring the KD distribution do not involve weak measurements~\cite{Bollen2010,Hofmann2012a,Buscemi_2014,Thekkadath2017,Lostaglio2023kirkwooddirac,Wagner23}. An early characterisation of a photon's transverse wavefunction relied on balanced homodyne detection~\cite{Bollen2010}. This procedure determined the state with the use of a reference system (a local oscillator) in a superposition of position eigenstates~\cite{Bollen2010}. 
The interference signal was proportional to the KD distribution. 

\begin{figure}
\includegraphics[scale=0.2]{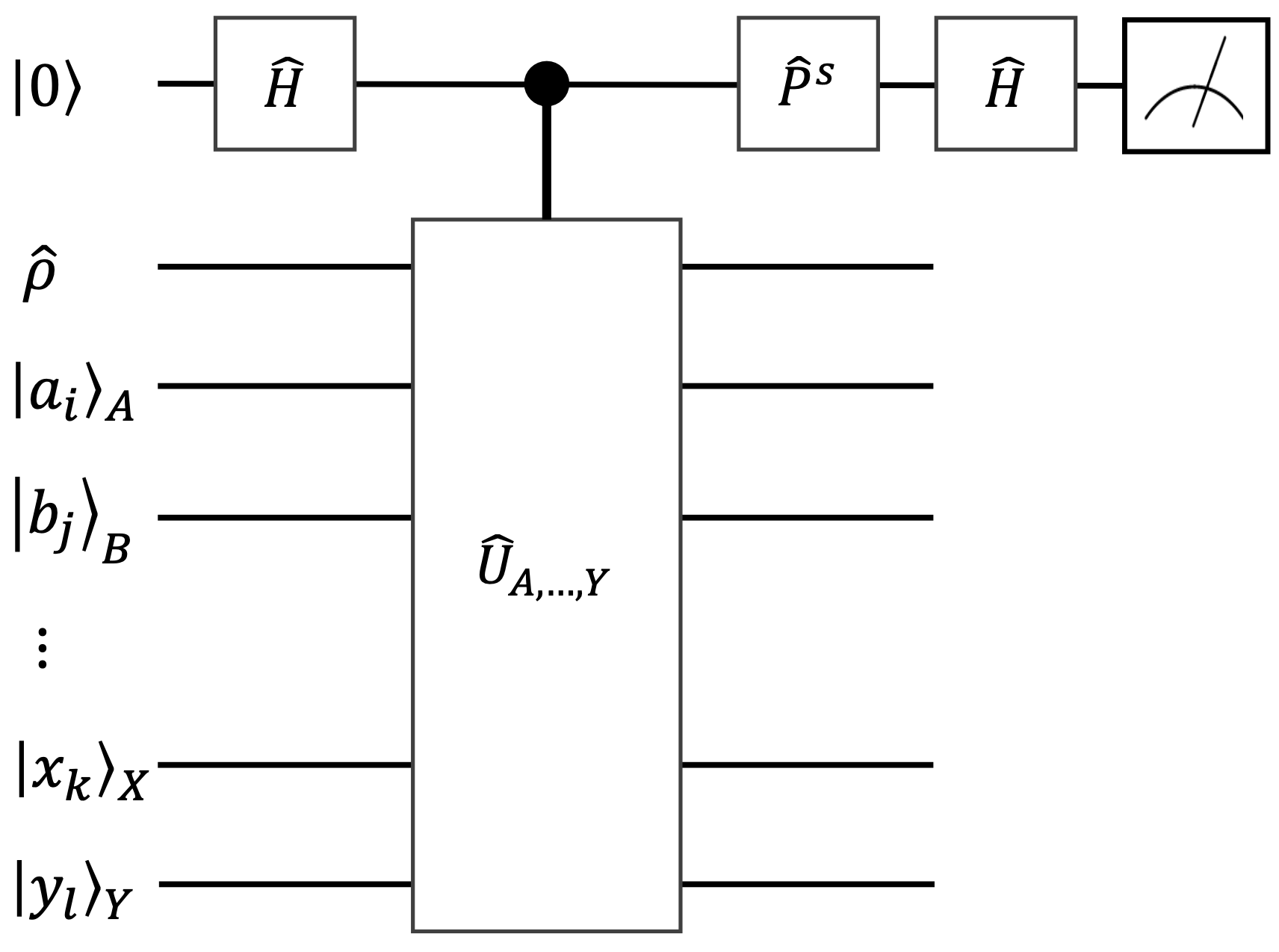}

\caption{
\label{fig:KDMeasure} 
\textbf{Quantum circuit for measuring the KD distribution.} $\hat{H}$ denotes the Hadamard gate, and $\hat{P} = \mathrm{diag}(1,i)$. The variable $s = 0,1$ determines whether $\hat{P} = \mathrm{diag}(1,i)$ is implemented. The central unitary gate is a conditioned version of $\hat{U}_{A,\ldots,Y}$, 
which cycles the quantum states such that $\ket{\psi} \ket{\phi_A}_A \ket{\phi_B}_B \cdots \ket{\phi_Y}_Y 
\rightarrow  \ket{\phi_Y} \ket{\psi}_A \ket{\phi_A}_B \cdots \ket{\phi_X}_Y$. (For a $k$-extended KD distribution, $\hat{U}_{A,\ldots,Y}$ can be implemented with $\mathcal{O}(k)$ conditioned SWAP gates.) If $s=0$, the probability of 
finding the top qubit in $\ket{0}$ yields the KD quasi-probability's real part: 
$P^{s=0}_0 
= \left[ 1 + \Re{\braket{y_l|x_k}  \ldots \braket{b_j | a_i } \braket{a_i | \hat{\rho} | y_l} } 
\right] / 2$. 
Similarly, if  $s=1$, the imaginary part can be measured: 
$P^{s=1}_0 
= \left[ 1 + \Im{ \left( \braket{y_l|x_k}  \ldots \braket{b_j | a_i } \braket{a_i | \hat{\rho} | y_l} \right) } \right]  / 2$. 
This protocol was developed by~\citet{Wagner23} (see also \cite{buscemi2013direct,chiribella2023dimensionindependent, Parzygnat24}). The protocol is closely related to, but more general than, an earlier protocol~\cite{Hofmann2012a}, whose experimental implementation is reported on in~\cite{Lundeen2011}. The original protocol  features a controlled SWAP gate used 
for optimal-universal-symmetric quantum cloning (see main text for details).   }

\end{figure}

Another direct method is based on optimal quantum cloning. After being proposed in~\cite{Hofmann2012a}, the method was experimentally demonstrated in~\cite{Thekkadath2017}. According to this method, imperfect copies of the unknown quantum state are produced. These copies are optimal-universal-symmetric quantum clones~\cite{Hofmann2012a}, created via a controlled SWAP gate. See Fig.~\ref{fig:KDMeasure} for a distinct but related scheme. After this optimal cloning, one measures one copy's position while measuring the other copy's momentum.  The distribution over the possible joint outcomes gives the KD distribution's real part: $\mathbb{P}(x,p;\hat{\rho})=\mathrm{Re} \LParen Q(x,p;\hat{\rho}) \RParen$. If a phase in the cloning process is changed, the protocol gives the imaginary part: 
$\mathbb{P}(x,p;\hat{\rho})=\mathrm{Im} \LParen Q(x,p;\hat{\rho}) \RParen$.

The three direct-measurement methods---the weak-measurement method and these latter two---can be seen in a common light. In each method, every trial involves measurements that project, to some degree, onto a superposition of position and momentum eigenstates. The projection onto $\left|x,p\right\rangle \equiv\left|x\right\rangle +\exp(i\phi)\left\langle p|x\right\rangle \left|p\right\rangle $
occurs with probability $\mathbb{P}(x,p)=\left\langle x,p\right|\hat{\rho}\left|x,p\right\rangle =\left\langle x\right|\hat{\rho}\left|x\right\rangle +\left\langle p\right|\hat{\rho}\left|p\right\rangle +\mathrm{Re} \LParen \exp(i\phi)Q(x,p;\hat{\rho}) \RParen$.
By varying the phase $\phi$, one finds the KD distribution's real and imaginary parts.

We have described many ways in which direct measurement and its generalisations have been applied. Through direct measurements, one can observe elements of quantum physics, including wavefunctions, mixed states, processes and detection. In a practical sense, direct measurement simplifies the characterisation of these elements. In a fundamental sense, direct measurement allows us to observe these elements in their barest form as ensemble averages, sidestepping standard quantum tomography's shadows and inference.

\section{The KD distribution in quantum thermodynamics}
\label{Sec:QThermo}

The KD distribution has benefited
the modern quantum formulation of thermodynamics~\cite{allahverdyan_nonequilibrium_2014,Solinas_probing_2016, Miller_time-reversal_2017, Levy_quasiprobability_2020,Lostaglio_quantum_2018,Lostaglio_kirkwood_2022,Teixido-Bonfill_first_2020, francica_class_2022,francica_most_2022, Hernandez-Gomez_projective_2022, Diaz_quantum_2022, Upadhyaya_what_2023, Santini_work_2023,Pei_exploring_2023, gherardini2024quasiprobabilities}.
Below, we review the treatment of classical thermodynamic work as a stochastic quantity. We then show how 
describing quantum work probabilistically produces inconsistencies. Conversely, the KD distribution 
satisfies reasonable requirements for a distribution describing work and heat fluctuations.
Furthermore, a KD \Dave{distribution's non-positivity signals} contextuality---provable physical non-classicality.
Although we lack space for the following topics,
the KD distribution has found thermodynamic applications also in a consistent-histories framework for work fluctuations~\cite{Miller_time-reversal_2017}, an extension of quantum thermodynamics to quantum field theories~\cite{Teixido-Bonfill_first_2020}, full-counting statistics~\cite{Solinas_probing_2016, francica_class_2022} and quadratic fermionic models~\cite{Santini_work_2023}.

The first and second laws of thermodynamics concern work, heat, and entropy. These quantities, when exchanged by small systems, fluctuate by amounts comparable to their averages. Fluctuation theorems are breakthroughs in non-equilibrium statistical mechanics (e.g.~\cite{Jarzynski_equilibrium_1997, crooks_nonequilibrium_1998}).
In a prototypical example, a classical system 
has a Hamiltonian \Dave{$H(\lambda_t,x,p)$} dependent on a time-varying control parameter $\lambda_t$ and on a phase-space point \Dave{$(x,p)$.}
In a `forward protocol', the system is
prepared at a time $t=0$ in a canonical ensemble 
$\{\exp\LParen-\beta H(\lambda_0, x)\RParen/Z(0)\}$ at an inverse temperature $\beta = 1/k_{\rm B}T$. 
The partition function $Z(t)\equiv \int dx \, \exp\LParen-\beta H(\lambda_t, x)\RParen$ normalises the distribution.
The parameter is then ramped  
along an arbitrary trajectory $\lambda_t$ arbitrarily quickly: $\lambda_0 \rightarrow \lambda_\tau$.
In each trial of the protocol, 
the system absorbs a random amount $W$ of work.
After many trials, one can infer a probability distribution $\mathbb{P}_{\rm F}^{\rm cl}(W)$
over the possible $W$ values.
Similarly, define a `reverse protocol' in which the canonical ensemble $\{\exp\LParen-\beta H(\lambda_\tau, x)\RParen/Z(\tau)\}$ is prepared at $t=0$. 
The parameter is ramped as $\lambda_\tau\mapsto \lambda_0$ along the trajectory $\lambda_{\tau-t}$.
One can infer a reverse work distribution $\mathbb{P}_{\rm R}^{(\rm cl)}(W)$.
Each canonical ensemble has a free energy $- \frac{1}{\beta} \ln\LParen Z(t)\RParen$.
Define the difference $\Delta F \equiv -\frac{1}{\beta} \ln \LParen Z(\tau)/Z(0)\RParen$ between the initial and final Hamiltonians' equilibrium free energies.
In terms of $\Delta F$, \emph{Crooks' theorem} reveals a symmetry of the two distributions~\cite{crooks_nonequilibrium_1998}: 
\begin{equation}
\label{eq:crooks}
    \frac{ \mathbb{P}_{\rm F}^{\rm cl}(W)}{ 
             \mathbb{P}_{\rm R}^{\rm cl}(-W)}  
   = e^{\beta (W- \Delta F)}.
\end{equation}
To understand this equality intuitively, consider a large work value $W \gg \Delta F$. The exponential $e^{\beta (W- \Delta F)}$ is enormous. Hence 
$\mathbb{P}_{\rm F}^{\rm cl}(W) \gg \mathbb{P}_{\rm R}^{\rm cl}(-W)$: One is far more likely to pay a large amount $W$ of work, during a forward trial, than to recoup an amount $W$ during a reverse trial.
Crooks' theorem implies  \emph{Jarzynski's equality}, 
\begin{equation}
\label{Eq:JarzEq}
\langle e^{-\beta W} \rangle_{\mathrm{F}} = e^{-\beta \Delta F} ,
\end{equation}
 which implies the second-law-like inequality $\langle W \rangle_{\mathrm{F}} \ge \Delta F$~\cite{Jarzynski_equilibrium_1997, Collin_verification_2005}. [The averages are with respect to 
$\mathbb{P}_{\rm F}^{\rm cl}(W)$.]
Hence Crooks' theorem and Jarzynski's equality are said to strengthen the second law.
They are also called \emph{fluctuation theorems}. Using either, one can infer $\Delta F$ from out-of-equilibrium experiments~\cite{Liphardt_equilibrium_2002}. This fact is useful, as $\Delta F$ is difficult to measure~\cite{Pohorille_10_Good}.

\begin{figure}[b]
\includegraphics[width=1.0\textwidth]{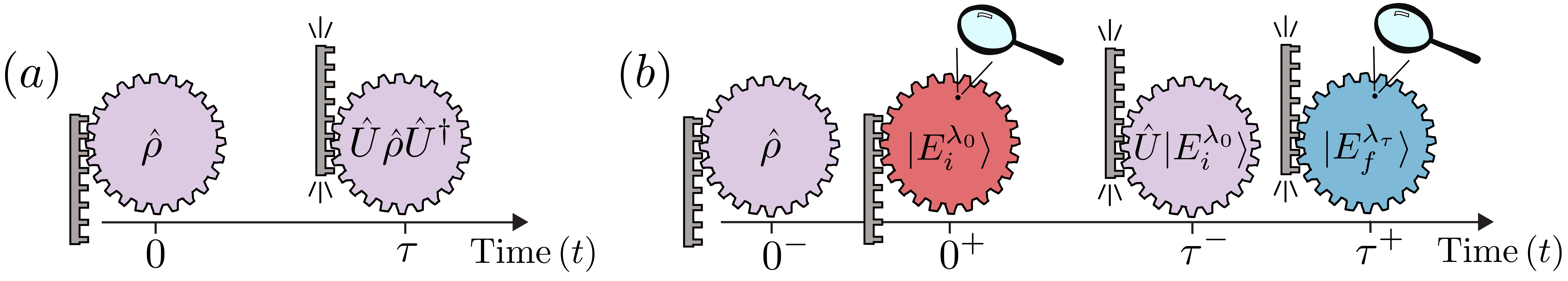}
\caption{\textbf{Work protocol and two-point-measurement scheme.} 
(a) Quantum work protocol: The system is prepared in a state $\hat{\rho}$. An external control system, represented by the linear gear, alters the time-dependent parameter $\lambda_t$ in the Hamiltonian $\hat{H}(\lambda_t)$, by performing work. The system evolves under the unitary $\hat{U}$. (b) Two-point-measurement scheme: Energy measurements interrupt the protocol before and after $\hat{U}$.
}
\label{fig:TPM}
\end{figure}

Measurement disturbance complicates a quantum extension of this protocol.
One approach, the \emph{two-point-measurement scheme}, involves the following forward protocol~(Fig.~\ref{fig:TPM}).
Consider a quantum system prepared in a canonical state $\hat{\rho} \propto \exp\LParen-\beta \hat{H}(\lambda_0)\RParen$ and evolved under a quantum Hamiltonian $\hat{H}(\lambda_t)$ dependent on a parameter $\lambda_t$.
$\hat{H}(\lambda_t)$ eigen-decomposes as $\hat{H}(\lambda_t)= \sum_j E_j(\lambda_t) \, \hat{\Pi}_j(\lambda_t)$, with eigenvalues $E_j(\lambda_t)$ and eigen-projectors $\hat{\Pi}_j(\lambda_t)$.
At time $t=0$, one measures the energy, obtaining an outcome $E_j(\lambda_0)$ and projecting the state onto $\hat{\Pi}_j(\lambda_0)/\Tr\LParen \hat{\Pi}_j(\lambda_0)\RParen$.
From $t=0$ to $t=\tau$, the unitary $\hat{U} = \mathcal{T} e^{-i \int_0^\tau \hat{H}(\lambda_t) dt/\hbar}$ evolves the system. $\mathcal{T}$ denotes the time-ordering operation.
At $t=\tau$, one measures the energy, obtaining an outcome $E_k(\lambda_\tau)$ and projecting the state onto $\hat{\Pi}_k(\lambda_\tau)/\Tr\LParen \hat{\Pi}_k(\lambda_\tau)\RParen$.
The joint probability of obtaining outcomes $E_j(\lambda_0)$ and $E_k(\lambda_\tau)$ is
\begin{equation}\label{eq:TPM}
    P_{j,k}(\hat{\rho}) 
    \equiv \Tr\bm{(} \hat{U}^\dagger \, \hat{\Pi}_k(\lambda_\tau) \, \hat{U} \, \hat{\Pi}_j(\lambda_0) \, \hat{\rho} \, \hat{\Pi}_j(\lambda_0) \bm{)}.
\end{equation}
The probability density associated with spending an amount $W$ of work during any given forward trial is 
\begin{align}
   \label{eq_fwd_work_tpm}
   \mathbb{P}_{\rm F}(W) 
   = \sum_{j,k} P_{j,k}(\hat{\rho})  \, \delta\LParen W - [E_k(\lambda_\tau) - E_j(\lambda_0)]\RParen .
\end{align}
One can define $\mathbb{P}_{\rm R}(W)$ similarly for a reverse protocol, then prove Crooks' theorem.

This two-point-measurement scheme faces a problem if $\hat{\rho}$ has coherence with respect to the initial energy eigenbasis [the $\hat{H}(\lambda_0)$ eigenbasis].
The initial measurement dephases $\hat{\rho}$, affecting the marginal probability $\sum_j P_{j,k}(\hat{\rho})$. 
One might expect this marginal to equal the probability 
$\Tr\LParen \hat{\Pi}_k(\lambda_\tau) \, \hat{U} \hat{\rho} \hat{U}^\dagger \RParen$ of obtaining the final-measurement outcome $E_k(\lambda_\tau)$, in the absence of any initial measurement. Yet the two probabilities do not equal each other:
$\sum_j  P_{j,k}(\hat{\rho}) 
\neq \Tr\LParen \hat{\Pi}_k(\lambda_\tau) \, \hat{U} \hat{\rho} \hat{U}^\dagger \RParen$. 
This lack of equality stems from measurement disturbance, caused by
the mutual non-commutativity of $\hat{\rho}$, $\hat{H}(\lambda_0)$ and $\hat{H}(\lambda_\tau)$.
Nevertheless, a KD distribution---a set of quasi-probabilities---has the desired marginal 
$\Tr\LParen \hat{\Pi}_k(\lambda_\tau) \, \hat{U} \, \hat{\rho} \, \hat{U}^\dagger\RParen$,
as well as the similarly desired marginal
$\Tr\LParen \hat{\Pi}_j(\lambda_0) \, \hat{\rho} \RParen$:
\begin{equation}\label{eq:work_KD}
    Q_{j,k}(\hat{\rho}) 
    \equiv \Tr\bm{(}\hat{U}^\dagger \, \hat{\Pi}_k(\lambda_\tau) \, \hat{U} \, \hat{\Pi}_j(\lambda_0) \, \hat{\rho} \bm{)}\, .
\end{equation}
In terms of this joint KD quasi-probability, one can define
the KD work distribution~\cite{allahverdyan_nonequilibrium_2014} 
\begin{align}
   \tilde{\mathbb{P}}_{\rm F}(W) 
   \equiv \sum_{j,k} Q_{j,k}(\hat{\rho}) \, 
   \delta\LParen W - [E_k(\lambda_\tau) - E_j(\lambda_0)]\RParen .
\end{align} 
A Jarzynski-like equality (Sec.~\ref{Sec:OTOC}) depends on an extended KD distribution over analogues of work~\cite{NYH_jarzynski_2017}.

The two-point-measurement joint distribution~\eqref{eq:TPM} and KD joint distribution~\eqref{eq:work_KD} share several properties.
We recover the marginal $P_{j,k}(\hat{\rho})$ from $Q_{j,k}(\hat{\rho})$ if $\hat{\rho}$ is block-diagonal with respect to $\hat{H}(\lambda_0)$'s eigenbasis. Therefore, $Q_{j,k}(\hat{\rho})$ leads to fluctuation theorems for such states $\hat{\rho}$. 
We recover $P_{j,k}(\hat{\rho})$ from $Q_{j,k}(\hat{\rho})$ also if one measures coarse-grained observables~\cite{Levy_quasiprobability_2020}. 
As shown in Sec.~\ref{SubSec:Standard}, $Q_{j,k}(\hat{\rho})$ has  marginals equal to the Born-rule predictions about preparing $\hat{\rho}$ and evolving the state under $\hat{U}$:
\begin{equation}
\label{eq:marginals}
    \sum_j Q_{j,k}(\hat{\rho}) 
    = \Tr\bm{(} \hat{\Pi}_k(\lambda_\tau) \, \hat{U} \hat{\rho} \hat{U}^\dagger \bm{)}, \quad  
    \sum_k Q_{j,k}(\hat{\rho}) 
    = \Tr\bm{(}\hat{\Pi}_j(\lambda_0) \, \hat{\rho} \bm{)} 
    \quad {\rm and} \quad 
    \sum_{j,k} Q_{j,k}(\hat{\rho}) =1 \, .
\end{equation}
Additionally, a work distribution should lead to an average work value equal to the usual definition 
$ \langle W \rangle 
\equiv \Tr\bm{(}\hat{H}(\lambda_\tau) \, \hat{U} \hat{\rho} \hat{U}^\dagger \bm{)} - \Tr\bm{(}\hat{H}(\lambda_0) \, \hat{\rho} \bm{)}$.
The two-point-measurement distribution [Eq.~\eqref{eq_fwd_work_tpm}] does not always satisfy this condition: 
$\int  \mathbb{P}_{\rm F}(W) \, dW \neq \langle W \rangle$, generally.
Yet the KD work distribution does, due to Eqs.~\eqref{eq:marginals}: 
$\int  \tilde{\mathbb{P}}_{\rm F}(W) \, dW = \langle W \rangle$. 
The KD average differs from the two-point-measurement average, and the KD variance differs from the two-point-measurement variance, whenever $\hat{\rho}$ has energy coherences~\cite{Diaz_quantum_2022}.
Furthermore, when $[\hat{U}^\dag \hat{H}(\lambda_\tau) \hat{U},  \hat{H}(\lambda_0)]  \neq 0$, the KD work variance can assume values only from a subset of the values achievable if work is a classical random variable~\cite{Lostaglio_kirkwood_2022}.

No-go theorems demonstrate 
the incompatibility of reasonable criteria for work probability distributions $\mathbb{P}(W)$~\cite{Perarnau-Llobet_no-go_2017, Hernandez-Gomez_projective_2022, Lostaglio_kirkwood_2022}.  
One criterion is convex-linearity in $\hat{\rho}$~\cite{Perarnau-Llobet_no-go_2017, Lostaglio_quantum_2018, Lostaglio_kirkwood_2022},
a property that $Q_{j,k}(\hat{\rho})$ satisfies.
Yet, if $[\hat{U}^\dag \hat{H}(\lambda_\tau) \hat{U},  \hat{H}(\lambda_0)]  \neq 0$, one cannot define a joint probability with convex-linearity and the marginal property~\eqref{eq:marginals}~\cite{Hernandez-Gomez_projective_2022}. Another no-go theorem replaces the marginal property with two alternative criteria~\cite{Perarnau-Llobet_no-go_2017}: (i) $\int  \mathbb{P}(W) \, dW$ equals the average energy change $\langle W \rangle$. (ii)
If $\hat{\rho}$ is block-diagonal with respect $\hat{H}(\lambda_0)$'s eigenbasis, then
$\mathbb{P}(W) = \mathbb{P}_{\rm F}(W)$.
No probability distribution satisfies the convex-linearity criterion in addition to criteria (i) and (ii)~\cite{Perarnau-Llobet_no-go_2017}.
However, the KD work distribution $\tilde{\mathbb{P}}_{\rm F}(W)$ 
is convex-linear in $\hat{\rho}$ and satisfies the marginal property and criteria (i) and (ii), as well as other reasonable criteria~\cite{Pei_exploring_2023}.
Alternative work-distribution criteria are satisfied only by a Terletsky--Margenau--Hill
distribution (a KD distribution's real part)~\cite{francica_most_2022, Pei_exploring_2023}.

In addition to satisfying the criteria above, KD distributions signal non-classicality in a quantum engine's linear-response work output.
\emph{Linear-response theory} describes how systems react
to small perturbations.
Consider a quantum system subject to
a static Hamiltonian $\hat{H}_0$ perturbed by a time-dependent potential $g \hat{V}(t)$, where $g \ll 1$.
Let the potential be cyclic: $\hat{V}(0) = \hat{V}(\tau) = 0$. 
In the interaction picture, the potential has the form $\hat{V}_I(t) = e^{i \hat{H}_0 t/\hbar} \hat{V}(t) e^{-i \hat{H}_0 t/\hbar}$ and the time average $ \hat{\bar{V}}_I(\tau)= \frac{1}{\tau}\int_{0}^\tau \hat{V}_I(t) dt$.
The system is prepared in $\hat{\rho}$ and evolved unitarily under $\hat{H}(t) = \hat{H}_0 + g \hat{V}(t)$. 
On average, the system produces an amount of work~\cite{Lostaglio_certifying_2020}
\begin{equation}
    \langle W \rangle = \frac{2 g \tau }{\hbar} \mathrm{Im}\left( \Tr\bm{(}\hat{H}_0 \hat{\bar{V}}_I(\tau) \hat{\rho} \bm{)}\right) + O\left(g^2\right)\,.
\end{equation}
If $\mathrm{Im}\left( \Tr\bm{(}\hat{H}_0 \hat{\bar{V}}_I (\tau) \hat{\rho} \bm{)}\right) \neq 0$, then the average work is first-order in the perturbation strength: $\langle W \rangle = O(g)$.
The lack of equality translates into a condition under which a KD distribution  signals contextuality (rigorous non-classicality), as detailed in Sec.~\ref{SubSec:GenContext}.  \Dave{In another work-extraction setting,  KD non-positivity signals that non-commutation can enhance the achievable power  \cite{pratapsi2024quantum}.}

Complementary to work expenditure is heat exchange. Consider heat baths ${\tt A}$ and ${\tt B}$, with respective Hamiltonians $\hat{H}^{\tt A}$ and $\hat{H}^{\tt B}$, isolated from their environment.
Let the initial state $\hat{\rho}^{\tt AB}$ be
a tensor product of thermal states
with inverse temperature $\beta^{\tt A}$ and $\beta^{\tt B} < \beta^{\tt A}$: $\hat{\rho}^{\tt AB} \propto \exp( - \beta^{\tt A} \hat{H}^{\tt A})\otimes \exp( - \beta^{\tt B} \hat{H}^{\tt B})$. 
Consider any unitary $\hat{U}$ that conserves the global energy: $[\hat{U}, \hat{H}^{\tt A} + \hat{H}^{\tt B}]=0$.
The second law of thermodynamics implies that heat flows from the hotter ${\tt B}$ to the colder ${\tt A}$ on average.
The net heat $\mathcal{Q}$ transferred from {\tt B} to {\tt A} satisfies an \emph{exchange} fluctuation theorem~\cite{Jarzynski2004}: 
${\mathbb{P}_{\rm F}(+\mathcal{Q})}  /  {\mathbb{P}_{\rm R}(-\mathcal{Q})} 
= \exp([\beta^{\tt B} - \beta^{\tt A}] \mathcal{Q})$.
This equation governs classical and quantum baths alike, if we define $\mathcal{Q}$ through projective energy measurements.

Non-classical heat flows may occur if $\hat{\rho}^{\tt AB}$ has thermal local 
marginals, $\Tr_{\tt B}( \hat{\rho}^{\tt AB})= \exp(- \beta^{\tt A} \hat{H}^{\tt A})/Z^{\tt A}$ and $\Tr_{\tt A}( \hat{\rho}^{\tt AB})= \exp( - \beta^{\tt B} \hat{H}^{\tt B})/Z^{\tt B}$, but
is correlated.
The correlations can drive heat from the colder ${\tt A}$ to the hotter ${\tt B}$ on average.
Quantum correlations can do so at a rate unachievable classically:
Let ${\tt A}$ and ${\tt B}$ have $d$-dimensional Hilbert spaces.
An amount of heat $\mathcal{Q} < \log(d) / (\beta^{\tt B} - \beta^{\tt A})$ constitutes an \emph{anomalous backflow}, requiring entanglement~\cite{Jennings_entanglement_2010, Jevtic_exchange_2015}. 
The required correlations disappear under strong energy measurements, \`a la the two-point-measurement scheme.
Measuring weakly preserves the anomalous backflow and suggests a KD heat quasi-probability distribution~\cite{Levy_quasiprobability_2020}. 
Negative and non-real KD quasi-probabilities signal non-classical heat~\cite{Levy_quasiprobability_2020} 
and work flows~\cite{Hernandez-Gomez_projective_2022}.
Aside from energy, 
${\tt A}$ and ${\tt B}$ can exchange other quantities, such as particles~\cite{majidy_noncommuting_2023}. 
The quantities need not commute with each other, as exemplified by the $x$-, $y$-, and $z$- components of spin~\cite{Yunger_Halpern_noncommuting_2020}.
An extended KD distribution characterises such non-commuting quantities' fluctuations, which can signal contextuality~\cite{Upadhyaya_what_2023}.

\section{The KD distribution behind the out-of-time-ordered correlator}
\label{Sec:OTOC}

The \textit{out-of-time-ordered correlator} (OTOC)~\cite{NYH_jarzynski_2017} is
a witness of many-body quantum chaos.
In addition to elucidating superconductors~\cite{Larkin_quasiclassical_1969}  and black holes~\cite{Shenker_Black_2014, Maldacena_Diving_2017, Yoshida_Efficient_2017, Gao_Traversable_2017}, the OTOC has been observed with nuclear magnetic resonance~\cite{Li_measuring_2017, Wei_exploring_2018, Nie_experimental_2020, Niknam_sensitivity_2020}, trapped ions~\cite{Britton_engineered_2012, Islam_emergence_2013, Garttner_measuring_2017,Landsman_verified_2019, Joshi_quantum_2020} and other platforms~\cite{Yan_observation_2013, Choi_observation_2017,Foxen_demonstrating_2020, Chen_detecting_2020, Pegahan_energy_2021, Blok_quantum_2021, Garcia_quantum_2021, Braumuller_probing_2022, Mi_information_2021, Geller_quantum_2022, Harris_benchmarking_2022, Blocher_measuring_2022, Wang_information_2022, Green_experimental_2022}.
The OTOC equals an average over an extended KD distribution~\cite{NYH_jarzynski_2017, NYH_quasiprobability_2018}.
We present the distribution after reviewing the OTOC.
The distribution 
\Dave{signals} a fluctuation-type theorem---resembling an extension of the second law of thermodynamics (Sec.~\ref{Sec:QThermo})---that contains the OTOC~\cite{NYH_jarzynski_2017}.
The fluctuation-type theorem suggested new techniques for measuring the OTOC experimentally~\cite{NYH_jarzynski_2017, NYH_quasiprobability_2018, Dressel_strengthening_2018, Mohseninia_optimizing_2019}.
Also, the extended KD distribution distinguishes between chaos and decoherence~\cite{Alonso_OTOC_2019}.
Further applications lie beyond the scope of this review: The extended KD distribution diagnoses 
a quantum kicked top's chaos
\cite{Alonso_diagnosing_2022} and features in an uncertainty relation for quantum chaos~\cite{Yunger_Halpern_entropic-2019}.

\begin{figure}[b]
\includegraphics[width=0.90\textwidth]{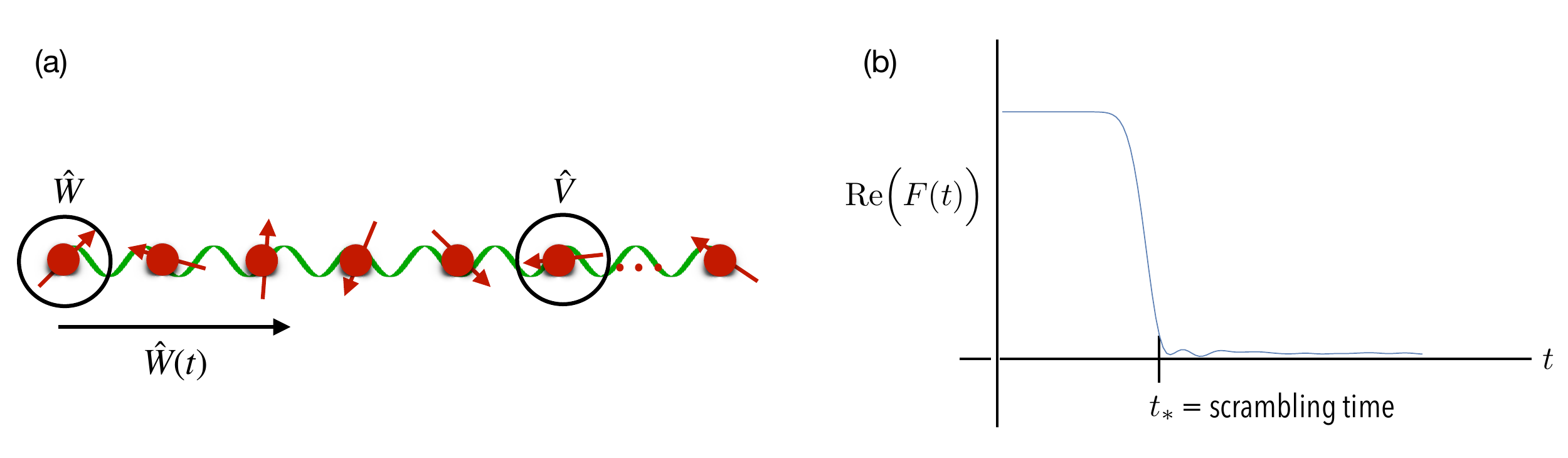}
\caption{ \textbf{Information scrambling.} (a) Operators $\hat{W}$ and $\hat{V}$ are localised on single spins. $\hat{W}$ evolves to $\hat{W}(t)$ in the Heisenberg picture.  Information scrambles as $\hat{W}(t)$ and $\hat{V}$ increasingly fail to commute. 
(b) The decay of the OTOC $F(t)$ signals scrambling. ${\rm Re}\LParen F(t)\RParen$ starts at 1 and  relaxes to $\approx 0$ around the scrambling time $t_*$.
}
\label{fig:scrambling}
\end{figure}

The OTOC signals \textit{scrambling}, the spreading of initially localised information through many-body entanglement~\cite{Swingle_Unscrambling_2018}.
Consider a quantum system of $N \gg 1$ degrees of freedom. Let a non-local, non-integrable (`chaotic') Hamiltonian $\hat{H}$ govern the system, which is in a state $\hat{\rho}$.
Denote expectation values by $\langle \bullet \rangle \equiv \Tr(\bullet \, \hat{\rho})$.
Figure~\ref{fig:scrambling} illustrates scrambling with an $N$-qubit chain in a thermal state $\hat{\rho}_\beta \equiv \exp(-\beta \hat{H})/\mathcal{Z}$. The inverse temperature is $\beta$, and the partition function $\mathcal{Z}$ normalises the state.
Denote by $\hat{W}$ and $\hat{V}$ unitary or Hermitian operators localised far apart.\footnote{Some OTOC results require $\hat{W}$ and $\hat{V}$ to be Hermitian; some, for the operators to be unitary; and some, for the operators to be both. The quasi-probability results highlighted here are general, holding if $\hat{W}$ and $\hat{V}$ are merely normal and thus diagonalisable. That is, $\hat{W}$ and $\hat{V}$ can be Hermitian or unitary.}
In our example, we denote by $\hat{\sigma}_a^{(j)}$ component $a = x, y, z$ of qubit $j$'s spin. 
The spin chain can have $\hat{W} = \hat{\sigma}_x^{(1)}$ and $\hat{V} = \hat{\sigma}_z^{(N)}$.
$\hat{W}$ could be a perturbation, whereas $\hat{V}$ could be an observable measured later, in an attempt to recover information about the perturbation.
The operators eigendecompose as ${\hat{W}} = \sum_{w_j} w_j \, \hat{\Pi}_{w_j}^{W}$ and $\hat{V} = \sum_{v_j} v_j \, \hat{\Pi}_{v_j}^{V}$.
Under the unitary $\hat{U} = \exp(-i \hat{H} t/\hbar)$, ${\hat{W}}$ evolves to ${\hat{W}}(t) \equiv \hat{U}^\dagger {\hat{W}} \hat{U}$ in the Heisenberg picture.
The more $\hat{W}(t)$ spreads, the more $[\hat{W}(t), \hat{V}]$ grows. Yet the simple correlator
$\langle [{\hat{W}}(t), \hat{V}]\rangle$ can vanish at late times, due to cancellations between $\hat{\rho}$ eigenstates in the average. The average  
$C(t) \equiv \langle | [{\hat{W}}(t), \hat{V}] |^2\rangle$, 
being of a non-negative squared magnitude, avoids this pitfall.
$C(t)$ more reliably certifies scrambling---the spreading of the information initially localised in $\hat{W}$.
If $\hat{W}$ and $\hat{V}$ are unitary, $C(t)$ decomposes as  $C(t) = 2[ 1 - 2 {\rm Re} \LParen F(t) \RParen ]$. The OTOC therein is defined as
\begin{equation}\label{eq:OTOC-def}
	F(t) \equiv \langle {\hat{W}}^\dagger (t) \hat{V}^\dagger {\hat{W}}(t) \hat{V} \rangle \equiv {\rm Tr} \LParen {\hat{W}}^\dagger (t) \hat{V}^\dagger {\hat{W}}(t) \hat{V} \hat{\rho}\RParen \, .
\end{equation}
A decay of $F(t)$ to $\approx 0$ signals scrambling, illustrated in Fig.~\ref{fig:scrambling}(a). 
The OTOC decays exponentially at early times under highly non-local interactions: ${\rm Re}\LParen F(t)\RParen \sim 1 - \frac{1}{N}\exp(\lambda_{\rm L} t)$.
$\lambda_{\rm L}$ 
resembles a Lyapunov exponent, which controls 
how phase-space trajectories diverge under classical chaos~\cite{Rozenbaum_lyapunov_2017}.

Yunger Halpern \emph{et al.} decomposed the OTOC in terms of an extended KD distribution~\cite{NYH_jarzynski_2017, NYH_quasiprobability_2018},
\begin{equation}
\label{Eq:OTOCKD}
	Q_{v_1, w_1, v_2, w_2} \LParen \hat{\rho}(t) \RParen
	= {\rm Tr}\left( \hat{\Pi}_{w_2}^{{W}(t)} \hat{\Pi}_{v_2}^{{V}} \hat{\Pi}_{w_1}^{{W}(t)} \hat{\Pi}_{v_1}^{{V}} \hat{\rho} \right)\, .
\end{equation}
In illustrative qubit-chain examples,
$Q \LParen  \hat{\rho}(t)  \RParen$  typically has a negative real component and a small imaginary component~\cite{NYH_quasiprobability_2018}.
The OTOC equals an average over $Q \LParen  \hat{\rho}(t)  \RParen$~\cite{NYH_quasiprobability_2018}:
\begin{equation}\label{eq:F(t)_as_avg}
	F(t) = \sum_{v_1, w_1, v_2, w_2} v_1 w_1 v_2^* w_2^* \,  
	Q_{v_1, w_1, v_2, w_2} \LParen  \hat{\rho}(t)  \RParen  
	\equiv \langle v_1 w_1 v_2^* w_2^* \rangle_{  Q \LParen \hat{\rho}(t)  \RParen  }\, .
\end{equation}
In the spin-chain example, since $W$ and $V$ are Pauli operators, each sum runs over the eigenvalues $\pm 1$.
The OTOC's decomposition motivated the 21st-century rediscovery of extended KD distributions~\cite{NYH_jarzynski_2017, NYH_quasiprobability_2018}.
More thermodynamically, the OTOC decomposes also in terms of the distribution's characteristic function, or the Fourier transform of the distribution:
$\left\langle 
e^{\beta_1 v_1  + \beta_1' w_1 + \beta_2 v_2^* +\beta_2'  w_2^*} 
\right\rangle_{ Q \LParen \hat{\rho}(t) \RParen }$.\footnote{This expression technically contains a Laplace transform, rather than a Fourier transform, since the $\beta$s are real numbers, rather than imaginary. However, the fluctuation-theorem community calls such transforms `Fourier'.} More precisely, the OTOC follows from differentiating the characteristic function:
\begin{equation}\label{eq:jarzynski-like}
	F(t) 
    = \frac{\partial^4}{\partial \beta_1 \, \partial \beta_1'\, \partial \beta_2 \, \partial \beta_2'} 
    \left\langle 
    e^{\beta_1 v_1  + \beta_1' w_1 + \beta_2 v_2^* +\beta_2'  w_2^*} 
    \right\rangle_{Q \LParen  \hat{\rho}(t)  \RParen} \, 
    \Big\vert_{\beta_1, \beta_1', \beta_2,\beta_2' = 0} \, , \quad \beta_1,\beta_1', \beta_2,\beta_2' \in \mathbb{R}.
\end{equation}
Equation~\eqref{eq:jarzynski-like} resembles Jarzynski's equality \Dave{[Eq. \eqref{Eq:JarzEq} in Sec.~\ref{Sec:QThermo}],} an extension of the second law of thermodynamics, in two ways. 
First, Eq.~\eqref{eq:jarzynski-like} and Jarzynski's equality suggest schemes for extracting difficult-to-measure quantities from more-easily-measurable quantities: Using Jarzynski's equality, we can infer a free-energy difference $\Delta F$ from a probability distribution over the possible amounts of the work spent on a non-equilbrium protocol~\cite{Jarzynski_equilibrium_1997, Talkner_fluctuation_2007}.
\nicole{Similarly, the Jarzynski-like equality~\eqref{eq:jarzynski-like} enables new methods for measuring $F(t)$ experimentally~\cite{NYH_jarzynski_2017}:
Eq.~\eqref{eq:OTOC-def} shows that the OTOC is neither a probability nor an expectation value.  
How to measure the  OTOC was therefore unclear for a while.
The initially proposed measurement schemes rely on 
interference~\cite{Swingle_Measuring_2016}, Ramsey interferometry~\cite{Yao_Interferometric_2016} and a quantum clock~\cite{Zhu_Measurement_2016}. Linking the OTOC to 
$Q  \LParen  \hat{\rho}(t)  \RParen$ 
unlocked new toolkits for measuring the OTOC
\cite{NYH_jarzynski_2017, NYH_quasiprobability_2018}.
First, the OTOC can be inferred from weak measurements~\cite{NYH_jarzynski_2017,NYH_quasiprobability_2018}.
Second, if $\hat{W}(t)$ and $\hat{V}$ square to the identity, $\hat{W}(t)^2 = \hat{V}^2 = \mathbbm{1}$, then one can infer the OTOC from arbitrary-strength measurements~\cite{Dressel_strengthening_2018, Mohseninia_optimizing_2019}.
Pauli operators illustrate this property: $\big( \hat{\sigma}_a^{(j)} \big)^2 = \hat{\mathbbm{1}}$.}

\Dave{The second way in which Eq.~\eqref{eq:jarzynski-like} resembles Jarzynski's equality is, both equations cast the object of interest [$\Delta F$ or $F(t)$] in terms of a moment-generating function.}
\nicole{Beyond these two parallels, }
Eq.~\eqref{eq:jarzynski-like} inspired a fluctuation theorem for the spreading of correlations from a system into an environment~\cite{Zhang_Quasiprobability_2022}.

A numerical study established $Q \LParen  \hat{\rho}(t)  \RParen$ as a scrambling witness more robust than $F(t)$~\cite{Alonso_OTOC_2019}.
An experimentalist may measure $F(t)$, observe a decay,  and conclude that their system likely contains many-body entanglement.
Yet this decay may stem from decoherence, rather than scrambling~\cite{Zhang_information_2019, Zanardi_information_2021, Touil_information_2021, Harris_benchmarking_2022}.
If used to detect scrambling, therefore, $F(t)$ can lead to false positives.
\begin{figure}[b]
\includegraphics[width=0.9\textwidth]{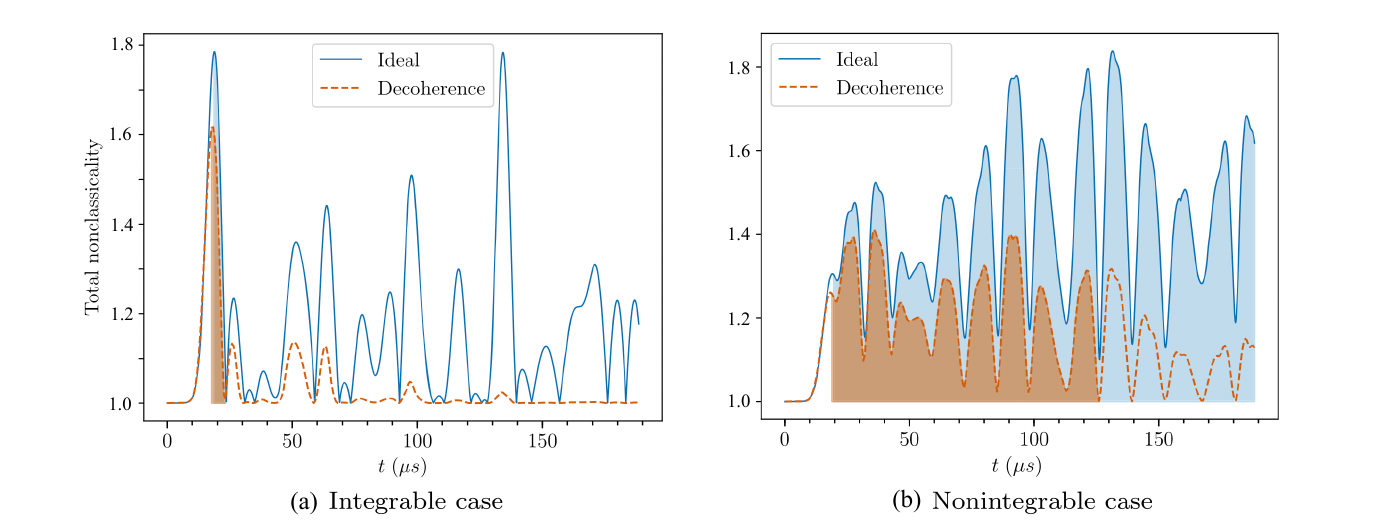}
\caption{\textbf{Total non-positivity $\mathcal{N}(t)$ vs. time.}
Figure (a) follows from evolution under an integrable Hamiltonian; and Fig. (b), from evolution under a scrambling (non-integrable) Hamiltonian.
Blue traces indicate the behaviour of the KD non-classicality, $\mathcal{N}(t)$, in the absence of decoherence; 
and brown traces, $\mathcal{N}(t)$'s behaviour in the presence of decoherence.
Shading under a trace highlights the time interval $t_{\rm int}$. Figures reproduced from~\cite{Alonso_OTOC_2019}.
}
\label{fig:nonintegrable-OTOC}
\end{figure}
$Q \LParen  \hat{\rho}(t)  \RParen$ overcomes this pitfall, distinguishing scrambling from integrable Hamiltonians despite decoherence~\cite{Alonso_OTOC_2019}.
Gonz\'alez Alonso \emph{et al.} simulated $N = 5$ superconducting qubits undergoing Lindblad dynamics.
The qubits evolve under a transverse-field Ising model with or without a longitudinal field:
\begin{equation}
\hat{H}= -J\sum_{j=1}^{N-1} \hat{\sigma}^{(j)}_z \hat{\sigma}^{(j+1)}_z - g \sum_{j=1}^N \hat{\sigma}^{(j)}_x - h \sum_{j=1}^N \hat{\sigma}^{(j)}_z.
\end{equation}
In the absence of the longitudinal field, if $(h/J, g/J) = (0.000, 1.05)$, $\hat{H}$ is integrable. In the presence of the longitudinal field, if $(h/J, g/J) = (0.500,1.05)$, $\hat{H}$ is scrambling (non-integrable).
The local operators used were $\hat{W} = \hat{\sigma}_1^z$ and $\hat{V} = \hat{\sigma}_N^z$. 
Figure~\ref{fig:nonintegrable-OTOC} shows the quasi-probability non-classicality, $\mathcal{N}(t) \equiv \sum | Q \LParen  \hat{\rho}(t)  \RParen | $ (Sec.~\ref{SubSec:KDNonclas}), plotted against time.
$\mathcal{N}(t)$ begins at 1, grows, peaks, returns to 1, and then keeps oscillating.
Denote by $t_{\rm int}$ the time interval from $\mathcal{N}(t)$'s first peak to the next 1. 
$t_{\rm int}$ distinguishes scrambling from integrable behaviour:
in the presence of decoherence, $t_{\rm int}$ is an order of magnitude longer under scrambling than under integrable dynamics.
$Q  \LParen  \hat{\rho}(t)  \RParen$ thus witnesses scrambling more robustly than $F(t)$ does.

\section{The KD distribution and the foundations of quantum mechanics}
\label{Sec:Foundations}

One way to define the notion of a classical experiment is to say that joint probability distributions describe the corresponding system's preparation, manipulation and measurement~\cite{Dirac26, Dirac26-2, Wigner32,  Dirac45, Cohen66, Hudson74, Srinivas75, Griffiths84, Hartle04, spekkens2008negativity, Allahverdyan15}. Under this definition, an experiment has been proven to be classical if, and only if, it is non-contextual~\cite{spekkens2008negativity} (we define this term below). Thus, if every quasi-probabilistic description of an experiment is non-positive, the experiment is non-classical. Every experiment admits of infinitely many quasi-probability representations, however. 
\Dave{Therefore, the existence of a single non-positive quasi-probability representation does not allow one to conclude that the experiment is non-classical.}
Nevertheless, operationally defined KD distributions' non-positivity has been linked to non-classical physics in several settings. In this section, we outline several examples.

\subsection{Generalised Leggett--Garg inequality}
\label{SubSec:LG}

Leggett and Garg designed a Gedankenexperiment for testing the limits of macroscopic coherence and notions of `quantumness'~\cite{Leggett85}.  In the spirit of Bell, Leggett and Garg formulated an inequality for  data acquired from sequential measurements. The inequality is satisfied under the assumptions of macroscopic realism and non-invasive detectability. However, quantum experiments generally can violate the inequality. Thus, the Leggett--Garg inequality has become a popular tool for distinguishing macro-realistic theories  from quantum mechanics.  Williams and Jordan~\cite{Williams08} have connected anomalous weak values~\cite{Aharanov88} with violations of the weak-measurement version of the Leggett--Garg inequality  ~\cite{PhysRevLett.96.200404,PhysRevLett.97.026805,emary2013leggett}. (For details about weak values and weak measurements, see Sec. \ref{sec:wv}.) 
Such violations have been observed experimentally~\cite{palacios2010experimental,goggin2011violation,knee2012violation,PhysRevLett.111.090506}. 
We now elaborate on this connection.

In experimental tests of the weak-measurement version of the Leggett--Garg inequality, three measurements are 
performed: A strong, followed by a weak, followed by another strong measurement. This set-up allows one to verify quantum behaviour using one experimental configuration, rather than the three or four configurations needed for the original Leggett--Garg approach~\cite{Leggett85}. 
The three outcomes are $r^{(1)}$, $r^{(2)}$ and $r^{(3)}$. One analyses the correlation functions $K_{ij} = \la r^{(i)} r^{(j)} \ra$.  The superscripts label when a measurement takes place. The measurement outcomes are scaled such that the statistical average $\la r_1 \ra = \la \psi |{\hat A}| \psi \ra $, where 
$\hat{A}$ denotes a measured observable and $|\psi\ra$ denotes the prepared quantum state. 
Analogous statements concern $r_2$ and $r_3$, together with observables ${\hat B}$ and ${\hat C}$. 

The simplest case involves a two-level system. The three observables are 
$\hat{A} = \ketbra{a_1}{a_1} - \ketbra{a_2}{a_2}$, 
$\hat{B} = \ketbra{b_1}{b_1} - \ketbra{b_2}{b_2}$ and 
$\hat{C} = \ketbra{c_1}{c_1} - \ketbra{c_2}{c_2}$.   
The observables' eigenvalues are $a_1=b_1=c_1=+1$ and $a_2=b_2=c_2=-1$. The generalised Leggett--Garg inequality is 
\be
{\cal L} \equiv  K_{12} + K_{23} - K_{13} \leq 1.
\ee
One can achieve the classical upper bound if
the first and the second measurements are completely correlated, as are the second and third: $K_{12} = K_{23} = 1$.  Classically, 
also the first and the third measurements would consequently be completely correlated: $K_{13} = 1$, so ${\cal L}=1$. Classically, no other configuration can supersede this value of  ${\cal L}=1$.

When the intermediate measurement is weak, we can express ${\cal L}$ in terms of the  KD distribution $Q_{j,k}(\ketbra{a_i}{a_i}) \equiv \braket{a_i|c_k} \braket{c_k|b_j} \braket{b_j|a_i} $. 
For simplicity, we suppose that the quantum system begins in an $\hat A$ eigenstate $|a_i\ra$. The first strong measurement yields the eigenvalue $a_i$. It is followed by a weak measurement of ${\hat B}$ and a strong measurement of ${\hat C}$.  In this weak-measurement limit, we can replace the correlation function with the approximation
\be
\label{eq_LG_help0}
{\cal L} \approx {\rm Re} \left( \la a_i | {\hat A} {\hat B} + {\hat B} {\hat C} - {\hat A} {\hat C} | a_i \ra \right) .
\ee
This is the form of the correlation function considered by Leggett and Garg~\cite{Leggett85}.
We stress that the ${\hat B}$ measurement's weakness allows us to neglect that measurement's influence on the ${\hat A} {\hat C}$
correlator. If the intermediate measurement is of finite strength, this assumption breaks down. Reference~\cite{Williams08} describes the measurement's influence on the corrrelator.
Let us insert the operators' spectral expansions into Eq.~\eqref{eq_LG_help0}. We obtain an expression dependent on the KD distribution:
\be
{\cal L} \approx \sum_{j,k} 
{\rm Re} \LParen Q_{j,k}(\ketbra{a_i}{a_i}) \RParen
\left( a_i b_j + b_j c_k - a_i c_k\right).
\ee
The rightmost parenthetical factor is upper-bounded by 1. Hence we see that, for the right-hand side to exceed the upper bound of 1, 
some joint KD quasi-probability $Q_{j,k}(\ketbra{a_i}{a_i})$ must have a negative real part. 
This bound violation was observed experimentally.
Furthermore, this connection between Leggett--Garg-type inequalities and quasi-probability negativity was observed also in Refs.~\cite{Bednorz12,Dressel11}.

The Leggett--Garg inequality further underscores the connection between weak values and the KD distribution. Weakly measuring the intermediate observable $\hat{B} $, given strong pre- and post-selective measurements, is equivalent to measuring the weak values $B_{\mathrm{w}}(a_i,c_k) \equiv \frac{\braket{c_k | \hat{B} | a_i}}{\braket{c_k|a_i}}$ [Eq. \ref{eq_weakhelp1}]. Thus, in the weak measurement limit, ${\cal L}$ can be expressed as
\begin{align}
    \label{eq_LG_help1}
    {\cal L} \approx \sum_{k} P(c_k|a_i)
    \Big[ (a_i + c_k) {\rm Re} \LParen B_{\mathrm{w}}(a_i,c_k) \RParen - a_i c_k\Big].
\end{align}
The $P(c_k|a_i) = |\la c_k | a_i\ra|^2$ denotes the conditional probability that, if the first measurement yields $a_i$ and the $\hat{B}$ measurement is weak, the final measurement yields $c_k$.

The right-hand side of Eq.~\eqref{eq_LG_help1} is a convex sum. Therefore, for the right-hand side
to exceed  the classical upper bound of $1$, the factor in square brackets must exceed $1$ for some $a_i$ and $c_k$. 
When $\text{Re} \LParen B_{\mathrm{w}} \RParen > 0$, this condition can  be met only when $a_i = c_k = +1$ (when $i=k=1$). Otherwise, the factor in curly brackets 
is negative. In such a case, the bound-violation condition simplifies to $2\text{Re} \LParen B_{\mathrm{w}}(a_1,c_1) - 1 \RParen > 1$, or $\text{Re} \LParen B_{\mathrm{w}}(a_1,c_1) \RParen > 1$. That is, the weak value's real part must exceed the greatest $\hat{B}$ eigenvalue.  
If, instead, $\text{Re} \LParen B_{\mathrm{w}} \RParen < 0$, the classical bound $\mathcal{L} \leq 1$ can be violated 
only when $a_i = c_k = -1$  (when $i=k=2$). In this case, the condition necessary for violating the Leggett–Garg inequality simplifies to $-2\text{Re} \LParen B_{\mathrm{w}}(a_2,c_2)-1 \RParen > 1$, or $\text{Re} \LParen B_{\mathrm{w}}(a_2,c_2) \RParen< -1$.
Again, the weak value must lie outside the $\hat{B}$ spectrum.

These relations establish  a one-to-one correspondence between a Leggett--Garg-inequality violation and an anomalous weak value. Expanding the conditions in terms of KD quasi-probabilities yields 
$\sum_j b_j \,\text{Re} \LParen Q_{j,k=1} (\ketbra{a_1}{a_1}) \RParen
>  \sum_j\text{Re} \LParen Q_{j,k=1}(\ketbra{a_1}{a_1}) \RParen$, implying 
$\text{Re} \LParen Q_{j=2,k=1}(\ketbra{a_1}{a_1}) \RParen < 0$, or 
$\sum_j b_j\,Q_{j,k=2}(\ketbra{a_2}{a_2})
  <-\sum_j\text{Re} \LParen Q_{j,k=2}(\ketbra{a_2}{a_2}) \RParen$, implying 
  $\text{Re} \LParen Q_{j,k=2}(\ketbra{a_2}{a_2}) \RParen  <  0$. 
We can now interpret, in terms of KD quasi-probabilities, the conditions necessary for violating the Leggett--Garg inequality's classical upper bound of 1, as has been observed experimentally.
At least one of two  KD quasi-probabilities, $Q_{j=2,k=1}(\ketbra{a_1}{a_1})$ or $Q_{j=1,k=2}(\ketbra{a_2}{a_2})$, must  have a negative real part. 
Each such negative real part causes a corresponding weak value, 
${\rm Re} \LParen B_{\mathrm{w}}(a_1,c_1) \RParen$ or 
${\rm Re} \LParen B_{\mathrm{w}}(a_2,c_2) \RParen$, to lie outside the $\hat{B}$ spectrum.

\subsection{The KD distribution and the consistent-histories interpretation of quantum mechanics}
\label{SubSec:Consistent}

The mathematical framework of quantum mechanics is well-established but not a complete physical theory.  The role of an \textit{interpretation}  of quantum mechanics \cite{lewis2016quantum} is to append to the mathematical formalism a theory of 
real underlying (ontic) physics that results in the quantum phenomena we observe in laboratories. Such interpretations include the many-worlds interpretation~\cite{Everett73, Dewitt15}, de Broglie--Bohm mechanics~\cite{Bohm52} and the consistent-histories interpretation~\cite{Griffiths84}. The latter has a deep connection with the KD distribution. We review this connection below.

The consistent-histories interpretation of quantum mechanics predicts the same measurement probabilities as textbook quantum mechanics. The Copenhagen interpretation \cite{Fay24}, however, suggests that quantum states have well-defined properties only when measured. In contrast, the consistent-histories interpretation provides  a framework that (sometimes) describes properties of quantum particles between observations. 
The interpretation was developed by several scholars~\cite{GellMan93, Omnes99, Hartle07, Hartle10}, most notably by Griffiths~\cite{Griffiths84, Hohenberg10}. 

In the consistent-histories interpretation, properties (or physical attributes) of quantum states are defined in accordance with von Neumann's theory~\cite{Neumann18, Griffiths84}. A property is associated with a projective operator onto a subspace of a Hilbert space.\footnote{The theory accommodates more-general positive-operator-valued measures, however. Via Steinspring dilation, a positive-operator-valued measure defined one one Hilbert space is equivalent to a projector-valued measure on a larger Hilbert space. 
(Non-unique dilations are problematic for ontic interpretations.)} For example, `position' is not a property. However,  `the particle is at position $x$' is.  Consider two sets of projective measurement operators, $\mathcal{A} = \{ \hat{\Pi}^{(1)}_i \}$ and $\mathcal{B} = \{ \hat{\Pi}^{(2)}_j \}$. 
Quantum theory struggles 
to describe quantum states' underlying because
the question \emph{does the state have properties $ \hat{\Pi}^{(1)}_1$ and $ \hat{\Pi}^{(2)}_3$?} seems difficult to answer when $[ \hat{\Pi}^{(1)}_1,  \hat{\Pi}^{(2)}_3]\neq 0$. Under certain consistency conditions, the consistent-histories interpretation facilitates a description of underlying physical properties of a quantum state's time-evolution through Hilbert space.  

Next, we review the technical framework for the consistent-histories interpretation. Afterwards, we connect the consistency criterion to a KD distribution. Finally, we provide an example.

Consider a pure initial quantum state $\hat{\rho}_{\mathrm{i}}$. Consider also a set of times $t_0 < t_1 < \ldots < t_k < t_{\mathrm{f}}$.  We work in the Heisenberg picture, in which quantum states are constant in time and observables evolve unitarily. Define $k$ sets of 
projector-valued measures
$\mathcal{A}^{(l)} = \{ \hat{\Pi}^{(l)}_{i_l}(t_{l}) \} $,   where $\sum_{i_l}  \hat{\Pi}^{(l)}_{i_l}(t_{l}) = \hat{\mathbbm{1}}$ and  $l = 1, 2, \ldots, k$. 
We convert a Schr\"odinger-picture observable $\hat{X}$ to a Heisenberg-picture observable through
$\hat{X}(t_i) = \hat{U}^{\dagger}(t_0,t_i) \hat{X} \hat{U}(t_0,t_i)$.
Finally, consider a final rank-$1$ observable  $\hat{\rho}_{\mathrm{f}}(t_{\mathrm{f}})$.\footnote{Generalisations of this setting are described in~\cite{Griffiths84, Hohenberg10}.} A history is defined as
\begin{equation}
\label{Eq:History}
H\left( \hat{\rho}_{\mathrm{i}},  \hat{\Pi}^{(1)}_{i_1},  \hat{\Pi}^{(2)}_{i_2}, \ldots,  \hat{\Pi}^{(k)}_{i_k}, \hat{\rho}_{\mathrm{f}}  \right)
\equiv
\hat{\rho}_{\mathrm{i}} \rightarrow   \hat{\Pi}^{(1)}_{i_1}(t_{1})  \rightarrow  \hat{\Pi}^{(2)}_{i_2}(t_{2}) \rightarrow \cdots \rightarrow  \hat{\Pi}^{(k)}_{i_k}(t_{k}) \rightarrow \hat{\rho}_{\mathrm{f}} (t_{\mathrm{f}}) .
\end{equation}
The history 
means that a quantum system
is initialised in a state $\hat{\rho}_{\mathrm{i}}$ at time $t_0$; at time $t_1$, 
the system has a property corresponding to $\hat{\Pi}^{(1)}_{i_1}(t_{1}) $;  at time $t_2$, 
the system has a property corresponding to  $ \hat{\Pi}^{(2)}_{i_2}(t_{2})$; and so on, until, at the final time $t_{\mathrm{f}}$, the 
system has a property corresponding to $\hat{\rho}_{\mathrm{f}}$.  Given an initial state $\hat{\rho}_{\mathrm{i}}$ and a final observable $\hat{\rho}_{\mathrm{f}}(t_{\mathrm{f}})$, one can write down several potential histories. Two histories, $H\left( \hat{\rho}_{\mathrm{i}},  \hat{\Pi}^{(1)}_{i_1},  \hat{\Pi}^{(2)}_{i_2}, \ldots,  \hat{\Pi}^{(k)}_{i_k}, \hat{\rho}_{\mathrm{f}} \right)$ and $H^{\star}\left( \hat{\rho}_{\mathrm{i}},  \hat{\Pi}^{(1)}_{i_1^{\star}},  \hat{\Pi}^{(2)}_{i_2^{\star}}, \ldots,  \hat{\Pi}^{(k)}_{i_k^{\star}}, \hat{\rho}_{\mathrm{f}} \right)$, are said to be consistent with respect to one another if, and only if, their (normalised) Hilbert--Schmidt inner product vanishes:
\begin{equation}
\tilde{Q}(H,H^{\star}) 
\equiv \frac{1}{\Tr \bm{(} \hat{\rho}_{\mathrm{f}} ( t_{\mathrm{f}} )  \hat{\rho}_{\mathrm{i}} \bm{)}} 
\Tr \left( \hat{\rho}_{\mathrm{f}} (  t_{\mathrm{f}} )\hat{\Pi}^{(k)}_{i_k}(t_{k})  \cdots  \hat{\Pi}^{(1)}_{i_1}(t_{1}) \hat{\rho}_{\mathrm{i}}  \hat{\Pi}^{(1)}_{i_1^{\star}} (t_{1}) \cdots \hat{\Pi}^{(k)}_{i_k^{\star}} (t_{k}) \hat{\rho}_{\mathrm{f}} (t_{\mathrm{f}}) \right) = 0  .
\end{equation} 
$\tilde{Q}(H,H^{\star})$ is a KD quasi-probability representing the strength of the quantum interference between two paths (histories) in Hilbert space.  

A \emph{family} $f = \{ H_i \}$ of histories 
consists of all the histories
\begin{equation}
\label{Eq:family}
f
\equiv
\left\{ \hat{\rho}_{\mathrm{i}} \rightarrow \mathcal{C} \left( \mathcal{A}^{(1)} \right)  \rightarrow \mathcal{C} \left( \mathcal{A}^{(2)} \right)  \rightarrow \cdots \rightarrow \mathcal{C} \left(    \mathcal{A}^{(k)} \right)  \rightarrow \hat{\rho}_{\mathrm{f}} \right\} ,
\end{equation}
where $\mathcal{C} \left(    \mathcal{A}^{(l)} \right) $ represents any projector in $\mathcal{A}^{(l)} $.  
A family $f = \{ H_i \}$ of histories is consistent if, and only if,
\begin{equation}
Q(H_i ,H_j) = 0   \; \mathrm{and} \; H_i ,H_j \in f  , \; \; \forall i \neq j  .
\end{equation}
If  $f$ is consistent, then $\tilde{Q}(H_i ,H_i) $ is a classical conditional joint probability distribution for all $H_i \in f$. Within the context of a consistent family, one can ask if an individual history has happened.  Given consistency, one can regard $\tilde{Q}(H_i ,H_i) $ 
as the probability that $H_i$ happened. 

Often, one is interested only in whether a single history happened. In this case, one need only consider the minimal family of which that history can be a member. The minimal family of  $H\left( \hat{\rho}_{\mathrm{i}},  \hat{\Pi}^{(1)}_{i_1},  \hat{\Pi}^{(2)}_{i_2}, \ldots,  \hat{\Pi}^{(k)}_{i_k}, \hat{\rho}_{\mathrm{f}} \right)$ is
\begin{equation}
f_{\mathrm{min}} \left( H \right) \equiv
\left\{ \hat{\rho}_{\mathrm{i}} \rightarrow \mathcal{C} \left( \left\{ \hat{\Pi}^{(1)}_{i_1}, \, \hat{\mathbbm{1}} - \hat{\Pi}^{(1)}_{i_1}  \right\} \right)  \rightarrow \mathcal{C} \left( \left\{ \hat{\Pi}^{(2)}_{i_2}, \, \hat{\mathbbm{1}} - \hat{\Pi}^{(2)}_{i_1}  \right\} \right) \rightarrow \cdots \rightarrow \mathcal{C} \left( \left\{ \hat{\Pi}^{(k)}_{i_k}, \, \hat{\mathbbm{1}} - \hat{\Pi}^{(k)}_{i_k}  \right\} \right) \rightarrow \hat{\rho}_{\mathrm{f}} \right\}.
\end{equation}
If $f_{\mathrm{min}} \left( H \right) $ is consistent, then the history $H$ is also said to be consistent.
In this case, one can meaningfully ask, \emph{did $H$ occur, or not?}  

\begin{figure}[b]
\includegraphics[width=0.3\textwidth]{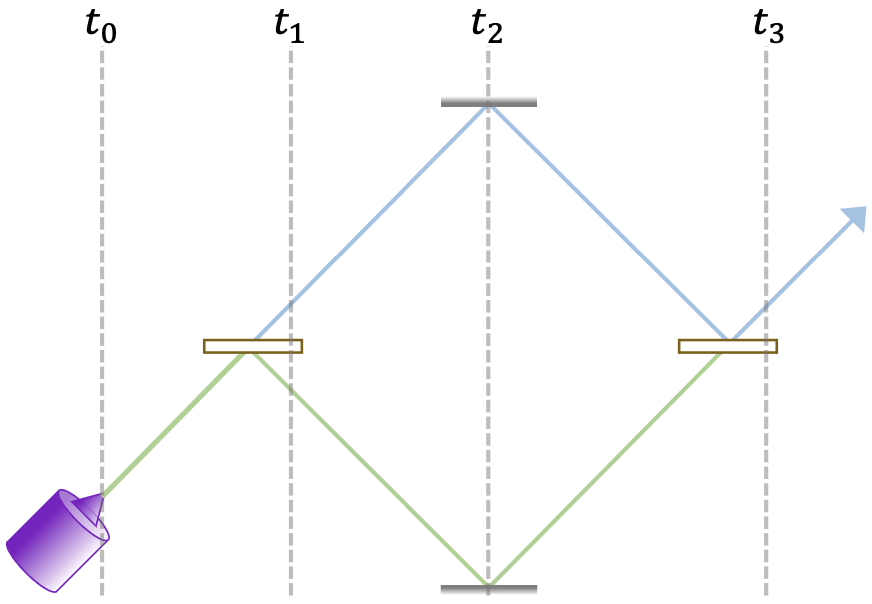}
\caption{\textbf{Mach--Zehnder histories.} 
A Mach--Zehnder interferometer. The quantum particle is initialised in the lower path. Hollow rectangles represent 50--50  beam-splitters; shaded rectangles represent mirrors. 
}
\label{fig:nonintegrable-OTOCMZ}
\end{figure}

To illustrate the consistent-histories interpretation, we provide an example. Consider a Mach--Zehnder interferometer~\cite{Zehnder91, Mach92} as shown in Fig.~\ref{fig:nonintegrable-OTOCMZ} . We associate the lower paths (green) with the state $\ket{0}$ and the upper paths (blue) with the state $\ket{1}$.  At time $t_0$, a quantum particle is prepared in the state  $\rho_{\mathrm{i}} = \ketbra{0}{0}$. Just before time $t_1$, a 50--50  beam-splitter implements the unitary operator $\hat{U}_{\mathrm{BS}}(t_0, t_1) =e^{ - i \frac{\pi}{4} \hat{\sigma}_y  }$.   Just before time $t_3$, another 50--50  beam-splitter implements $\hat{U}_{\mathrm{BS}}(t_2, t_3)=e^{ - i \frac{\pi}{4} \hat{\sigma}_y  } $.  
In a final event, the particle has
the property  $\rho_f(t_4) = \ketbra{1}{1}$ at time $t_4$.  

We wish to say something about the particle's position property at time $t_2$, between the two beam-splitters.  According to the Copenhagen interpretation, we cannot say anything about this
property at $t_2$, because no measurement was made.  
Let us now analyse the problem from the consistent-histories perspective.  
We want to ask, at $t_2$, if the particle had the property $\hat{\Pi}^0 \equiv \ketbra{0}{0}$ (was on the lower path) or had the property $\hat{\Pi}^1 \equiv \ketbra{1}{1} = \hat{\mathbbm{1}} -\hat{\Pi}^0 $ (was on the upper path). Define the histories
\begin{align}
H_0 &
\equiv
\hat{\rho}_{\mathrm{i}} \rightarrow  \hat{\Pi}^0 (t_2)  \rightarrow \hat{\rho}_{\mathrm{f}} (t_3)  \; \; \mathrm{and} \\
H_1 &
\equiv
\hat{\rho}_{\mathrm{i}} \rightarrow  \hat{\Pi}^1 (t_2)  \rightarrow \hat{\rho}_{\mathrm{f}} (t_3) .
\end{align}
$\{ H_0, H_1 \}$ forms the minimal family for $H_0$ and for $H_1$; we call this family $f^{(0/1)}$.
We can calculate $\tilde{Q}(H_0,H_1)=\tilde{Q}(H_1,H_0)=\frac{1}{4}$. Thus, according to the consistent-histories interpretation, one cannot ask, in a meaningful way, if the particle was in the upper or lower path at $t_2$. 

Instead, let us define 
$\ket{\pm}=\frac{1}{\sqrt{2}} \left( \ket{0} \pm \ket{1} \right)$.
We now ask whether, at $t_2$, the particle had the property $\hat{\Pi}^+ \equiv \ketbra{+}{+}$  or the property $\hat{\Pi}^- \equiv  \ketbra{-}{-} = \hat{\mathbbm{1}} - \hat{\Pi}^+$.
Each of the histories
\begin{align}
H_+ &
\equiv
\hat{\rho}_{\mathrm{i}} \rightarrow  \hat{\Pi}^+ (t_2)  \rightarrow \hat{\rho}_{\mathrm{f}} (t_3)  \; \; \mathrm{and} \\
H_- &
\equiv
\hat{\rho}_{\mathrm{i}} \rightarrow  \hat{\Pi}^- (t_2)  \rightarrow \hat{\rho}_{\mathrm{f}} (t_3) 
\end{align}
has the minimal family $f^{(+/-)} \equiv \{ H_+, H_-\}$.
We can calculate $\tilde{Q}(H_+,H_-)=\tilde{Q}(H_-,H_+)=0$. Thus, according to the consistent-histories interpretation, one can ask meaningfully if the particle had the property of a $\ket{+}$-type or a $\ket{-}$-type superposition at $t_2$.  In fact, $\tilde{Q}(H_+,H_+)=1$. Therefore, the consistent-histories interpretation 
implies that a particle initially in $\rho_{\mathrm{i}} = \ketbra{0}{0}$ and finally in $\rho_{\mathrm{f}} = \ketbra{1}{1}$ definitely had the property $\hat{\Pi}^+ = \ketbra{+}{+}$ at $t_2$. 

\subsection{\Dave{Contextuality}}
\label{SubSec:Context}

\nicole{
According to a useful notion of non-classicality, an experiment is 
called non-classical when explaining it requires
radical departures from the classical world-view~\cite{jennings2016no}.  
Bell non-locality~\cite{brunner2014bell} is arguably the strongest notion of non-classicality. However, it can be provably identified only in experiments involving space-like-separated systems.  A more widely applicable notion is contextuality, first developed by Kochen and Specker \cite{Specker60, Kochen68, Budroni22}. Spekkens generalised the notion later~\cite{Quanta22,spekkens2005contextuality}.  Below, we review  Kochen--Specker contextuality and generalised contextuality
and their connections with KD distributions. 
}

\subsubsection{\Dave{Kochen--Specker contextuality}}
\label{SubSec:KSContext}

\nicole{
In classical physics, 
one can ascribe exact physical properties to objects. 
For example, even if all measuring rods' imperfections lead to uncertainty in the measurement of a room's length, one assumes that the room has an exact length.
However, a property of a quantum object,
as registered by a measurement, may depend on the 
measurement's context---on which other properties were measured simultaneously.  This dependence is \emph{contextuality}. 
}

\Dave{
The initial demonstration of Kochen--Specker contextuality involved sharp measurements of rank-$1$ projectors.  Consider a set of rank-$1$ projector observables:  
$\mathcal{G} \equiv \{  \hat{A}^{(1)},  \hat{A}^{(2)}, \ldots,  \hat{A}^{(n)} \}$. 
\nicole{Denote the $i^{\rm th}$ eigenvalue of projector $j$ by $a_{i_j}^{(j)}  \in  \{0, 1 \}$.} 
A Kochen--Specker context $c_{\mathrm{KS}}$ is a subset of compatible observables in $\mathcal{G}$, i.e., $c_{\mathrm{KS}} \subseteq \mathcal{G}$.  
\Dave{For any given context $c_{\mathrm{KS}}$,  there is a probability distribution over the 
eigenvalues $a_{i_j}^{(j)}$
of each observable $\hat{A}^{(j)} \in c_{\mathrm{KS}}$.}   
\Dave{As all observables in $c_{\mathrm{KS}}$ are compatible,} one \Dave{can construct} a joint probability \nicole{consistent with the observables' individual probabilities}: 
$P_{c_{\mathrm{KS}}} \left( \{ a^{(j)}_{i_j} \; | \; \hat{A}^{(j)} \in c_{\mathrm{KS}}  \} \right)$. } \Dave{An experiment is Kochen--Specker non-contextual if there exists a global probability assignment over the products of the outcomes of all  the observables in $\mathcal{G}$, such that it is compatible with the outcome probabilities $P_{c_{\mathrm{KS}}} ( \{ a^{(j)}_{i_j} \; | \; \hat{A}^{(j)} \in c_{\mathrm{KS}} )$ of any considered context $c_{\mathrm{KS}}$. If not, the experiment is Kochen--Specker contextual.}

\nicole{One might wish to ascribe to a quantum system exact physical properties, corresponding to the observables, as in classical physics. Which necessary properties of $P_{c_{\mathrm{KS}}}\left( \{ a^{(j)}_{i_j} \; | \; \hat{A}^{(j)} \in c_{\mathrm{KS}}  \} \right)$ enable such an ascription?}
\nicole{Consider contexts $c_{\mathrm{KS}}$ and $c_{\mathrm{KS}}^{\prime}$ associated to probability distributions $P_{c_{\mathrm{KS}}}$ and $P_{c^{\prime}_{\mathrm{KS}}}$.
Consider marginalising $P_{c_{\mathrm{KS}}}$  over the outcomes of the observables absent from $c_{\mathrm{KS}}^{\prime}$.  We denote the marginal by 
$\tilde{P}_{c_{\mathrm{KS}}} $, and we define
$\tilde{P}_{c_{\mathrm{KS}}^{\prime}} $ analogously.
\nicole{In terms of these distributions, we can state a condition necessary for the existence of a global probability assignment over the observables in $\mathcal{G}$~\cite{Abramsky_2011}:}
\begin{equation}
\tilde{P}_{c_{\mathrm{KS}}} \left(  \left\{ a_{i_j}^{(j)} \; | \; \hat{A}^{(j)} \in c_{\mathrm{KS}} \cap c_{\mathrm{KS}}^{\prime}  \right\}  \right) 
= \tilde{P}_{c^{\prime}_{\mathrm{KS}}}  \left(  \left\{ a_{i_j}^{(j)} \; | \; \hat{A}^{(j)} \in c_{\mathrm{KS}} \cap c_{\mathrm{KS}}^{\prime}  \right\}  \right) , \quad \Dave{\forall c_{\mathrm{KS}}, c_{\mathrm{KS}}^{\prime}} .
\end{equation}
}

\nicole{
We now specify the forms assumed by the joint probabilities $P_{c_{\mathrm{KS}}}\left( \{ a^{(j)}_{i_j} \; | \; \hat{A}^{(j)} \in c_{\mathrm{KS}}  \} \right)$ in any Kochen--Specker--non-contextual  model.
Consider
 measuring any observable.
The outcome follows deterministically from a hidden variable, regardless of the context, in any non-contextual Kochen--Specker theory. This hidden variable is called an \emph{ontic state} and denoted by $\lambda \in \Lambda$.
$\Lambda$ is a discrete or continuous set of ontic states. 
Now, consider measuring all the observables in a context $c_{\mathrm{KS}}$.
We obtain the outcomes \Dave{$ \{ a_{i_j}^{(j)} \; | \; \hat{A}^{(j)} \in c_{\mathrm{KS}}  \}$, in} any non-contextual Kochen--Specker theory, with a probability
\begin{equation}
P_{c_{\mathrm{KS}}}  \left( \left \{ a_{i_j}^{(j)} \; | \; \hat{A}^{(j)} \in c_{\mathrm{KS}}  \right\}  \right) 
= \sum_{\lambda} P(\lambda) P \left( \left\{ a_{i_j}^{(j)} \;   | \; \hat{A}^{(j)} \in c_{\mathrm{KS}}  \right\}  | \lambda \right) .
\end{equation}
The conditional probability factorises:
\begin{equation}
\label{eq_P_lambda} 
P \left( \left\{ a_{i_j}^{(j)} \;   | \; \hat{A}^{(j)} \in c_{\mathrm{KS}}  \right\}  | \lambda \right) = \prod_{ a_{i_j}^{(j)} \; :   \; \hat{A}^{(j)} \in c_{\mathrm{KS}} } P (  a_{i_j}^{(j)}  | \lambda  )  \in \{ 0,1\} .
\end{equation}
\nicole{\Dave{The reason $P_{c_{\mathrm{KS}}}\left( \{ a^{(j)}_{i_j} \; | \; \hat{A}^{(j)} \in c_{\mathrm{KS}}  \} \right)$ can be expressed as a product} is that $P (  a_{i_j}^{(j)}  | \lambda ) \in \{ 0 , 1\}$, because}   Kochen--Specker theory concerns deterministic hidden-variable models. 
}

\nicole{
Kochen and Specker introduced example contexts in which a quantum system has \Dave{properties}  represented by \Dave{projectors.} If the system is in one of certain quantum states, no definite (0 or 1) value can be ascribed to \Dave{those properties~\cite{Kochen68}}. 
Hence quantum theory is contextual in the Kochen--Specker sense. Other researchers simplified the initial proofs~\cite{Peres90, Mermin93, Simon01,Larsson02,Cabello08, Klyachko08}.  \Dave{Whilst quantum theory is Kochen--Specker contextual,  certain quantum experiments can be explained by Kochen--Specker non-contextual ontic models.  In such experiments, one can ascribe definite properties to projective observables of a system, irrespectively of how these properties may be measured.  With respect to the theory of Kochen and Specker, these experiments are deemed classical.}
}

\nicole{Kirkwood--Dirac distributions offer promise for elucidating Kochen--Specker contextuality and the logical paradoxes stemming from it. 
\Dave{Many} aspects of Kochen--Specker contextuality
connect to weak values \Dave{(see Sec. \ref{Sec:WeakValues}) and thus  indirectly }
to KD distributions \cite{HofmannParadoxes15, HofmannContext20, HofmannNonclassical23, Ji2024quantitative, Hance2023}.  
These connections invite further exploration.
Here, we review the Kochen--Specker inequality proved by Klyachko, Can, Binicio\u{g}lu and Shumovsky (KCBS)~\cite{Klyachko08, Budroni22}. We highlight its close connection with KD distributions. 
}

\begin{figure}[b]
\includegraphics[width=0.35\textwidth]{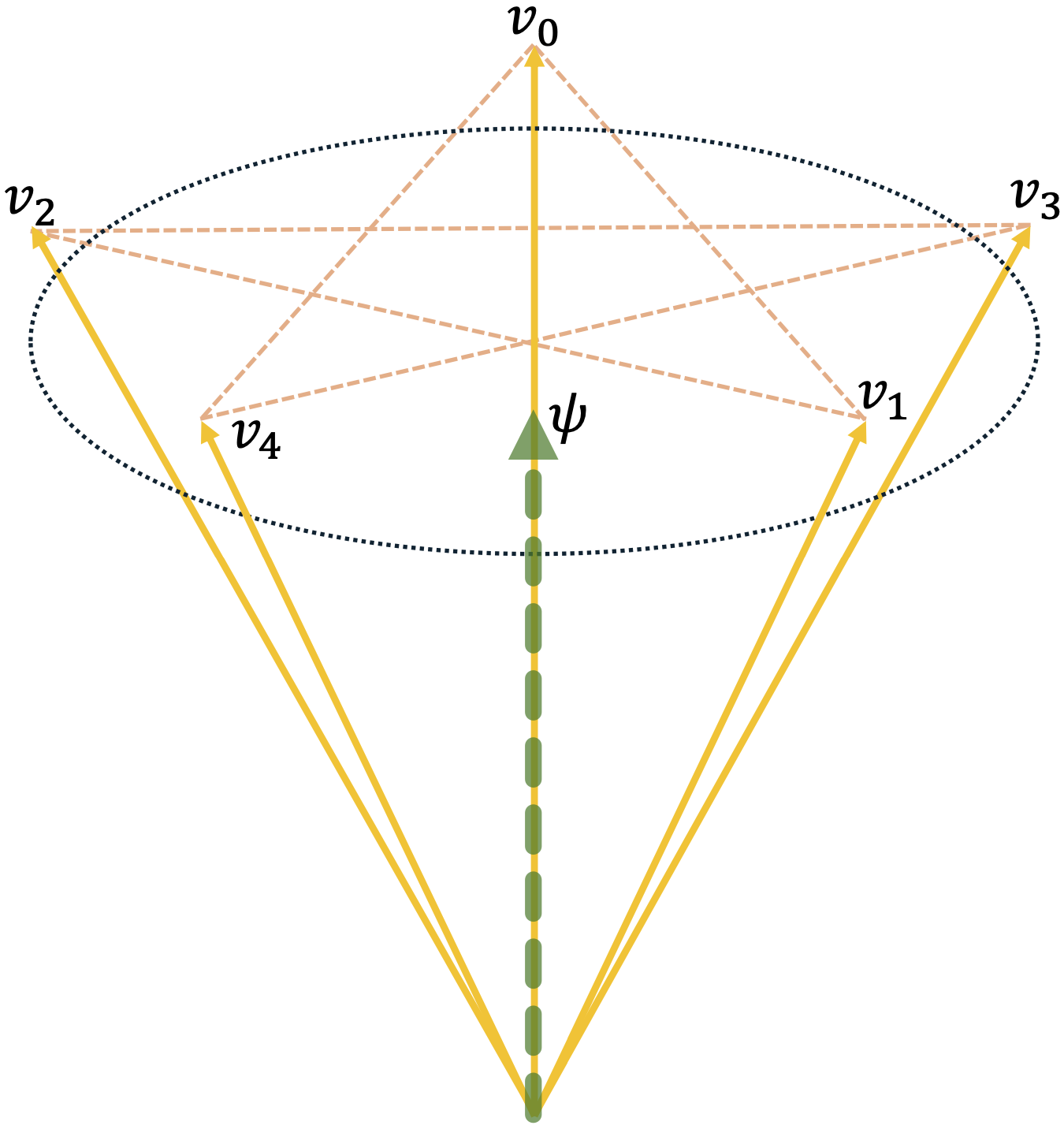}
\caption{\textbf{KCBS pentagram.} \nicole{
The solid yellow arrows on $\mathbb{R}^3$ represent the 5
real vectors $\ket{v_j}$, wherein $j \in \{0, 1, 2, 3, 4 \}$.
A dashed green line represents the quantum state $\ket{\psi}=(0,0,1)^{\top}$.}
}
\label{fig:Pentagram}
\end{figure}

\nicole{
The setup includes a Hilbert space of dimensionality $d=3$, the least 
dimensionality in which one can prove Kochen--Specker contextuality~\cite{Budroni22}.  Consider $5$ real qutrit vectors
$\ket{ v_j }$, wherein $j \in \{0, 1, 2, 3, 4 \}$.  
Relative to a computational basis, $\ket{v_j}$ is represented by a list of three real numbers:
\begin{equation}
   \label{eq_Components}
   \ket{v_j} = \bm{\left(}  \sin(\theta) \sin(\phi_j)  , \: \sin(\theta) \cos(\phi_j), \: \cos(\theta) 
   \bm{\right)}^{\top },
\end{equation}
where $\cos(\theta) = \frac{1}{\sqrt[4]{5}}$ and 
$\phi_j = \left( 2j + 1 \mod 5 \right)  \times \frac{2 \pi}{5} $.
We can regard Eq.~\eqref{eq_Components} as specifying the $x$-, $y$-, and $z$-coordinates of a point in three-dimensional space. In this way, we depict $\ket{ v_{0,\ldots,4} }$ with solid yellow lines in the KCBS pentagram of Fig. \ref{fig:Pentagram}.}

\nicole{From the vectors, we form 5 observables:
\begin{equation}
\hat{A}^{(j)} \equiv 2 \ketbra{v_j}{v_j} - \hat{\mathbbm{1}} .
\end{equation}
\Dave{Above, we defined the $\hat{A}^{(j)}$ as projectors, whose eigenvalues lie in $\{ 0,1\}$.  To simplify later algebra, we have rescaled and shifted the spectra: }
The eigenvalues $a^{(j)}_{i_j}$ of $\hat{A}^{(j)}$ are $+1$, $-1$, and $-1$, for all $j$.
The observables share another important property: Consider any two vectors connected by a salmon dashed line in Fig. \ref{fig:Pentagram}. The corresponding vectors are orthogonal: 
$[\hat{A}^{(j)} , \hat{A}^{(j+1\mod 5)}]=0$ for all $j$. 
}

\nicole{
KCBS introduced a set of 5 contexts, each formed from two commuting (compatible) observables:
$\{ ( \hat{A}^{(0)} , \hat{A}^{(1)} ) , \:  ( \hat{A}^{(1)} , \hat{A}^{(2)} ) , \:  ( \hat{A}^{(2)} , \hat{A}^{(3)} ) , \:  ( \hat{A}^{(3)} , \: \hat{A}^{(4)} ) , \:  ( \hat{A}^{(4)} , \hat{A}^{(0)} )  \}$.  
Using these contexts, one can define a correlation function:
\begin{equation}
\label{Eq:KS-S}
S \equiv \braket{\hat{A}^{(0)}  \hat{A}^{(1)}} + \braket{\hat{A}^{(1)}  \hat{A}^{(2)}} + \braket{\hat{A}^{(2)}  \hat{A}^{(3)}} + \braket{\hat{A}^{(3)}  \hat{A}^{(4)}} + \braket{\hat{A}^{(4)}  \hat{A}^{(0)}} .
\end{equation}
KCBS bounded $S$, using properties of the observables' eigenvalues, as follows. 
Recall that each eigenvalue $a^{(j)}_{i_j} = \pm 1$. Therefore,
$( a^{(j)}_{i_j} )^2 = 1$, and
$\left(  a^{(0)}_{i_0} a^{(1)}_{i_1}  \right)
   \left( a^{(1)}_{i_1} a^{(2)}_{i_2} \right)
   \left( a^{(2)}_{i_2} a^{(3)}_{i_3} \right)
   \left( a^{(3)}_{i_3} a^{(4)}_{i_4} \right)
   \left( a^{(4)}_{i_4} a^{(0)}_{i_0} \right)
   =  1 .$ ~\cite{Klyachko08}
Furthermore, since each $a^{(j)}_{i_j} = \pm 1$, at least one parenthesised factor equals 1. Hence, if we exchange parentheses for pluses,
\begin{equation}
a^{(0)}_{i_0} a^{(1)}_{i_1} + a^{(1)}_{i_1} a^{(2)}_{i_2} + a^{(2)}_{i_2} a^{(3)}_{i_3} + a^{(3)}_{i_3} a^{(4)}_{i_4} + a^{(4)}_{i_4} a^{(0)}_{i_0} \geq -3 .
\end{equation}
The sum saturates the bound if 4 of the terms (the greatest possible number of terms) equal $-1$.
Let us combine this inequality with Eq.~\eqref{eq_P_lambda}, Eq.~\eqref{Eq:KS-S}, and the upper bound of 1 on each probability:
\begin{equation}
   \label{eq_S_bound}
S = \sum_{i_0, \ldots, i_4} P(a^{(0)}_{i_0}, a^{(1)}_{i_1}, a^{(2)}_{i_2},a^{(3)}_{i_3} ,a^{(4)}_{i_4} \, | \, \lambda) \left( a^{(0)}_{i_0} a^{(1)}_{i_1} + a^{(1)}_{i_1} a^{(2)}_{i_2} + a^{(2)}_{i_2} a^{(3)}_{i_3} + a^{(3)}_{i_3} a^{(4)}_{i_4} + a^{(4)}_{i_4} a^{(0)}_{i_0} \right) \geq -3 
\end{equation}
in every non-contextual deterministic hidden-variable model.
}

\nicole{
One can rephrase Eq.~\eqref{Eq:KS-S} in terms of a KD distribution:
in terms of the $5$-extended KD quasi-probability
\begin{equation}
Q_{i_0,i_1,i_2,i_3, i_4} (\hat{\rho}) \equiv  \braket{a_{i_4}^{(4)} | a_{i_{3}}^{(3)}} \braket{a_{i_{3}}^{(3)} | a_{i_{2}}^{(2)}}  \braket{a_{i_{2}}^{(2)} | a_{i_{1}}^{(1)}}   \braket{a_{i_{1}}^{(1)} | a_{i_{0}}^{(0)} }  \bra{a_{i_{0}}^{(0)}} \hat{\rho} \ket{a_{i_{4}}^{(4)}} ,
\end{equation}
the correlator is
\begin{equation}
S\bm{(} Q(\hat{\rho}) \bm{)} =  \sum_{i_0,\ldots, i_4} Q_{i_0,\ldots, i_4} (\hat{\rho})  \left( a^{(0)}_{i_0} a^{(1)}_{i_1} + a^{(1)}_{i_1} a^{(2)}_{i_2} + a^{(2)}_{i_2} a^{(3)}_{i_3} + a^{(3)}_{i_3} a^{(4)}_{i_4} + a^{(4)}_{i_4} a^{(0)}_{i_0} \right)  .
\end{equation}
Restricting $Q(\hat{\rho})$ to be a probability distribution, one reproduces the non-contextuality bound Eq.~\eqref{eq_S_bound}:
\begin{equation}
\max_{Q_{i_0,\ldots, i_4}  \in [0,1]} S\bm{(} Q(\hat{\rho}) \bm{)}  \geq -3 .
\end{equation}
Thus, if the KD distribution $Q_{i_0,\ldots, i_4} (\hat{\rho})$ is positive, 
then $S \geq -3$, and a Kochen--Specker non-contextual model describes the state. A non-positive KD distribution may break the bound. For example, define 
$\ket{\psi} = (1,0,0)^{\top}$ and $\hat{\rho} = \ketbra{\psi}{\psi}$. 
The quasi-probability $Q_{i_0,\ldots, i_4} (\hat{\rho})$ assumes negative values and violates the non-contextual inequality:
\begin{equation}
S\bm{(} Q(\hat{\rho}) \bm{)}  = 5 - 4 \sqrt{5} \approx -3.9 .
\end{equation} 
To summarise, non-positive quasi-probabilities in the KD distribution $Q_{i_0,\ldots, i_4} (\hat{\rho})$ are necessary for breaking the Kochen--Specker non-contextual inequality of \emph{Klyachko et al.}
}

\subsubsection{\Dave{Generalised contextuality}}
\label{SubSec:GenContext}
\nicole{
Kochen--Spekker contextuality can be ascribed only to \Dave{deterministic hidden-variable models of quantum experiments that involve projective measurements.} 
Spekkens lifted these restrictions in two ways, in developing generalised contextuality \cite{spekkens2005contextuality}. First, any ontological model may have generalised contextuality---not only deterministic hidden-variable models of quantum mechanics. Second, preparations, general measurements, and transformations can be contextual---not only projective measurements.
}
Below, we introduce generalised contextuality. We then explain that, if a particular protocol for measuring the KD distribution yields non-positive values, the protocol is contextual.
Afterwards, we discuss how  KD distributions  witness \Dave{generalised} contextuality in enhanced work extraction and in breaking bounds on thermodynamic currents.

Non-contextuality, in its generalised form~\cite{spekkens2005contextuality,Quanta22}, is the property that a physical theory does not introduce distinct 
mathematical representations of operationally indistinguishable phenomena within a class of theories known as \emph{ontological models}. In other words, if two things cannot be 
distinguished, 
they are the same thing within the ontological model.
For example, Einstein, in his `On the Electrodynamics of Moving Bodies'~\cite{einstein1905electrodynamics}, criticised the way in which electrodynamics (as understood at the time via the notion of ether) assigns distinct physical descriptions to a metallic ring moving towards a magnet and to a magnet moving towards a ring. After all, the two experiments are impossible to tell apart by any measurement. One can argue that the theory-selection principle 
being applied is non-contextuality~\cite{spekkens2019ontological}.

The word \emph{ontological} labels a theory of an underlying physical reality. The underlying physical reality may be inaccessible (hidden).  An experiment consists of preparation procedures \Dave{$\mathscr{P}$,} transformation procedures \Dave{$\mathscr{T}$,} and measurement procedures \Dave{$\mathscr{M}$.} These generate a set $\{P(k|\mathscr{P},{\mathscr{T}},{\mathscr{M}}) \}_{\mathscr{P},{\mathscr{T}},{\mathscr{M}}}$ of outcomes statistics. $p(k|\mathscr{P},{\mathscr{T}},{\mathscr{M}})$ denotes the probability of observing outcome $k$, given that we prepared the system according to $\mathscr{P}$, transformed it according to ${\mathscr{T}}$ and measured it according to ${\mathscr{M}}$.
An ontological model associates to operational procedures (preparations, transformations and measurements) probabilistic representations over a measurable state space $\Lambda$~\cite{Quanta22},  the set of \emph{physical states}.
We model a preparation procedure $\mathscr{P}$ by sampling
$\lambda \in \Lambda$ according to a probability distribution $P_{\mathscr{P}}(\lambda)$. Each  transformation ${\mathscr{T}}$ is modelled by a matrix; each element $P_{\mathscr{T}}(\lambda'|\lambda)$ equals the probability of jumping from $\lambda$ to $\lambda'$. Each measurement procedure ${\mathscr{M}}$ is modelled by another matrix; each element $P_{\mathscr{M}}(k|\lambda')$ equals the probability of obtaining outcome $k$, given the ontic state $\lambda'$.  

The maps ${\mathscr{P}} \mapsto P_{\mathscr{P}}$, ${\mathscr{T}} \mapsto P_{\mathscr{T}}$ and ${\mathscr{M}} \mapsto P_{\mathscr{M}}$ must be convex-linear. This property ensures a reasonable probabilistic interpretation. For example, imagine tossing a fair coin to decide whether we follow the preparation procedure ${\mathscr{P}}_1$ or ${\mathscr{P}}_2$. The ontological model must assign the probability 
$[P_{{\mathscr{P}}_1}(\lambda) + P_{{\mathscr{P}}_2}(\lambda) ]/2$ to the physical state $\lambda$. 
From the propagation of probabilities, an ontological model predicts that outcome $k$ will occur with a probability
 \begin{equation}
 \label{eq:ontologicalmodelprediction}
     P(k|{\mathscr{P}},{\mathscr{T}},{\mathscr{M}}) = \int_{\Lambda} d\lambda \int_{\Lambda} d\lambda'  \, P_{\mathscr{M}}(k|\lambda') \, P_{\mathscr{T}}(\lambda'|\lambda) \, P_{\mathscr{P}}(\lambda)\, .
 \end{equation} 
See Ref.~\cite{leifer2014quantum} for an introduction to ontological models.
 
Ontological models, unless otherwise restricted, can reproduce nearly anything. 
For example, consider a quantum experiment. In the quantum formalism, ${\mathscr{P}}$ is represented by a pure state $|\psi\rangle \in \mathbb{C}^d$, ${\mathscr{T}}$ by a  $d \times d$ unitary $\hat{U}$ and ${\mathscr{M}}$ by a projective measurement $\mathcal{M} = \{ \ketbra{k}{k} \}_{k=1}^d$. 
By the Born rule,
$P(k|{\mathscr{P}},{\mathscr{T}},{\mathscr{M}}) = |\langle k| \hat{U} |\psi\rangle|^2$.
The right-hand side of Eq.~\eqref{eq:ontologicalmodelprediction} can reproduce this function if $\Lambda$ consists of normalised pure states.  
In this model, whose ontic states are normalised pure states, the following assignments are made: $P_\psi(\lambda)  = \delta(\lambda- \psi)$, 
$P_{\mathscr{T}}(\lambda'|\lambda) = \delta(\lambda' - \hat{U} \psi)$ and 
$P_{\mathscr{M}}(k|\lambda') = |\langle k| \lambda'\rangle|^2$~\cite{beltrametti1995classical}.
On can represent, within the ontological model, a general preparation procedure ${\mathscr{P}}$ with a probability  measure on $\Lambda$:
 \begin{equation}\label{eq:pPdef}
 P_{\mathscr{P}}(\lambda)=\sum_n P_n\delta(\lambda-\psi_n)=\sum_n P_n P_{\psi_n}(\lambda),
 \quad \text{such that} \quad 
 \sum_n P_n=1.
 \end{equation}
Hence 
a general ontological model
can model any quantum experiment.
Ontological models 
need not be deterministic. For example, in the above model, $\lambda$ determines the measurement outcome probabilistically.

Non-contextuality can 
restrict ontological models to satisfy
a notion of classicality. 
Imagine preparing two copies of a system according to two different procedures, ${\mathscr{P}}$ and ${\mathscr{P}}'$. Suppose that we obtain the same measurement statistics, no matter which measurements ${\mathscr{M}}$ we conduct: $P(k|{\mathscr{P}},{\mathscr{M}})=P(k|{\mathscr{P}}',{\mathscr{M}})$ for all $k$ and ${\mathscr{M}}$. This condition, known as an \emph{operational equivalence}, is denoted by ${\mathscr{P}} \sim {\mathscr{P}}'$~\cite{spekkens2005contextuality}. Non-contextuality requires that the ontological model assign the same distribution over $\lambda$ to both preparations: $P_{\mathscr{P}}(\lambda)=P_{{\mathscr{P}}'}(\lambda)$ for all $\lambda$. For example, consider forming a non-contextual  description of the preparation of a density operator $\hat{\rho}$. 
We must represent $\hat{\rho}$ by the same $P_{\hat{\rho}}(\lambda)$, independently of the pure-state mixture used to prepare $\hat{\rho}$. 
The previous example of an ontological model for quantum mechanics is contextual (is not non-contextual). Indeed, 
$P_{\mathscr{P}}$ can differ from $P_{{\mathscr{P}}'}$ [see Eq.~\eqref{eq:pPdef}], while 
\begin{equation}
    \hat{\rho}_{\mathscr{P}} = \sum_n P_n \ketbra{\psi_n}{\psi_n}
    = \hat{\rho}_{{\mathscr{P}}'}
    =\sum_n P'_n \ketbra{\psi'_n}{\psi'_n}.
\end{equation}
In other words, in contextual models, different procedures ${\mathscr{P}}$ can yield different $P_{\mathscr{P}}$, while nevertheless yielding the same density matrix [see Eq.~\eqref{eq:pPdef}] and hence the same experimental outcomes. In non-contextual models ${\mathscr{P}} \sim {\mathscr{P}}'$ implies $P_{\mathscr{P}}(\lambda)=P_{{\mathscr{P}}'}(\lambda)$.
Similarly, consider two transformations that realise the same quantum channel $\mathcal{C}$.
Non-contextuality requires that the transformations be assigned the same $P_{\mathcal{C}}(\lambda'|\lambda)$. 
A similar rule governs two measurements associated to the same positive-operator-valued measure $\mathcal{M}$. The measurements are assigned the same $P_{\mathcal{M}}(k|\lambda)$ in any non-contextual model.
In short, generalised non-contextuality is the property
that operational equivalence implies ontological identity.

\nicole{
Generalised non-contextuality relates to Kochen--Specker contextuality (outlined above) as follows.  Consider a sharp measurement represented, in quantum theory, by the projector-valued measures 
$\mathcal{M} = \{ \hat{\Pi}_k \}$.  
\nicole{Sharp measurements are commonly assumed to obey outcome determinism in classical theories~\cite{spekkens2005contextuality}. That is, the ontic state $\lambda$ is assumed to dictate a measurement's outcome deterministically:}
$P_{\mathcal{M}}(k|\lambda)  \in \{ 0,1\}$.  
Consider sharp measurements whose projector-valued measures, $\mathcal{M}_1$ and $\mathcal{M}_2$, share a projector $\hat{\Pi}_{k^{\star}}$: 
$\hat{\Pi}_{k^{\star}} \in  \mathcal{M}_1 \cap \mathcal{M}_2$.  
By the above definitions of generalised measurement non-contextuality, 
\begin{equation}
P_{\mathcal{M}_1}(k^{\star}|\lambda) = P_{\mathcal{M}_2}(k^{\star}|\lambda) .
\end{equation}
As $k^{\star}$ labels a projector $\hat{\Pi}_{k^{\star}}$,  outcome determinism ensures that \cite{Lostaglio_quantum_2018}
\begin{equation}
P_{\mathcal{M}_1}(k^{\star}|\lambda) = P_{\mathcal{M}_2}(k^{\star}|\lambda) \in \{ 0, 1\} .
\end{equation}
The first equality implies that the probability of observing $k^{\star}$ does not depend on the context within which the projector is measured (does not depend on whether $\mathcal{M}_1$ or $\mathcal{M}_2$ is measured).   The second equality states that the $k^\star$ outcome's probability equals $0$ or $1$, depending on the ontic state $\lambda$.  
The present problem satisfies the definition of Kochen--Specker non-contextuality \Dave{[Eq. \eqref{eq_P_lambda},  etc.].}
To summarise,  under the assumption of outcome determinism and sharp measurements,  generalised measurement non-contextuality implies Kochen--Specker non-contextuality. 
}

From an experimental viewpoint, operational equivalence need not be checked exactly. In fact, they cannot be.
Fortunately, contextuality proofs can accommodate experimental imperfections, e.g.,~\cite{schmid2018contextual, lostaglio2020contextual, kunjwal2019anomalous}. 
Experimental contextuality tests must 
check
operational equivalences for a tomographically complete set of preparations and measurements. To achieve this complete testing
one can  rely on quantum mechanics to tell us what these tomographically complete sets are. Alternatively, we can perform theory-agnostic generalised-probabilistic-theory tomography~\cite{mazurek2016experimental,mazurek2021experimentally}. For other questions related to generalised contextuality and its experimental testing, see Ref.~\cite{schmid2023addressing}. 

If an experiment $\{P(k|{\mathscr{P}},{\mathscr{T}},{\mathscr{M}}) \}_{{\mathscr{P}},{\mathscr{T}},{\mathscr{M}}}$ is \emph{contextual}, 
no non-contextual ontological model satisfies Eq.~\eqref{eq:ontologicalmodelprediction}. 
Contextuality has been identified in (specific aspects of) experiments that involve state discrimination~\cite{schmid2018contextual,flatt2022contextual}, approximate cloning~\cite{lostaglio2020contextual}, uncertainty relations~\cite{catani2022nonclassical}, interference~\cite{flatt2022contextual}, randomness certification~\cite{i2022quantum} and more. However, we focus  on the link between contextuality and quasi-probabilities---specifically, the KD distribution.
 
A positive quasi-probability representation of quantum mechanics (or of a subtheory of quantum mechanics) assigns to each density operator $\hat{\rho}$, quantum channel $\mathcal{C}$ and positive-operator-valued measure $\mathcal{M}= \{ \hat{M}_k \}$   probability distributions $P_{\hat{\rho}}(\lambda)$, $P_{\mathcal{C}}(\lambda'|\lambda)$ and $P_{\mathcal{M}}(k|\lambda)$ over a measurable space $\Lambda$~\cite{Ferrie2011}. Density operators, channels and positive-operator-valued measures can be understood as equivalence classes of preparation procedures, transformations and measurements. (The \emph{equivalence} refers to the above sense of operational equivalence.) 
Hence, two conditions are equivalent: The existence of a non-contextual ontological model for an experiment and the existence of a positive quasi-probability representation of the experiment~\cite{spekkens2008negativity}.  
In other words, contextuality is equivalent to the negativity or non-reality of \emph{every} quasi-probability representation of the experiment. 
A particular quasi-probability's negativity
does not suffice for proving contextuality, generally. (However, the condition can suffice in some subtheories of quantum mechanics~\cite{schmid2022uniqueness}.) Nevertheless, there exist surprisingly direct connections between the KD distribution and generalised contextuality.

Suppose that 
a KD distribution has negative real part:
 \begin{equation}\label{eq:work_KD_2}
    \textrm{Re}\,  \bm{(} Q_{j,k}(\hat{\rho}) \bm{)} 
    = \textrm{Re} \bm{(} \Tr( \hat{\Pi}_k^{(B)} \hat{\Pi}_j^{(A)} \hat{\rho} ) \bm{)} <0 .
\end{equation}
The following sequential-measurement scheme (the weak-value experiment described in Sec.~\ref{SubSec:WVDefs}) is then contextual~\cite{Pusey14, kunjwal2019anomalous}:
weakly measure $\{ \hat{\Pi}_j^{(A)} \}$ 
(measuring the meter's position eigenbasis),  
then strongly measure $\{\hat{\Pi}_k^{(B)}\}$.
No non-contextual ontological model can explain the set $\{P(k|{\mathscr{P}},{\mathscr{T}},{\mathscr{M}})\}_{{\mathscr{P}},{\mathscr{T}},{\mathscr{M}}}$ of outcome statistics 
(including the statistics that
support the operational equivalences)~\cite{Pusey14}. 
Therefore, 
KD negativity implies the negativity of every quasi-probability representation of this experiment. 
Similarly, suppose that $\textrm{Im}\, \bm{(} Q_{j,k}(\hat{\rho}) \bm{)} \neq0$. An analogous scheme, in which the meter's momentum eigenbasis is measured, is contextual~\cite{kunjwal2019anomalous}. 
Even if a continuous meter's outcomes are coarse-grained
(even if a qubit replaces the meter), the results hold~\cite{kunjwal2019anomalous}.  
In conclusion, KD non-positivity and contextuality characterise an experiment that reports an anomalous weak value~\cite{Pusey14, kunjwal2019anomalous}.

The KD distribution also witnesses contextuality in the context of work extraction (and injection), in the linear-response regime discussed in Sec.~\ref{Sec:QThermo}~\cite{Lostaglio_certifying_2020}. We have already mentioned that the work scales as~$O(g)$ only when a the imaginary part of 
an average with respect to a KD distribution
is non-zero. Otherwise, the work scales as $O(g^2)$. More precisely,
\begin{equation}
    \langle W \rangle = \frac{2 g \tau }{\hbar} \mathrm{Im}\left( \Tr\bm{(}H_0 \bar{V_I}(\tau) \hat{\rho} \bm{)}\right) + O\left(g^2\right)\,.
\end{equation}
This expression connects the work's $O(g)$ behaviour with a KD quantity's  being non-zero: $\mathrm{Im}\left(  \Tr\bm{(}H_0 \bar{V_I}(\tau) \hat{\rho} \bm{)}\right) \neq 0$.
This result says nothing directly about contextuality. However, a work-extraction transformation ${\mathscr{T}}$ (a unitary in quantum mechanics), in the presence of an extra condition called \emph{stochastic reversibility}, can generate $O(g)$ behaviour only if the experiment is  contextual~\cite{Lostaglio_certifying_2020}. Stochastic reversibility is the condition under which one can reverse a transformation ${\mathscr{T}}$, up to first order in $g$, via probabilistic mixture with another operation ${\mathscr{T}}^*$. That is, there exist transformations ${\mathscr{T}}^*$ and ${\mathscr{T}}'$ such that
\begin{equation}
    \frac{1}{2}{\mathscr{T}} + \frac{1}{2}{\mathscr{T}}^* \sim (1-p_d) I + p_d {\mathscr{T}}',
\end{equation}
where $I$ is the trivial (identity) transformation and $p_d=O(g^2)$.
Qubit systems always satisfy the above operational equivalence, according to quantum mechanics.

Complementary to work is heat.
As discussed in 
\nicole{Sec.}~\ref{Sec:QThermo}, KD distributions 
satisfy many desirable criteria for distributions characterising measurements of fluctuating work, heat and entropy. Moreover, we have seen that KD non-positivity witnesses contextuality in experiments that 
report anomalous weak values. The connection between such experiments and contextuality has led to theoretical bounds on the average heat~\cite{Levy_quasiprobability_2020} and average work~\cite{Hernandez-Gomez_projective_2022} that can flow during a sequential-measurement protocol admitting of a non-contextual ontological model. These bounds can be broken only if the sequential-measurement process is contextual.
Like heat and work, entropy production is a thermodynamic quantity that fluctuates from trial to trial. A notion of stochastic entropy production can be defined via KD distributions~\cite{Upadhyaya_what_2023}. A non-real stochastic entropy production signals the contextuality of a 
thermodynamic process that involves sequential measurements~\cite{Upadhyaya_what_2023}.

\section{Mathematical structure of the KD-positive states and properties of KD non-positivity}
\label{Sec:NonPos}

If  $Q \left( \hat{\rho} \right)$  has neither negative nor non-real components,  it is  a classical probability distribution.  A KD distribution's ability to assume negative and non-real values  allows the distribution to describe quantum experiments that cannot be described and analysed with classical probability theories. Recall that we refer to negative or non-real values simply as  \emph{non-positive}, and that we call $Q \left( \hat{\rho} \right) $ \emph{positive} if all its entries are positive or zero. Consider a KD distribution  defined with respect to two operators's eigenbases. For the KD distribution to contain non-positive values, a necessary but insufficient condition is that the operators do not commute~\cite{ArvidssonShukur2021}. 
Thus, KD non-positivity can be seen as a property related to, but stricter than, non-commutation.

As we saw in previous sections, experiments described by non-positive KD distributions can often generate data that cannot 
be generated via sampling from classical joint probability distributions. It is, therefore,  important to understand when a KD distribution is non-positive. In this section, we summarise the literature on this topic.

\subsection{KD positivity}

Consider a $d$-dimensional Hilbert space $\Hcal$, as well as $k$ projective positive-operator-valued measures $\mathcal{M}^{(l)} \equiv \{ \hat{M}_{i_l}^{(l)} \}$, where $l = 1,2,\ldots,k $. We call a  state $\hat{\rho}$ KD-positive if, for all  indices, $Q_{i_1,\ldots, i_k}(\hat{\rho}) \geq 0$.  In this case, $Q(\hat\rho)$ defines a probability distribution over the set of  indices.   KD positivity of $\hat\rho$ depends on the choice of  positive-operator-valued measures $\mathcal{M}^{(l)} $. In what follows, we will not indicate this dependence, the choice being clear from the context.  

We denote the set of all KD-positive states by $\EcalKDC$.
$\EcalKDC$ is a convex, bounded, closed set. We denote by $\EcalKDCext$ the set of  extreme points\footnote{A convex set's extreme points are the points that do not lie on any open line segment  that connects two points in the set.} of $\EcalKDC$. Further, we denote the set of  pure KD-positive states by $\EcalKDCpu$. We denote by $\conv{\mathcal{E}}$ the set of all convex combinations of the elements of a set $\mathcal{E}$.  A convex combination is defined as a weighted finite  sum of elements  $ \hat{\rho}_j\in\mathcal E$, of the form $\sum_j \alpha_j \hat{\rho}_i$, such that $\sum_j \alpha_j =1$ and $\alpha_j \ge 0$. One has that $\EcalKDCpu \subseteq \EcalKDCext$, and thus
\begin{equation}
\conv{\EcalKDCpu} \subseteq \conv{\EcalKDCext}=\EcalKDC .
\end{equation}
The last equality follows from the Krein-Milman theorem~\cite{hiriart-urrutylemarechal2001}.

The following  lemma links positivity of the KD distribution of all states 
to commutativity of the  positive-operator-valued measures $\mathcal{M}^{(l)}  = \{ \hat{M}_{i_l}^{(l)} \}$ used in the distribution's construction. 
 \begin{Lemma}\label{lem:pos_comm}
 The two following statements are equivalent:
 \begin{itemize}
 	\item For all quantum states $\hat{\rho}$,  the KD distribution is positive: $Q(\hat{\rho})\geq 0$.
 	\item All positive-operator-valued-measure operators pairwise commute: $\left[ \hat{M}^{(l)}_{i_l} ,  \,  \hat{M}^{(l^{\prime})}_{i_{l^{\prime}}}  \right] = 0$ for all $l,l^{\prime}$ and all ${i_l}, {i_{l^{\prime}}}$. 
 \end{itemize}
 \end{Lemma}
\nicole{The lemma prevents any state $\hat{\rho}$ from having
negative or non-real 
KD quasiprobabilities 
when all the positive operators $\hat M_{i_l}^{(l)}$  commute pairwise. }
By contraposition, the lemma also guarantees the existence of a quantum state $\hat{\rho}$ for which $Q(\hat{\rho})$ is not a probability distribution, when there exists  at least one pair $(l,l^{\prime})$ and one pair $(i_l,i_{l^{\prime}})$ such that $\hat{M}_{i_l}^{(l)}$ and $\hat{M}_{i_{l^{\prime}}}^{(l^{\prime})}$ do not commute. So non-commutativity of measurement operators, a typical quantum feature linked to incompatibility~\cite{Designolle_2019}, is a prerequisite for the existence of at least one state $\hat{\rho}$ that is not KD-positive. Further links between incompatibility and non-commutativity of observables, uncertainty of states with respect to observables, and non-positivity of the KD distribution are elaborated upon in Sec.~\ref{subsec:KDposwitn} (see~\cite{ArvidssonShukur2021, DeBievre2021, debievre2023}).
\begin{proof}
To begin, we prove that the lemma's first point implies the second point.
For ease of notation, and without loss of generality, we focus on  $l=1$ and $l^{\prime}=2$. Since $Q(\hat\rho) \geqslant 0$ for all quantum states $\hat{\rho}$, 
\begin{equation}
 \forall (i_1,i_2), \;  \mathrm{Tr} \left( \hat{M}_{i_2}^{(2)}\hat{M}_{i_1}^{(1)}\hat{\rho} \right) \geqslant 0 .
\end{equation}
This statement governs 
all pure states, in particular. Therefore, $\hat{M}_{i_2}^{(2)}\hat{M}_{i_1}^{(1)} \geqslant 0$ for all $ (i_1,i_2)$. Consequently, $\hat{M}_{i_2}^{(2)}\hat{M}_{i_1}^{(1)}$ is self-adjoint. Thus,
\begin{equation}
\hat{M}_{i_1}^{(1)} \hat{M}_{i_2}^{(2)} = \hat{M}_{i_2}^{(2)} \hat{M}_{i_1}^{(1)} ,
\end{equation}
concluding the proof.

Now, we prove that the lemma's second point implies the first. Recall that, if $\hat{X}$ and $\hat{Y}$ are non-negative operators (if $\hat{X} \geqslant 0$ and $\hat{Y} \geqslant 0$) and if they commute, then $\hat{X} \hat{Y} \geqslant 0$.  Therefore, under the hypothesis of the lemma's second point, for all $(i_1, \dots, i_k)$,  $\hat{M}_{i_k}^{(k)}\cdots \hat{M}_{i_1}^{(1)} \geqslant 0$. 
 \end{proof}

\subsection{Characterising the set of KD-positive states}
\label{sec:CarKD+}

A precise, simple characterisation of the KD-positive states is not easy to come by. Most efforts have 
focused on the standard KD distribution [Eq. \eqref{Eq:KDStand2}], for which $k=2$.  We now discuss these efforts in some detail. We further suppose that the measurement operators 
$\hat{M}_{i_l}^{(l)}$,
with $l=1,2$, form complete sets 
$\big\{ \hat{\Pi}_{i_l}^{(l)} \big\}$
of 1-dimensional orthogonal projectors in a $d$-dimensional Hilbert space. In other words, there exist two orthonormal bases, $\{ \cket{a_i} \}$ and$ \{ \cket{b_j} \}$, such that $\hat{\Pi}_{i}^{(1)} = \ketbra{a_i}{a_i}$ and 
$\hat{\Pi}_{j}^{(2)} = \ketbra{b_j}{b_j}$.
For simplicity, we set $i\equiv i_1$ and $j \equiv i_2$.  We  introduce
\begin{equation}
    \Acal=\{ \ketbra{a_i}{a_i} \} \quad \mathrm{and} \quad   \Bcal=\{ \ketbra{b_j}{b_j} \}.
\end{equation}
The union of the 
projectors onto the basis vectors forms a subset  of the pure KD-positive states, which 
form a subset of the extremal KD-positive states:
\begin{equation}
\label{eq:Extincl}
\Acal\cup\Bcal\subseteq \EcalKDCpu\subseteq \EcalKDCext.
\end{equation}
Indeed, one can straightforwardly check that all the basis states are  KD-positive. In view of Eq.~\eqref{eq:Extincl}, in the simplest situation,
\begin{equation}\label{eq:Extegal}
\Acal\cup\Bcal = \EcalKDCpu = \EcalKDCext.
\end{equation}
In this situation, $\EcalKDC=\convAB$, and $\EcalKDC$ forms  a  polytope.  

We define a unitary transition matrix $\hat{U}$ whose entries are $\hat{U}_{i,j} \equiv \bracket{a_i}{b_j}$. The following result sums up multiple situations
in which Eq.~\eqref{eq:Extegal} holds~\cite{langrenez2023characterizing}.
\begin{Theorem} 
Suppose that $\mab \equiv \min_{i,j} \left|\bracket{a_i}{b_j}\right|> 0$. Equation~\eqref{eq:Extegal} is true under any one of the following conditions:
\begin{enumerate}
	\item $d=2$.
	\item $d$ is a prime number, and the transition matrix $\hat{U}$ equals the discrete-Fourier-transform matrix.
	\item In any dimension $d$, for a set of transition matrices  $\hat{U}$ that has probability  $1$  in the set of all unitary matrices. 
\end{enumerate}
\end{Theorem}

In the three scenarios covered by the theorem, the only extreme KD-positive states are the pure KD-positive states and the only pure KD-positive states are the basis states belonging to $\Acal$ or $\Bcal$.  A more detailed statement, as well as proofs of parts (1) and (2), appear in~\cite{langrenez2023characterizing}. There, part (3) of the theorem is conjectured. The proof of part (3) appears in~\cite{ADBLSLT24}.

The theorem's condition $\mab>0$ can be interpreted as a weak form of incompatibility for the measurements associated with $\Acal$ and $\Bcal$~\cite{debievre2023}.  As mentioned above, $\mab>0$ also implies that the KD distribution is informationally complete: It determines $\hat{\rho}$ uniquely [see Eq.~\eqref{Eq:StateDecom}]. 

\begin{figure}
\begin{center}
\begin{tikzpicture}[scale=1.05]
	\fill[color=white!85!black] (2,0) -- ({sqrt(4-1.2*1.2)},1.2) --({-sqrt(4-1.4*1.4)},1.4)-- (-2,0)-- (-{sqrt(4-1.5*1.5)},-1.5)--({sqrt(4-1.3*1.3)},-1.3)--(2,0);
	\draw[color=white!85!black] (2,0) -- ({sqrt(4-1.2*1.2)},1.2) --({-sqrt(4-1.4*1.4)},1.4)-- (-2,0)-- (-{sqrt(4-1.5*1.5)},-1.5)--({sqrt(4-1.3*1.3)},-1.3)--(2,0);
	\node at (0,0) {$\convAB$};
	\draw (0,0) circle (2cm);
	\fill (2,0) circle (0.05); 
	\node[right] at (2,0){$\cket{a_1}\bra{a_1}$};
	\fill ({sqrt(4-1.2*1.2)},1.2) circle (0.05); 
	\node[right] at  ({sqrt(4-1.2*1.2)},1.2){$\cket{a_2}\bra{a_2}$};
	\fill ({sqrt(4-1.3*1.3)},-1.3) circle (0.05); 
	\node[right] at ({sqrt(4-1.3*1.3)},-1.3){$\cket{a_3}\bra{a_3}$};
	\fill (-2,0) circle (0.05); 
	\node[left] at (-2,0){$\cket{b_1}\bra{b_1}$};
	\fill ({-sqrt(4-1.4*1.4)},1.4) circle (0.05); 
	\node[left] at ({-sqrt(4-1.4*1.4)},1.4){$\cket{b_2}\bra{b_2}$};
	\fill (-{sqrt(4-1.5*1.5)},-1.5) circle (0.05); 
	\node[left] at (-{sqrt(4-1.5*1.5)},-1.5){$\cket{b_3}\bra{b_3}$};
	\fill (-0.2,-1.7) circle (0.05); 
	\node[right] at  (-0.2,-1.85){$D$};
	\fill (0,2) circle (0.05); 
	\node[above] at  (0,2){$C$};
\end{tikzpicture}
\hspace{1cm}
\begin{tikzpicture}[scale=1.05]
	\fill[pattern=horizontal lines, pattern color = orange!80!white] (2,0) -- ({sqrt(4-1.2*1.2)},1.2)--(0,2)--({-sqrt(4-1.4*1.4)},1.4)-- (-2,0)-- (-{sqrt(4-1.5*1.5)},-1.5)--({sqrt(4-1.3*1.3)},-1.3)--(2,0);
	\draw[color=orange!80!white] (2,0) -- ({sqrt(4-1.2*1.2)},1.2)--(0,2)--({-sqrt(4-1.4*1.4)},1.4)-- (-2,0)-- (-{sqrt(4-1.5*1.5)},-1.5)--({sqrt(4-1.3*1.3)},-1.3)--(2,0);
	\node at (0,0) {$\conv{\EcalKDCpu}$};
	\draw (0,0) circle (2cm);
	\fill (2,0) circle (0.05); 
	\node[right] at (2,0){$\cket{a_1}\bra{a_1}$};
	\fill ({sqrt(4-1.2*1.2)},1.2) circle (0.05); 
	\node[right] at  ({sqrt(4-1.2*1.2)},1.2){$\cket{a_2}\bra{a_2}$};
	\fill ({sqrt(4-1.3*1.3)},-1.3) circle (0.05); 
	\node[right] at ({sqrt(4-1.3*1.3)},-1.3){$\cket{a_3}\bra{a_3}$};
	\fill (-2,0) circle (0.05); 
	\node[left] at (-2,0){$\cket{b_1}\bra{b_1}$};
	\fill ({-sqrt(4-1.4*1.4)},1.4) circle (0.05); 
	\node[left] at ({-sqrt(4-1.4*1.4)},1.4){$\cket{b_2}\bra{b_2}$};
	\fill (-{sqrt(4-1.5*1.5)},-1.5) circle (0.05); 
	\node[left] at (-{sqrt(4-1.5*1.5)},-1.5){$\cket{b_3}\bra{b_3}$};
	\fill (-0.2,-1.7) circle (0.05); 
	\node[right] at  (-0.2,-1.85){$D$};
	\fill (0,2) circle (0.05); 
	\node[above] at  (0,2){$C$};
\end{tikzpicture}
\end{center}
\begin{center}
\begin{tikzpicture}[scale=1.05]
	\fill[pattern=dots, pattern color=green!80!black] (2,0) -- ({sqrt(4-1.2*1.2)},1.2) --(0,2)--({-sqrt(4-1.4*1.4)},1.4)-- (-2,0)-- (-{sqrt(4-1.5*1.5)},-1.5)--(-0.2,-1.7)--({sqrt(4-1.3*1.3)},-1.3)--(2,0);
	\draw[color=green!80!black](2,0) -- ({sqrt(4-1.2*1.2)},1.2) --(0,2)--({-sqrt(4-1.4*1.4)},1.4)-- (-2,0)-- (-{sqrt(4-1.5*1.5)},-1.5)--(-0.2,-1.7)--({sqrt(4-1.3*1.3)},-1.3)--(2,0);
	\node at (0,0) {$\EcalKDC$};
	\draw (0,0) circle (2cm);
	\fill (2,0) circle (0.05); 
	\node[right] at (2,0){$\cket{a_1}\bra{a_1}$};
	\fill ({sqrt(4-1.2*1.2)},1.2) circle (0.05); 
	\node[right] at  ({sqrt(4-1.2*1.2)},1.2){$\cket{a_2}\bra{a_2}$};
	\fill ({sqrt(4-1.3*1.3)},-1.3) circle (0.05); 
	\node[right] at ({sqrt(4-1.3*1.3)},-1.3){$\cket{a_3}\bra{a_3}$};
	\fill (-2,0) circle (0.05); 
	\node[left] at (-2,0){$\cket{b_1}\bra{b_1}$};
	\fill ({-sqrt(4-1.4*1.4)},1.4) circle (0.05); 
	\node[left] at ({-sqrt(4-1.4*1.4)},1.4){$\cket{b_2}\bra{b_2}$};
	\fill (-{sqrt(4-1.5*1.5)},-1.5) circle (0.05); 
	\node[left] at (-{sqrt(4-1.5*1.5)},-1.5){$\cket{b_3}\bra{b_3}$};
	\fill (-0.2,-1.7) circle (0.05); 
	\node[right] at  (-0.2,-1.85){$D$};
	\fill (0,2) circle (0.05); 
	\node[above] at  (0,2){$C$};
\end{tikzpicture}
    \caption{  \textbf{Geometry of the set of KD-positive states.}
    Schematic representation of the situation in which $\Acal\cup\Bcal\subsetneq \EcalKDCpu\subsetneq \EcalKDCext$ and the Hilbert space is 3-dimensional. The black circle represents the set of pure quantum states. The point $C$ represents a pure KD-positive state different from the basis states. The point $D$ represents a mixed extreme state of $\EcalKDC$.  The (grey) shaded area represents the set $\convAB$. The (orange) horizontally hatched area represents the set $\conv{\EcalKDCpu}$. The (green) dotted area represents the set $\EcalKDC$.}
    \label{fig:inclusions}
\end{center}
\end{figure}
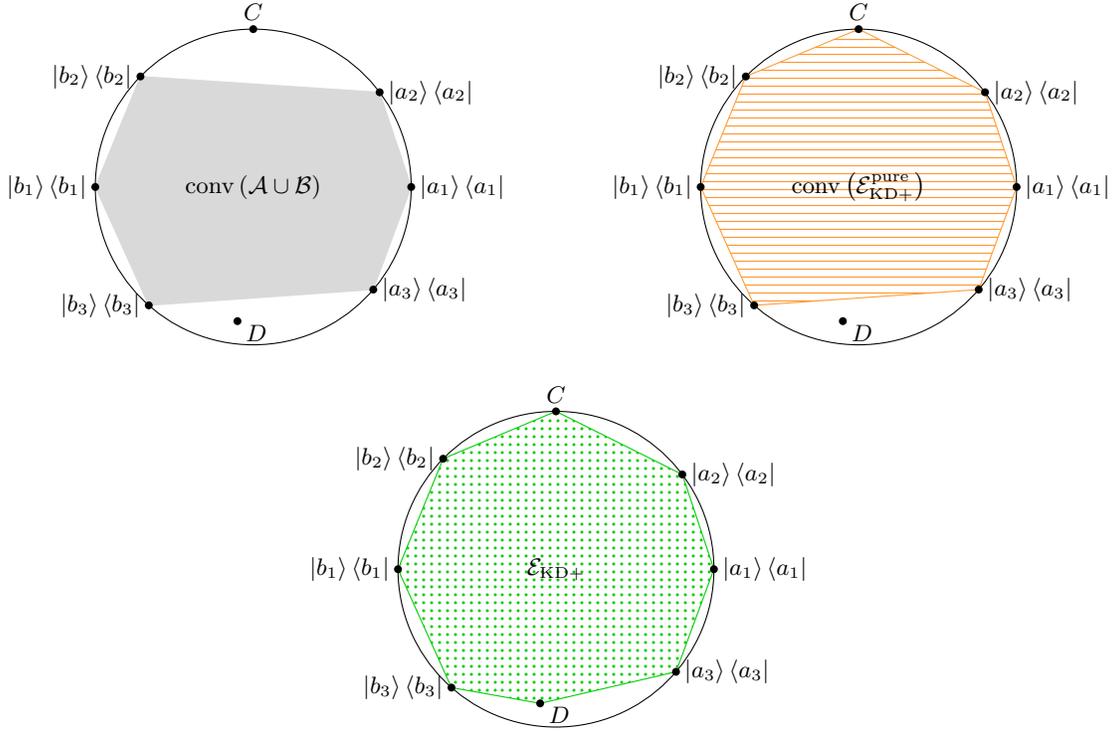

In specific cases, the structure of $\EcalKDC$ as a convex set can be considerably more complicated than when the equalities of Eq.~\eqref{eq:Extegal} hold.
Figure~\ref{fig:inclusions} shows a schematic representation of the situation where 
\begin{equation}\label{eq:ineq}
\Acal\cup\Bcal\subsetneq \EcalKDCpu\subsetneq \EcalKDCext \, .
\end{equation}
In this case, some pure KD-positive states are not basis states, and some extreme KD-positive states are mixed.
The latter type of states are KD-positive, yet cannot be expressed as convex combinations of KD-positive pure states.  Such states have been used to investigate fundamental aspects of  generalised contextuality (see Sec. \ref{SubSec:GenContext}). Certain experiments that measure these states' KD distributions can be described by non-contextual models, yet the experiments can verify contextuality \cite{thio2024}.  
The situation described in Eq. \eqref{eq:ineq} arises in dimension $d=3$, e.g., for spin-1 systems: 
One can choose for $\{ |a_i\rangle \} $ the eigenbasis of the angular-momentum operator $\hat{J}_z$ in the $z$-direction and,  for $\{ |b_j\rangle \} $,  the eigenbasis of the  angular-momentum operator $\hat{J}_{\hat n}$ for an appropriate rotation axis $\hat n$~\cite{langrenez2023characterizing}. More information about $\EcalKDCpu$, 
when the transition matrix is a Hadamard matrix ($\left|\bracket{a_i}{b_j}\right|^2 = \frac{1}{d} = \mab$ for all $i,j$), appears in~\cite{DeBievre2021,debievre2023,xu2022b}.

\subsection{Measure of KD non-positivity}
\label{SubSec:KDNonclas}

If, and only if, a KD distribution is a classical joint probability distribution,   $\sum_{i_1,\ldots,i_k}  \left| Q_{i_1,\ldots,i_k} (\hat{\rho}) \right| =  1$. Thus, a popular measure  of a KD distribution's total non-positivity is
\begin{equation}
\label{Eq:TotNonPos}
\mathcal{N} \LParen Q_{i_1,\ldots,i_k} (\hat{\rho})  \RParen 
= \sum_{i_1,\ldots,i_k}  \left| Q_{i_1,\ldots,i_k} (\hat{\rho}) \right| \geq 1,
\end{equation}
as defined in~\cite{Alonso_OTOC_2019}.
We now list a few properties of this non-positivity measure. Details appear in~\cite{ArvidssonShukur2021}.
\begin{enumerate}

\item{ \textit{Positivity:} A KD distribution is a classical probability distribution if, and only if,  $\mathcal{N} \LParen Q_{i_1,\ldots,i_k} (\hat{\rho})  \RParen = 1$.}

\item{ \textit{Stricter than non-commutation:}   If $\mathcal N\LParen Q_{i_1,\ldots,i_k} (\hat{\rho})  \RParen > 1$, then $[\hat{A},\hat{B}] \neq 0$,  $[\hat{A},\hat{\rho}] \neq 0$ and  $[\hat{B},\hat{\rho}] \neq 0$. However, even if $[\hat{A},\hat{B}] \neq 0$,  $[\hat{A},\hat{\rho}] \neq 0$ and  $[\hat{B},\hat{\rho}] \neq 0$, the corresponding KD distribution may be positive, such that $\mathcal N\LParen Q_{i_1,\ldots,i_k} (\hat{\rho})  \RParen = 1$.
}
\item{ \textit{Convexity:} The total non-positivity is a convex function with respect to mixed states: 
$\mathcal{N} \LParen Q_{i_1,\ldots,i_k} \left( \sum_t p_t \hat{\rho}_t \right)  \RParen 
\leq \sum_t p_t \,  \mathcal{N} \LParen Q_{i_1,\ldots,i_k} \left(  \hat{\rho}_t \right)  \RParen $. }

\item{ \textit{Extension dependence:} A KD distribution $Q_{i_1,\ldots,i_l} (\hat{\rho})$ may follow from the marginalisation of another KD distribution, $Q_{i_1,\ldots,i_k} (\hat{\rho})$, where $k>l$. In this case, $\mathcal{N} \LParen Q_{i_1,\ldots,i_l} (\hat{\rho})  \RParen > 1$ implies that  $\mathcal{N} \LParen Q_{i_1,\ldots,i_k} (\hat{\rho})  \RParen \geq \mathcal{N} \LParen Q_{i_1,\ldots,i_l} (\hat{\rho})  \RParen >  1$. }

\item{ \textit{Coarse-graining:} 
A KD distribution $Q_{i_1,\ldots,i_k} (\hat{\rho})$ may 
represent a state $\hat{\rho}$ in terms of measurement operators that result from coarse-graining the operators in another KD distribution, $Q_{i_1^\prime,\ldots,i_k^{\prime}}^{\prime} (\hat{\rho})$, where $k^{\prime} > k$. In this case,  
$\mathcal{N} \LParen Q_{i_1,\ldots,i_k} (\hat{\rho})  \RParen
\leq  \mathcal{N} \LParen Q_{i_1^\prime,\ldots,i_{k^{\prime}}}^{\prime} (\hat{\rho})  \RParen  $.  }

\item{\textit{Maximum value:} Consider 
the KD distributions defined in terms of 
projective positive-operator-valued measures $\mathcal{M}^{(1)}, \ldots, \mathcal{M}^{(k)}$ (see Sec. \ref{SubSec:GeneralKD}).  For such KD distributions, 
the maximum value of $\mathcal{N}$ depends on both the Hilbert-space dimension $d$ and the extendedness $k$:}
\begin{equation}
\label{Eq:MaxKDNeq}
\max_{\hat{\rho},  \mathcal{M}^{(1)}, \ldots, \mathcal{M}^{(k)}} 
\left\{ \mathcal{N} \LParen Q_{i_1,\ldots,i_k} (\hat{\rho})  \RParen \right\} = d^{(k-1)/2} .
\end{equation}

\end{enumerate}

To elucidate when the maximum, Eq.~\eqref{Eq:MaxKDNeq}, attains, we review \emph{mutually unbiased bases.}
Bases $\{ \ket{a_i}\} $ and $ \{ \ket{b_j}\} $ are \emph{mutually unbiased} if 
\begin{equation}
\label{Eq:MutUnbias}
    |\braket{a_i|b_j}| = 1/\sqrt{d} \, ,
\end{equation}
for all $i,j$. For details, see Ref.~\cite{Durt10}.
Suppose that a basis $ \{  \ket{a^{(l)}_{i_l}} \} $
is mutually unbiased with respect to
$\{  \ket{a^{(l \oplus 1)}_{i_{l \oplus 1}}} \} $,
for all $l = 1, 2, \ldots, k$.
The $\oplus$ represents addition modulo $k$.
Suppose, further, that $\hat{\rho}$ is pure and that $ \braket{a^{(1)}_{i_1}|\hat{\rho}|a^{(1)}_{i_1}}  =  \braket{a^{(k)}_{i_k}|\hat{\rho}|a^{(k)}_{i_k}}  = 1/d $ for all $i_1$ and $i_k$.
This situation achieves the maximum Eq.~\eqref{Eq:MaxKDNeq}. 
Conceptually, the neighbouring vectors in Eq.~\eqref{Eq:ExtKD} should be mutually unbiased.

\subsection{KD non-positivity and coherence}
\label{SubSec:KDCohere}

\nicole{Quantum superpositions play an important role in the study of non-classicality. Furthermore, superpositions accompany coherence.}
Therefore, we will sketch a connection between KD non-positivity and quantum coherence.

\nicole{We begin with background information about coherence.}
Consider a basis $\{ \ket{a_i} \}$.  In terms of it, we can express a quantum state $\hat{\rho}$ as 
\begin{equation}
\hat{\rho} = \sum_{i,j} \rho_{i,j} \ketbra{a_i}{a_j} ,
\end{equation}
wherein $\rho_{i,j} \equiv \braket{a_i | \hat{\rho}| a_j}$.  $\hat{\rho}$ is incoherent with respect to the basis $\{ \ket{a_i} \}$ if $\rho_{i,j}=0$ for all $i\neq j$.  If $\rho_{i,j}\neq 0$ for any $i \neq j$, then $\hat{\rho}$ is coherent with respect to $\{ \ket{a_i}\}$.  A popular measure of quantum coherence is based on the $\ell_1$ norm.  \nicole{Denote by $\mathcal{I}_{\{\ket{a_i}\}}$ the set of all $\{ \ket{a_i}\}$-incoherent states.  }  
The $\ell_1$ coherence  of $\hat{\rho}$ with respect to $\{ \ket{a_i} \}$ is
\begin{equation}
C_{\ell_1} \left( \hat{\rho}; \{ \ket{a_i} \}\right) 
\equiv \min_{\hat{\sigma} \in \mathcal{I}_{\{\ket{a_i}\}}} 
\left\{  || \hat{\rho} - \hat{\sigma}||_{\ell_1}  \right\} = \sum_{i\neq j} |\rho_{i,j}| .
\end{equation}

The KD distribution $Q_{i,j} (\hat{\rho}) = \braket{b_j|a_i}\braket{a_i|\hat{\rho}|b_j}$ quantifies coherence  \cite{Budiyono23, budiyono2023characterizing, Budiyono_2023}.  Consider \nicole{summing the absolute values of the quasi-probabilities imaginary components, then maximising}
over all bases $\{\ket{b_j}\}$:
\begin{equation}
C_{\mathrm{KD}} \left( \hat{\rho}; \{ \ket{a_i} \}\right)  
\equiv \max_{\{\ket{b_j}\}}  
\left\{ \mathcal{N} \LParen  \Im Q (\hat{\rho}) \RParen  \right\} 
= \max_{\{\ket{b_j}\}} \sum_{i,j} 
\left\{  | \Im   \braket{b_j|a_i}\braket{a_i|\hat{\rho}|b_j} |  \right\} .
\end{equation}
$C_{\mathrm{KD}} \left( \hat{\rho}; \{ \ket{a_i} \}\right)$ witnesses coherence faithfully: $C_{\mathrm{KD}} \left( \hat{\rho}; \{ \ket{a_i} \}\right) = 0$ if and only if $\hat{\rho}$ is incoherent in $\{ \ket{a_i}\}$. References~\cite{Budiyono23, budiyono2023characterizing, Budiyono_2023} demonstrate further relationships between  $C_{\mathrm{KD}} \left( \hat{\rho}; \{ \ket{a_i} \}\right)$ and coherence. 

Another KD distribution is more directly related to $C_{\ell_1} \left( \hat{\rho}; \{ \ket{a_i} \}\right)$. Consider the extended KD distribution $Q^{\star}_{i,j,k}(\hat{\rho}) = \braket{a_j|b_k} \braket{b_k|a_i}\braket{a_i|\hat{\rho}|a_j}$, wherein $\{ \ket{b_k}\}$ is mutually unbiased with respect to $\{ \ket{a_i} \}$: $|\braket{a_i|b_k}|=1/\sqrt{d}$, for all $i$ and $j$.  
\nicole{The total non-positivity [Eq. \eqref{Eq:TotNonPos}] minus 1 equals the $\ell_1$ measure of coherence:}
\begin{equation}
\mathcal{N} \LParen Q^{\star}(\hat{\rho}) \RParen -1 
= -1 + \sum_{i,j,k} |  \braket{a_j|b_k} \braket{b_k|a_i}\braket{a_i|\hat{\rho}|a_j} | =  -1 + \sum_{i,j} | \braket{a_i|\hat{\rho}|a_j} | = C_{\ell_1} \left( \hat{\rho}; \{ \ket{a_i} \}\right) .
\end{equation}

\subsection{KD-positivity witnesses, uncertainty and (complete) incompatibility}
\label{subsec:KDposwitn}

As pointed out in Sec.~\ref{SubSec:Useage}, whether a KD distribution is positive or non-positive can be of operational importance.  A witness of this property is the total non-positivity, defined in Eq.~\eqref{Eq:TotNonPos}.
A state  $\hat{\rho} $ is KD-positive ($\hat{\rho} \in \EcalKDC$) if and only if $\mathcal{N} \left[ Q_{i_1,\ldots,i_k} (\hat{\rho}) \right]=1$. Thus,  
$\mathcal{N} \LParen Q_{i_1,\ldots,i_k} (\hat{\rho})  \RParen $ 
faithfully witnesses KD non-positivity, much as the Wigner negative volume 
faithfully witnesses Wigner non-positivity~\cite{kenfackzyczkowski2004}. However, to compute $\mathcal{N} \LParen Q_{i_1,\ldots,i_k} (\hat{\rho})  \RParen $, one must know the full KD distribution. One 
might therefore seek a witness whose computation requires less information.  Below, we review such witnesses.

\setkeys{Gin}{draft =false}
\begin{figure}
    \centering
    \includegraphics[scale=0.5]{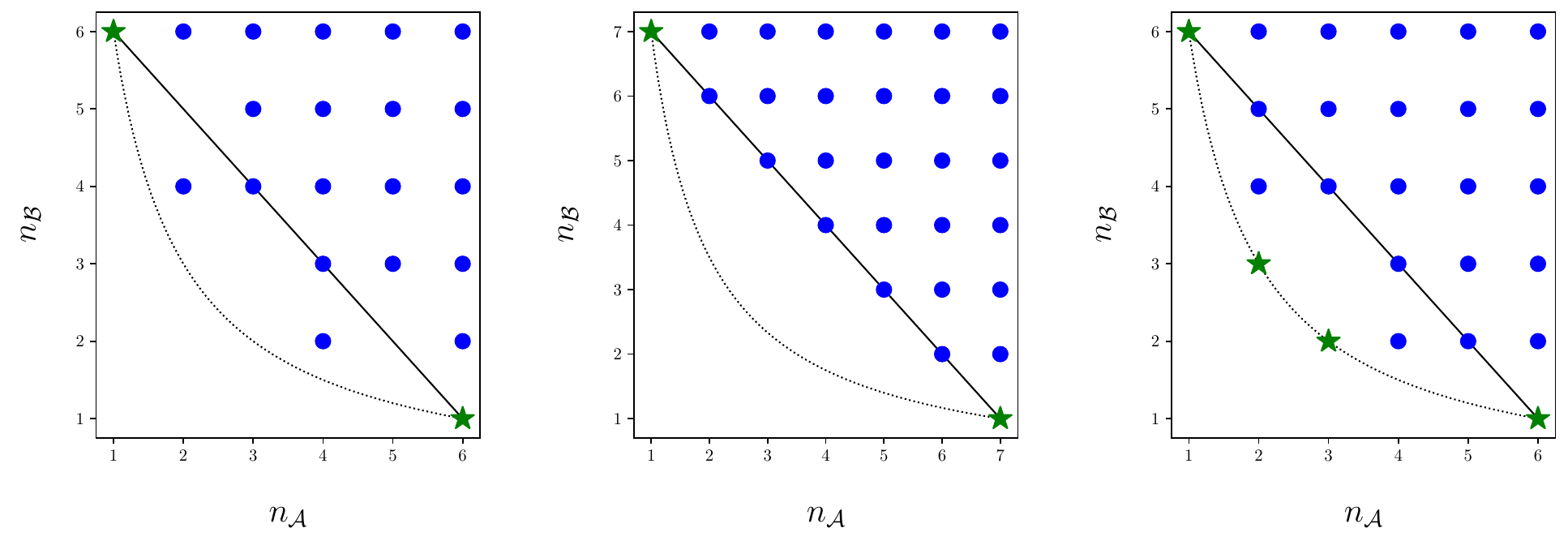}
    \caption{ \textbf{Uncertainty diagrams for three mutually unbiased bases.} Solid line: $\na (\psi) + \nb (\psi) =d+1$. Dotted curve: 
    $\na(\psi)\nb(\psi)=  d$. Blue dots correspond to KD-non-positive states, and green stars correspond to KD-positive states.  The left diagram is for the Tao matrix~\cite{Tao04} in dimension 6; the middle diagram is for the discrete-Fourier-transform matrix in dimension 7; and the right diagram is for the discrete Fourier-transform matrix in dimension 6. 
    }
    \label{Fig:UncDiag}
\end{figure}
\setkeys{Gin}{draft}

From a theoretical perspective, an interesting  witness is a state's support uncertainty, which we now introduce and study further.   We, again, restrict our analysis to the standard KD distribution with respect to orthonormal bases $\{ \ket{a_i} \}$ and $\{ \ket{b_j} \}$.  Let $\sharp X$ denote the cardinality of the set $X$. Let $\cket{\psi}\in\Hcal$ denote a pure state. We define as follows the support uncertainties of $\cket{\psi}$ in the bases $\{ \ket{a_i} \}$ and  $\{ \ket{b_j} \}$:
\begin{equation}
n_{\Acal}(\psi) \equiv \sharp \left\{ i\in [1,d] \mid \bracket{a_i}{\psi}\neq 0 \right\}, 
\quad \mathrm{ and } \quad 
n_{\Bcal}(\psi) \equiv \sharp \left\{ j\in [1,d] \mid \bracket{b_j}{\psi}\neq 0 \right\}.
\end{equation}
To interpret these definitions, we imagine expanding $\cket{\psi}$ in the basis $\{ |a_i\rangle \}$
[$\{ |b_j\rangle \} $].
The number of non-vanishing expansion coefficients is
$n_\Acal(\psi)$ [$n_\Bcal(\psi)$].  
$n_\Acal(\psi)$ and $n_\Bcal(\psi)$ enable the construction of an inequality~\cite{donohostark1989} that can be viewed as an uncertainty relation:
\begin{equation}\label{eq:donstark}
    \na(\psi) \, \nb(\psi)\geq M^{-2}, 
    \quad \text{wherein} \quad
    M \equiv \max_{i,j} \left\{  |\langle a_i|b_j\rangle|  \right\} . 
\end{equation}
The more $|\psi\rangle$ is localised in the $\{ \ket{a_i} \}$ basis, the less the state must be localised in the $\{ \ket{b_j} \}$ basis, and vice versa.

Figure~\ref{Fig:UncDiag} illustrates the lower bound Ineq.~\eqref{eq:donstark} with three examples of mutually unbiased bases.  
The set of points $\bm{(} \na(\psi), \nb(\psi) \bm{)}$ in the  $\na$-$\nb$ plane maps out an uncertainty diagram for the two bases. It is known that, for any bases $\{ |a_i\rangle \} $ and $\{ |b_j\rangle \}$, any state $|\psi\rangle$ that saturates Ineq.~\eqref{eq:donstark} is KD-positive~\cite{DeBievre2021}. The reverse implication is not generally true, however~\cite{DeBievre2021}. If the bases are mutually unbiased ($M = \mab = 1 / \sqrt{d}$), nevertheless, saturating Ineq.~\eqref{eq:donstark} is equivalent to $|\psi\rangle$'s being KD-positive~\cite{Xu2021}. 
The three panels of Fig.~\ref{Fig:UncDiag} reflect this relation.
As a result of this relation, 
for mutually unbiased bases in prime dimension, the only pure KD-positive states are 
the bases' elements. For further results about the geometric structure of the set of KD-positive states, see Sec.~\ref{sec:CarKD+} and~\cite{langrenez2023characterizing}.

Let us further explore the links amongst incompatibility, uncertainty and KD non-positivity in the general case---when the bases are not necessarily mutually unbiased bases. A different, additive uncertainty relation is more pertinent than Ineq.~\eqref{eq:donstark}.
The support uncertainty~\cite{DeBievre2021} is defined by
\begin{equation}\label{eq:def}
\nab (\psi) \equiv n_{\Acal}(\psi) + n_{\Bcal}(\psi).
\end{equation}
This quantity is linked to KD~positivity through the following theorem~\cite{DeBievre2021}.
\begin{Theorem}\label{thm:supuncpure}
Suppose that $\mab \equiv \min_{i,j} \left\{ \left|\bracket{a_i}{b_j}\right|  \right\} > 0$. If $\ketbra{\psi}{\psi}\in\EcalKDC$, then $\nab (\psi) \leqslant d+1$. 
\end{Theorem}
The theorem implies an uncertainty relation: If the uncertainty $\nab (\psi) > d+1$, then the KD distribution must be non-positive. 
The bound of Theorem \ref{thm:supuncpure} holds even when $\mab=0$, such that the matrix with elements $\bracket{a_i}{b_j}$ has zeroes, if there are not too many zeroes~\cite{DeBievre2021}.   In the absence of constraints on the two bases, other than that none of the $|b_j\rangle$s are parallel to any of the $|a_i\rangle$s, 
the upper bound becomes $\nab (\psi) \leqslant \frac32 d$~\cite{ArvidssonShukur2021}. Intermediate estimates of the type  $\nab (\psi) \leqslant  d+s$, with $1\leq s\leq \frac{d}2 \, ,$ were recently proven  in~\cite{xu2022a} under suitable conditions on the structure of the transition matrix $\hat U$. Reference~\cite{ArvidssonShukur2021} encompasses also scenarios in which the KD distribution is expressed in terms of  projectors other than rank-$1$ projectors.

Theorem~\ref{thm:supuncpure} implies that the support uncertainty is a KD-positivity witness for pure states: If  $\nab (\psi)$ is large, then $\ket{\psi}$ is not KD-positive.  However, $\nab (\psi)$ is not always a faithful witness: There may exist states $|\psi\rangle$ for which $\nab(\psi)\leq d+ 1$ but that are nevertheless KD-non-positive. See the three panels of Fig.~\ref{Fig:UncDiag} for an example. The support uncertainty nevertheless has an advantage over the total non-positivity: Limited information about the pure state $\cket{\psi}$ suffices to determine the support uncertainty. 
An extension of this result to mixed states appears in~\cite{ArvDeBLan24}. The extension involves the convex roof of the support uncertainty.

Theorem~\ref{thm:supuncpure}  states  that pure KD-positive states have low support uncertainty $\nab$: No pure KD-positive state can lie above the line $\nab=d+1$.  Figure~\ref{Fig:UncDiag} illustrates this observation. Further examples appear in~\cite{DeBievre2021,debievre2023}. Theorem~\ref{thm:supuncpure}, as such, evokes quantum optics. There, the Glauber--Sudarshan function of pure states is positive only for coherent states, which minimise $(\Delta x)^2+(\Delta p)^2$ (the sum of two conjugate quadratures' squared uncertainties).  
Unlike in quantum optics, however $d+1$ does not necessarily \emph{minimise} the support uncertainty $\nab$. 
Figure~\ref{Fig:UncDiag} evidences this fact.
Nevertheless, the analogy with quantum optics can be sharpened through \emph{complete incompatibility}, introduced in~\cite{DeBievre2021,debievre2023}, to which we now turn our attention.

Two observables are said to be incompatible when they do not commute. This notion is weak: In the context of Eq.~\eqref{eq:ABobs}, it means that at least one of the elements in $\{ \ket{a_i} \}$ and one of the elements in $\{ \ket{b_j} \}$ satisfy $0<|\langle a_i|b_j\rangle|<1$. This criterion is satisfied if the bases are not permutations of one another. A slightly stronger requirement is that 
$M_{\Acal, \Bcal} \equiv \max_{ij}  \{ |\langle a_i|b_j\rangle| \} <1,$
which follows from
$m_{\Acal \Bcal} \equiv \min_{ij} \{ |\langle a_i|b_j\rangle| \} >0.$
This latter statement guarantees that all $|a_i\rangle$s are distinct from all $|b_j\rangle$s. 
The latter statement also entails that, if a measurement of $\{ \ket{a_i} \}$ yields the outcome $a_i$, then a subsequent measurement of $\{ \ket{b_j} \}$ can yield any value $b_j$. The uncertainty in this outcome is maximal for  mutually unbiased bases, for which $\Mab = \mab=\frac{1}{\sqrt d}$ [see Eq.~\eqref{Eq:MutUnbias}]. For this reason, mutually unbiased bases are sometimes called \emph{maximally incompatible}. To introduce complete incompatibility of the  bases $\{ \ket{a_i} \}$ and $\{ \ket{b_j} \}$,  we proceed as follows.

We consider measurements of $\PiAcal(S)$ and $\PiBcal(S)$. 
We regard these measurements as coarse-grained measurements of the observables $\hat{A}$ and $\hat{B}$. For \Dave{all subsets} $S, T \subset \{1,2,\dots, d\}$, we let 
\begin{equation}
\PiAcal(S)=\sum_{i\in S} \ketbra{a_i}{a_i} \quad \mathrm{and} \quad \PiBcal(T)
=\sum_{j\in T} \ketbra{b_j}{b_j} . 
\end{equation}
\Dave{In what follows, we shall designate by $\sharp S$ the number of elements belonging to $S$.}

Suppose that a system is prepared in a state $\cket{\psi}$, a measurement of the projector $\PiAcal(S)$  yields the outcome $1$, and a subsequent measurement of $\PiBcal(T)$ also yields $1$. The system is then in the non-normalised state $\PiBcal(T) \, \PiAcal(S)\cket{\psi}$. Suppose, now, that a subsequent measurement of $\PiAcal(S)$ yields the outcome $1$ with probability $1$. In other words, suppose that the measurement of $\PiBcal(T)$ has not disturbed the outcome of the previous measurement of $\PiAcal(S)$. This
condition is met only if $\PiBcal(T)\PiAcal(S) \ket{\psi} \in \PiAcal(S)\Hcal$---in other words, provided that
\begin{equation}
\label{Eq:COINCCond}
\PiAcal(S)\Hcal\cap\PiBcal(T)\Hcal\not=\{0\}.
\end{equation}
$\PiAcal(S)\Hcal$ denotes the $\PiAcal(S)$ eigenspace associated with the eigenvalue $1$. $\PiBcal(T)\Hcal$ is defined similarly. Subsequent measurements of $\PiAcal(S)$ and $\PiBcal(T)$ will then consistently yield the outcome $1$. 
Two successive measurements can be compatible in this way
even if $\PiAcal(S)$ and $\PiBcal(T)$ do not commute; see~\cite{debievre2023} for examples.

Bases $\Acal$ and $\Bcal$ are said to be completely incompatible  when the above-described situation [Eq.~\eqref{Eq:COINCCond}] is never realised~\cite{DeBievre2021, debievre2023}. That is, $\Acal$ and $\Bcal$ are are completely compatible when, for all $S,T$ with \Dave{$\sharp S+\sharp T\leq d$}, 
\begin{equation}
    \PiAcal(S)\Hcal\cap\PiBcal(T)\Hcal=\{0\}.
\end{equation}
The restriction \Dave{$\sharp S+\sharp T\leq d$} is needed since, whenever $\Dave{\sharp S+\sharp T}> d$, $\PiAcal(S)\Hcal\cap\PiBcal(T)\Hcal\not=\{0\}$ for dimensional reasons: Sufficiently coarse-grained measurements  are always  compatible in the above sense [Eq.~\eqref{Eq:COINCCond}].

For completely incompatible bases, for all $\cket{\psi}\in\Hcal$,
\begin{equation}
\nab(\psi)\geq d+1.
\end{equation}
Therefore, if the bases $\Acal$ and $\Bcal$ are completely incompatible and if $\ketbra{\psi}{\psi} \in\EcalKDC$, then $\nab(\psi)=d+1$. In other words, when the bases are completely incompatible, the KD-positive states have minimal support uncertainties. 
The middle panel of Fig.~\ref{Fig:UncDiag} realises this situation, but the other two panels do not.

There is an open, dense set of bases that are completely incompatible~\cite{debievre2023}. However, it is not  generally straightforward to determine if two given bases are  completely incompatible. For example, consider two bases whose transition matrix is the discrete Fourier transform. These bases 
are completely incompatible only when the dimension $d$ is prime~\cite{langrenez2023characterizing}. Consequently, not all mutually unbiased bases are completely incompatible. In other words, \textit{maximal} incompatibility does not imply \textit{complete} incompatibility.  
Figure~\ref{Fig:UncDiag} illustrates this phenomenon. Reference~\cite{xu2022a} proposes a further extension of the notion of complete incompatibility. In summary,  the KD distribution can be a useful tool for designing and studying notions of  incompatibility that extend beyond  non-commutation.

\section{Conclusion and outlook}

Throughout this Article,  we have provided a comprehensive review of  
use cases of the KD distribution. In Sec.~\ref{Sec:DefProp}, we 
defined the KD distribution and showed that it obeys a quasi-probabilistic version of Bayes' theorem. In Sec.\ref{SubSec:Disturb}, we showed that non-real 
KD quasi-probabilities \Dave{signal} the disturbance of measurement-outcome probabilities.
Similarly, as we outlined in Sec.~\ref{SubSec:StandardMetrology}, one’s ability to conduct quantum metrology, to learn unknown parameters encoded in a quantum state, hinges on non-real KD quasi-probabilities. In post-selected quantum metrology, one passes several quantum 
particles through a filter that distils their metrological information into the particles which pass the filter. In Sec.\ref{SubSec:PSMetrology}, we showed that the rate of information distillation could be arbitrarily large if a KD distribution has negative components. In Sec.~\ref{Sec:WeakValues}, we reviewed weak values, pre- and post-selected observable averages of quantum states. We showed that non-positive KD distributions can lead to a weak value that lies outside the measured observable's spectrum. We also reviewed how such anomalous weak values can amplify metrological sensitivity to small unknown parameters. In Sec.~\ref{Sec:StateMeas}, we introduced the continuous-variable KD distribution. Then, we reviewed how measurements of continuous-variable KD distributions have been used to directly measure quantum wavefunctions. 

In classical thermodynamics, probability distributions describe statistic work and heat exchanges. In Sec.~\ref{Sec:QThermo}, we showed how KD distributions can describe statistic exchanges in quantum thermodynamics. In Sec.~\ref{Sec:OTOC}, we introduced the out-of-time-ordered correlator (OTOC), a popular witness of many-body quantum chaos. First, we showed that the OTOC equals the average over a KD distribution. Second, whilst the OTOC struggles to distinguish information scrambling from decoherence, the KD distribution’s non-positivity can witness scrambling more reliably. 

We also summarised the KD distribution's importance in the foundations of quantum mechanics. In Sec.~\ref{SubSec:LG}, we showed how non-positive KD quasi-probabilities are required to violate Leggett--Garg inequalities (temporal Bell inequalities). In Sec.~\ref{SubSec:Consistent}, we described how the KD distribution is used in the consistent-histories interpretation of quantum mechanics. In Sec.~\ref{SubSec:Context}, we reviewed a rigorous notion of non-classicality: Contextuality. Non-positive KD quasi-probabilities can enable non-classical advantages in the operation of engines and  weak measurements.

Given the KD distribution's diverse use cases, it is unsurprising that the distribution has been subject to growing mathematical research. We summarised, in Sec.~\ref{Sec:NonPos}, the current knowledge about the KD distribution’s mathematical properties. A basic necessary, but insufficient, condition for KD non-positivity is that $ \Acal=\{ \ketbra{a_i}{a_i} \}  $  differ from $ \Bcal=\{ \ketbra{b_j}{b_j} \} $. One could further ask, when is a quantum state $\hat{\rho}$ KD-positive? As outlined in Sec.~\ref{sec:CarKD+}, 
there are several scenarios in which $\hat{\rho}$ is KD-positive 
if, and only if, $\hat{\rho}$ is a convex combination of $\ketbra{a_i}{a_i}$s and $\ketbra{b_j}{b_j}$s.
For some pairs of bases, however, KD-positivity is not equivalent to this convex-combination property. 
Often, KD non-positivity \Dave{signal}s non-classical advantages in quantum experiments. In Sec.~\ref{SubSec:KDNonclas}, we summarised properties of a KD distribution's total non-positivity. In Sec.~\ref{subsec:KDposwitn}, we showed how to construct KD non-positivity witnesses that do not require full knowledge of the KD distribution. Also, we discussed how the KD distribution relates to a quantum state's uncertainties with respect to two observables.

Research on the KD distribution is ongoing. We conclude this review by listing a few promising outlook directions. 

\begin{itemize}
    \item \textbf{Quantum metrology:} We described the relationships between discrete-variable quantum metrology and  
    non-positive KD distributions. Such relations have barely been explored in continuous-variable systems. Negative Wigner quasi-probabilities can \Dave{signal} advantages in continuous-variable metrology. An open question is whether the negativity requirement extends to KD representations of the experiments.

    \item  \textbf{Weak values:}
    The KD distribution offers a means of understanding weak values' anomalous behaviours.
    A negative KD quasi-probability is required for the conditioned average of a weakly measured observable to 
    lie outside the observable's spectrum. When is the KD distribution, or one of its extensions, the most appropriate quasi-probability distribution for characterising more-general conditioned sequences of weakened measurements? 
    For example, as the weak-measurement strength increases, the KD quasi-probabilities will transform smoothly into projective-measurement probabilities. However, the intermediate-measurement-strength regime requires more study.

    \item  \textbf{Direct measurement:} Direct measurement provides an effective characterisation of quantum states, KD distributions, processes and detectors. An experimental opportunity is to capitalise on direct measurement's advantage over standard quantum tomography 
    in applications to 
    complex quantum systems.
    Possible targets include molecules and molecular processes, the multi-particle entangled states prepared by quantum circuits and quantum materials.

    \item \textbf{Quantum thermodynamics:} Opportunities for future work include the marriage of KD distributions with conserved quantities (charges) that fail to commute with each other~\cite{Lostaglio_2017, Guryanova16, YHN16}. Such charges were overlooked for decades but engendered a growing subfield recently~\cite{majidy_noncommuting_2023}. Example charges include the $x$-, $y$-, and $z$-components of spin~\cite{Yunger_Halpern_noncommuting_2020,NYH2022,Kranzl23}. The charges' noncommutation suggests them as observables whose eigenbases can define KD distributions naturally suited to quantum thermodynamics. Initial work on applying KD distributions to noncommuting thermodynamic charges has begun (Sections~\ref{Sec:QThermo} and~\ref{SubSec:GenContext}) but merits expansion beyond currents.

    \item \textbf{Foundations of quantum mechanics:} One of the most rigorous notions of quantum phenomena is contextuality. 
    If all quasi-probabilistic representation of an experiment are non-positive, the experiment is contextual~\cite{spekkens2008negativity}.
    Investigating infinitely many quasi-probability distributions is a formidable task. Could some distributions be more important than others? Often, the KD distribution is tailored to the operations that form an experiment.
    One could imagine that KD non-positivity may propagate to other quasi-probability representations.

    \item \textbf{Mathematical properties:} As reviewed above, the geometric structure of the convex set of KD-positive states is known in many cases. Nevertheless, for several operational tasks, we lack figures of merit for determining which KD-non-positive states are the most useful. Such metrics call for development.

\end{itemize}

\section{Acknowledgements}
The authors thank Holger Hofmann for useful discussions, in particular for pointing out the relation between imaginary components of the KD distribution and measurement disturbance. The authors also thank Noah Lupu-Gladstein and Batuhan Yilmaz for sharing their data and preparing Fig.~\ref{fig:MetroThree}. Further, the authors thank Crispin Barnes for input and guidance about the consistent-histories interpretation of quantum mechanics; \nicole{Jean-Pierre Gazeau for input on frames, quantisation and dequantisation;} and Wilfred Salmon for suggestions regarding gate-based measurements of the KD distribution.

This work was supported in part by the Agence Nationale de la Recherche under grant ANR-11-LABX-0007-01 (Labex CEMPI), by the Nord-Pas de Calais Regional Council and the European Regional Development Fund through the Contrat de Projets \'Etat-R\'egion (CPER), by the CNRS through the MITI interdisciplinary programs, and by the John Templeton Foundation (award no. 62422).

JSL acknowledges the support of the Natural Sciences and Engineering Research Council of Canada (NSERC), Canada Research Chairs, the Transformative Quantum Technologies Canada First Excellence Research Fund.

DRMAS was supported by Girton College.
\medskip

\bibliography{KD_Review_bib}

\end{document}